\documentclass[desactivate]{aa}  %% para quitar la numeracion de lineas en los margenes
\usepackage{graphicx}
\usepackage{amsmath}
\usepackage[varg]{txfonts}
\usepackage{appendix}
\usepackage{placeins} % for \FloatBarrier command
\newcommand{\iram}{IRAM-30\,m}

\newcommand{\lsim}{\raisebox{-.4ex}{$\stackrel{<}{\scriptstyle \sim}$}}
\newcommand{\gsim}{\raisebox{-.4ex}{$\stackrel{>}{\scriptstyle \sim}$}}
\newcommand{\farc}{\mbox{$.\!\!^{\prime\prime}$}}

\newcommand{\mloss}{\mbox{$\dot{M}$}}
\newcommand{\mpagb}{\mbox{$\dot{M}_{\rm pAGB}$}}
\newcommand{\my}{\mbox{$M_{\odot}$~yr$^{-1}$}}
\newcommand{\ls}{\mbox{$L_{\odot}$}}
\newcommand{\msun}{\mbox{$M_{\odot}$}}

\newcommand{\rsun}{\mbox{$R_{\odot}$}}
\newcommand{\rs}{\mbox{$R_{\star}$}}

\newcommand{\lstar}{\mbox{$L_{\star}$}}

\newcommand{\rout}{\mbox{$R_{\rm out}$}}
\newcommand{\req}{\mbox{$R_{\rm eq}$}}
\newcommand{\kms}{\mbox{km\,s$^{-1}$}}

\newcommand{\vgrad}{\mbox{$\nabla \upsilon$}}
\newcommand{\vexp}{\mbox{$V_{\rm exp}$}}

\newcommand{\vrot}{\mbox{$V_{\rm rot}$}}
 
\newcommand{\vsys}{\mbox{$V_{\rm sys}$}} 
\newcommand{\vlsr}{\mbox{$V_{\rm LSR}$}} 
\newcommand{\porb}{\mbox{$P_{\rm orb}$}}

\newcommand{\h}{$^{\rm h}$}
\newcommand{\m}{$^{\rm m}$}

\newcommand{\te}{\mbox{$T_{\rm e}$}}
\newcommand{\tex}{\mbox{$T_{\rm ex}$}}

\newcommand{\teff}{\mbox{$T_{\rm eff}$}}

\newcommand{\tkin}{\mbox{$T_{\rm kin}$}}      %% temperature (K) 
\newcommand{\tdyn}{\mbox{$t_{\rm kin}$}}      %% time (yr)
\newcommand{\dens}{\mbox{$n_{\rm H_2}$}}
\newcommand{\dense}{\mbox{$n_{\rm e}$}}
\newcommand{\nh}{\mbox{$N_{\rm H_2}$}}
\newcommand{\nco}{\mbox{$N_{\rm CO}$}}

\newcommand{\intil}{\mbox{$\int{I_{\rm L}d\upsilonup}$}}
\newcommand{\il}{\mbox{$I_{\rm L}$}}

\newcommand{\nlyc}{\mbox{$N_{\rm LyC}$}}
\newcommand{\tlon}{\mbox{$T_{l}^{\rm on}$}}
\newcommand{\tloff}{\mbox{$T_{l}^{\rm off}$}}
\newcommand{\tc}{\mbox{$T_{\rm c}$}}

\newcommand{\hal}{\mbox{H$\alpha$}}

\newcommand{\htal}{\mbox{H30$\alpha$}}
\newcommand{\htnal}{\mbox{H39$\alpha$}}
\newcommand{\hcg}{\mbox{H55$\gamma$}}
\newcommand{\hce}{\mbox{H51$\epsilon$}}
\newcommand{\hsd}{\mbox{H63$\delta$}}

\newcommand{\jb}{\mbox{Jy\,beam$^{-1}$}}

\newcommand{\oh}{\mbox{OH\,231.8$+$4.2}}

\newcommand{\micron}{\mbox{$\mu$m}}
\newcommand{\coral}{\mbox{Co$^{3}$RaL}}

\newcommand{\docem}{$^{12}$CO}
\newcommand{\trecem}{$^{13}$CO}

\def\snu#1{\ifmmode {S_\nu\,\propto\,\nu^{#1}}
          \else \hbox{$S_\nu$\,$\propto$\,$\nu^{#1}$}\fi}
\def\cm#1{\ifmmode {\,{\rm cm^{-#1}}}                  % cm-1, cm-2, cm-3, ...
        \else \hbox{$\,${\rm cm$^{\rm -#1}$}}\fi}
\def\raw {\ifmmode\rightarrow\else$\rightarrow$\fi}
\def\ex#1{\ifmmode {\times 10^{#1}}         % x10$^{-1}$, x10$^{-2}$, etc
        \else \hbox{{$\times 10^{\rm #1}$}}\fi}

\begin{document}

%\title{ALMA Observations of the Compact Ionized Collimated Wind at the core of M\,2-9}
%\title{Unveiling the Compact Collimated Ionized Wind in the Core of M,2-9 with ALMA}
\title{Uncovering the structure and kinematics of the ionized core of M\,2-9 with ALMA}

%   \subtitle{I. Overviewing the $\kappa$-mechanism}

   \author{C.~S\'anchez Contreras\inst{1}
    \and D. Tafoya\inst{2}
    \and J.P.~Fonfr{\'i}a\inst{3}
    \and J.~Alcolea\inst{3}
    \and A.~Castro-Carrizo\inst{4}
    \and V.~Bujarrabal\inst{5}
     %  et al. \inst{2}\fnmsep\thanks{xx}
          }

   \institute{Centro de Astrobiolog{\'i}a (CAB), CSIC-INTA, ESAC-campus, Camino Bajo del Castillo s/n, E-28692, 
  Villanueva de la Ca\~nada, Madrid, Spain. \email{csanchez@cab.inta-csic.es}
  \and Department of Space, Earth, Environment, Chalmers University of Technology, Onsala Space Observatory, 439 92 Onsala, Sweden
  \and Observatorio Astron\'omico Nacional (IGN), Alfonso XII
  No 3, 28014 Madrid, Spain
  \and Institut de Radioastronomie Millimetrique, 300 rue de la Piscine, 38406 Saint Martin d’Heres, France
  \and Observatorio Astron\'omico Nacional
  (IGN), Ap 112, 28803 Alcal\'a de Henares, Madrid, Spain
   }
   
   \date{Received 26/07/2024; accepted 31/10/2024}

% \abstract{}{}{}{}{}
% 5 {} token are mandatory
 
  \abstract{We present interferometric observations at 1 and 3\,mm
    with the Atacama Large Millimeter Array (ALMA) of the free-free
    continuum and mm-wavelength recombination line (mRRL) emission of
    the ionized core (within $\lsim$130\,au) of the young Planetary
    Nebula (PN) candidate M\,2-9.
    %, also known as ``The Butterfly'' Nebula.
    These inner regions are concealed in the vast majority of similar
    objects.  A spectral index for the mm-to-cm continuum of $\sim$0.9
    indicates predominantly free-free emission from an ionized wind, with a minor
    contribution from warm dust. The mm-continuum emission in M\,2-9
    reveals an elongated structure along the main symmetry axis of the
    large-scale bipolar nebula with a C-shaped curvature surrounded by
    a broad-waisted component. This structure is consistent with an
    ionized bent jet and a perpendicular compact dusty disk. The
    presence of a compact equatorial disk (of radius $\sim$50\,au) is
    also supported by red-shifted CO and \trecem\ absorption profiles
    observed from the base of the receding north lobe against the
    compact background continuum. The redshift observed in the CO
    absorption profiles likely signifies gas infall movements from the
    disk toward a central source. The mRRLs exhibit velocity gradients
    along the axis, implying systematic expansion in the C-shaped
    bipolar outflow. The highest expansion velocities ($\sim$80\,\kms)
    are found in two diagonally opposed compact regions along the
    axis, referred to as the high-velocity spots/shells (HVS),
    indicating either rapid wind acceleration or shocks at radial
    distances of $\sim$0\farc02-0\farc04 ($\sim$15-25\,au) from the
    center. A subtle velocity gradient perpendicular to the lobes is
    also found, suggestive of rotation. Our ALMA
    observations detect increased brightness and broadness in the
    mRRLs compared to previously observed profiles, implying variations in wind
    kinematics and physical conditions on timescales of less than two
    years, in agreement with extremely short kinematic ages
    ($\lsim$0.5-1\,yr) derived from observed velocity gradients in the
    compact ionized wind.  Radiative transfer modeling indicates an
    average electron temperature of $\sim$15000\,K and reveals a
    non-uniform density structure within the ionized wind, with
    electron densities ranging from \dense$\approx$10$^6$ to
    10$^8$\,\cm3.
    %The average mass-loss
    %rate of the ionized wind is deduced to be of
    %\mloss$\approx$10$^{-7}$\,\my.
    These results potentially reflect a
    complex bipolar structure resulting from the interaction of a
    tenuous companion-launched jet and the dense primary star's wind.}


%    These observations have been analysed using
%    the non-LTE radiative transfer code co3Ral, providing new insights
%    into the kinematics and physical conditions of M\,2-9's ongoing
%    mass ejections down to $\sim$20\,au spatial scales.

   \maketitle
%
%________________________________________________________________

\section{Introduction}
\label{intro}

The onset of asphericity and polar acceleration in Planetary Nebulae
(PNe) is still not fully understood, but these phenomena are already
active in the early stages of evolution beyond the Asymptotic Giant
Branch (AGB). To comprehend the complex and rapid ($\approx$1000\,yr)
evolution of nebulae from the AGB to the PN phase, it is then crucial
to study pre-PNe (pPNe) and young PNe (yPNe). Previous studies of pPNe
suggest that the presence of multiple lobes and high velocities may be
attributed to the impact of collimated fast winds (CFWs or jets) on
the slowly expanding circumstellar envelopes formed during the AGB
phase \citep[see e.g.\,][for a comprehensive review]{bal02}. However,
direct characterization of the post-AGB jets and their launch regions,
within a few hundred astronomical units, is challenging due to their
small angular sizes and significant obscuration caused by optically
thick circumstellar dust shells or disks.
   
During the mid-stages of their post-AGB evolution towards the PN
phase, the central stars of pPNe begin to ionize their surroundings,
typically reaching a B-type spectral classification.  The just
emerging (nascent) central ionized cores of pPNe/yPNe can be traced
using radio continuum and recombination line emissions, offering the advantage of
minimal dust extinction effects. A recent pilot study of millimeter
radio recombination lines (mRRLs) in a sample of pPNe/yPNe with the
\iram\ radiotelescope \citep[][hereafter CSC17]{san17} shows that
mRRLs are optimal tracers to probe the deepest regions at the heart
($\lsim$150 au) of these objects, from where CFWs
%, presumably responsible for the onset of asphericity and polar
%acceleration in these late evolutionary stages,
are launched.  One key finding from
the study by CSC17 is the determination of mass-loss rates for young
($\sim$15--30\,yr old) post-AGB ejections
(\mpagb$\sim$10$^{-6}$--$10^{-7}$\,\my) much higher than those
currently used in stellar evolution models \citep[e.g.][]{sch83,blo95,vas93,mil16}.
%Such elevated rates would
%lead to a transition from the AGB phase to the PN phase at a
%significantly faster pace than previously believed, particularly
%noticeable fow low-mass ($\sim$1\,\msun) progenitors.
These large rates imply a much faster transition from the AGB phase to
the PN phase, particularly for low-mass ($\sim$1\,\msun) progenitors.
Subsequent mRRL studies conducted with the Atacama Large Millimeter
Array (ALMA), achieving a remarkable angular resolution down to
$\sim$0\farc02, have also led to the first discovery of a rotating
fast ($\sim$100\,\kms) bipolar wind and disk system at the core of
MWC\,922, a B[e]-type post main-sequence star with a distinctive
X-shaped nebula \citep{san19}. This finding highlights the
significance of mRRL observations in uncovering such intriguing
systems.

In this work, we present an ALMA-based study
of the nascent compact
ionized region at the center of the pPN/yPN candidate M\,2-9.  We studied its inner
regions down to 0\farc03 ($\sim$20\,au) scales using 1 and
3\,mm observations of the continuum and the H30$\alpha$ and
H39$\alpha$ mRRLs using the ALMA interferometer. This paper is
organized as follows.  In Sect.\,\ref{m29}, we provide an introductory
overview of M\,2-9, and in Sect.\,\ref{obs} we describe the observations. In
Sect.\,\ref{res_cont} and Sect.\,\ref{res-lines}, the observational results
from the continuum and from the lines, respectively, are reported. The
analysis of the data, which includes line and continuum non-LTE
radiative transfer modeling, is presented in Sect.\,\ref{analysis}. The
results are further examined and discussed in Section
\ref{discussion}, and a summary of our main findings and conclusions
is provided in Section \ref{summary}.

\section{M\,2-9}
\label{m29}

M2-9, also known as ``The Butterfly'' or the ``Twin Jet'' Nebula, is
one of the most iconic and well-studied pPN/yPNe candidates. At optical
wavelengths (Fig.\,\ref{f-hstcont}), it displays a distinctive morphology with a bright
compact core and nested bilobed structures that extend over large
($\sim$1\arcmin) angular scales in a north-south orientation
\citep{sch97,cly15}. 
%% (Schwarz et al. 1997; Clyne et al. 2015).
The nebula's brightness and
morphology have been dynamically changing on timescales of a few years
\citep{all72,vdB74,koh80,doy00}. 
The progressive east-to-west displacement of the
primary emission features ($knots$) within the inner lobes represents an
intriguing feature not observed in other pPNe/yPNe.  This phenomenon
lacks a comprehensive explanation but has been linked with potential
causes such as a rapidly rotating (radiation and/or particle) beam exciting
the inner cavity walls of M\,2-9's lobes \citep[][and references
  therein]{doy00,cor11}. 
The rotating pattern of the knots is considered indirect evidence for the presence
of a central binary system with an orbital period of $\sim$90\,yr. 
%is attributed to a
%rapidly rotating, collimated jet traveling at high speeds (exceeding
%10,000\,\kms) from the central binary system (with orbital period
%\porb$\sim$90\,yr) that excites the walls of the inner cavity of
%M\,2-9 \citep{cor11}. 
%% Corradi et al. dicen que hay jets de particulas
%% moviendose a velocidades de 10,000km/s y que impactan en las
%% paredes de los lobulos, para explicar la formacion y el movimiento
%% lateral de los knots brillantes de emision inscritos en los
%% lobulos: 
%"... collimated spray of high velocity particles (jet) from the central
%source, which excites the walls of the inner cavity of M 2–9,
%rather than by a ionizing photon beam. The speed of such a jet would be remarkable: between 11 000 and 16 000 km s−1 ."

The lobes of M\,2-9, like many other pPNe/yPNe, display expansive
kinematics, with velocities ranging from $\approx$20\,\kms\ in the
inner regions of the lobes to $\approx$160\,\kms\ at their tips
\citep{sch97,sol00,tor10,cly15}. In the central core, broad
\hal\ emission with wings spanning $\sim$1600\,\kms\ is
observed. However, the exact nature of these broad wings and, thus the
gas kinematics at the nucleus, remain uncertain, as the \hal\ wings
are significantly broadened by Raman scattering
\citep{tor10,arr03,lee01}.
%(Torres-Peimbert et al. 2010; Arrieta \& Torres-Peimbert
%2003, Lee et al. 2001).
Recent hydrodynamic modeling by \cite{bal18} indicates that a
low-density and mildly collimated jet with gas travelling at
$\sim$200\,\kms\ from the core and impacting on a pre-existing
envelope/wind can account for certain complex characteristics and
kinematics within M2–9's lobes, with higher jet velocities being
unlikely.
%Moreover. expansion velocities in the core as indicated by forbidden
%lines are also much lower ($\sim$60-100\,\kms).
%% Solf 2000: micro-jets from the nucleus (at subarc-scales from the
%% center) seen in forbidden lines, e.g. [NII], but with widths
%% implying expansion velocities of ~150km/s (and not larger),
%% i.e. much narrower than the Halpha wings. The latter are known to
%% be broadened by Raman scattering (Torres-Peimbert & Arriera 2003,
%% Lee xx).

The central star of
%% es a la que se atribuyen el espectro de lineas de emision en el optico de la region
%% ionizada central, Calvet & Cohen dicen que es B1 usando el radio continuo de Purton et al.
%% NO DICEN QUE SEA DE secuencia principal!! 
M\,2-9 is estimated to be a B1 or late O-type with a temperature of
$\sim$25-35\,kK \citep{cal78,swi79}. The
presence of significant extinction and high-infrared excess in the
direction of the nucleus of M\,2-9 indicates the presence of a dusty
torus that obstructs direct visibility of the central star and its putative companion.
%% A la discussion: 
%% Se ha especulado que esta compañera puede ser una WD o una secuencia principal.
%% La B, es la luminosa, y la que determina los rasgos espectrales de la region HII en el optico.
%% NO hay rastro en el espectro de una estrella fria (AGB), que tendria bandas moleculares 
However, this torus allows the central source's radiation to
illuminate the lobes of the nebula. The binarity of the nucleus has
been indirectly confirmed through interferometric CO mm-wavelength
emission maps, revealing two off-centered rings aligned with the
equatorial plane of the nebula \citep{cc12,cc17}. The rings are
interpreted as the result of two short mass ejections occurring at
different positions in the binary orbit, inclined by
$i$$\sim$17\degr\ with respect to the line of sight. 
These mass ejections, which happened
about $\sim$1400\,yr and $\sim$900\,yr ago, also resulted in the
formation of two symmetric hourglass-shaped expanding structures that
are observed to emerge from the CO rings and overlap with the base of
the large-scale optical lobes \citep{cc17}.

%% AQUI O EN LA DISCUSSION??
%{\bf añadir que no se conoce bien el sistema binario: interpretacion mas frecuente hay una mass-lossing (AGB o postAGB) star que es responsable de laa eyeccion de la nebulosa, y una comp. WD que es la que ioniza ese material y puede contribuir al proceso de shaping. }

M\,2-9 exhibits thermal radio-continuum emission, and previous studies
have identified two components required to adequately fit the mm- and
cm-wavelength data \citep{kwo85,lim00,lim03,dlF22}. The first
component is a compact ionized wind at the center, generating
free-free emission following a power-law relationship with frequency
S$_{\nu}$ $\propto$ $\nu$ $^{\sim [0.6-0.85]}$.  The second component
consists of high-density ionized condensations located in the extended 
lobes, contributing to a nearly flat continuum emission distribution.
At mm-wavelengths, the contribution of the extended lobes to the
observed continuum is minimal, and the dominant sources are the
free-free emission from the compact ionized core and thermal emission
from dust \citep[CSC17,][]{san98}. In their study, CSC17 reported the detection of
mm-wavelength emission from the
H30$\alpha$ and H39$\alpha$ lines, 
%(and, tentatively, from the
%H31$\alpha$ and H41$\alpha$ transitions),
consistent with an ionized core-wind that has been ejected over the
past $<$15\,yr at an average rate of $\sim$3.5\ex{-7}\,\my\ and with a
mean expansion velocity of \vexp$\sim$22\,\kms.
%by one of the stars
%in the central binary (CSC17).
In a recent study by \cite{dlF22} using multi-epoch $\sim$23-43\,GHz
continuum maps, it was observed that the elongated ionized core-wind
of M\,2-9 exhibits a subtle C-shaped curvature, whose orientation is found to vary over time.
%These
%findings led these authors to suggest that the observed curvature may
%be a consequence of the interaction between the ionized wind, which is
%hypothesized to be launched by the secondary compact component
%(presumably a white dwarf star), and the wind from the mass-losing
%primary star, leading to either a pushing-away or bending effect on
%the former.
%% que resulta en un cambio de curvatura con el movimiento orbital,
%% segun la compañera que empuja el viento este al este o al oeste del
%% viento ionizado.  ¿podria solo un efecto de presion de radiacion
%% provocar esa curvatura? o es necesaria la interaccion mecanica
%% entre dos vientos? que otras cosas podrian resultar en esa forma?

The distance to M\,2-9 is highly uncertain, with values ranging from
50\,pc to 3\,kpc in the literature \citep[see references given
  in][]{san17}. In this study, we adopt a distance of 650 pc for
M\,2-9, which is based on the analysis of the proper motions of the
optical knots by \cite{sch97} and of the spatio-kinematics, including
also tentative proper motions, of the molecular rings
\citep{cc12}. This distance choice allows for a direct comparison with
the study conducted by CSC17 on the nascent ionized core of M\,2-9.

%The nebula of M 2–9 is oriented in the sky at position angle
%358◦ and is seen close to edge-on: an inclination of its symmetry
% ◦axis on the plane of sky of 11 (Goodrich 1991), 15 (Schwarz
%et al. 1997), and 17◦ (Solf 2000) was determined. 

\section{Observations and data reduction}
\label{obs}

%\input{t_obs.tex}
%                                             Two column Table 
%-------------------------------------------------------------
%
\begin{table}[ht!]
\small
  %\centering
\caption{Central frequency, bandwidth, velocity resolution, detected spectral lines and continuum fluxes in the different spectral windows (SPWs).} \label{tab:spws}
\begin{tabular}{r c c l c}
\hline
\hline
Center       &  Bandwidth     &   $\Delta$v  & Line & Continuum \\ 
(GHz)        &     (MHz)      &   (km/s)            &        & Flux (mJy) \\    
\hline                    
%\multicolumn{5}{c}{\sl M2-9\_b\_06\_TM1 (2017-Oct-15)}\\
\multicolumn{5}{c}{\sl Band 3. DATE: 2017-10-15 }\\ 
95.350   & 5900 &3   & &   105 \\ %NOT detected 
97.000   & 5801 &3   & & 106 \\ 
106.737 & 659   &0.3&  \htnal\ & 124 \\   %H39alpha 
107.100 & 657   &2.7& & 120 \\ 
109.000 & 1290 &2.7& &  120 \\ 
109.536 & 642   &0.7& \hcg\ &  123 \\ % very weak 
110.201 & 638   &0.7& \trecem\,1-0) &  122 \\     %13co1-0
%\vspace{0.1cm}
%\multicolumn{5}{c}{\sl M2-9\_a\_06\_TM1 (2017-Nov-12)}\\
\multicolumn{5}{c}{\sl Band 6. DATE: 2017-11-12}\\ 
215.400  &2612&1.5&  &  227 \\ %##spw3 -- 31 NOT Detected 
217.500  &2587&1.4& &   232 \\ %##spw4 -- 33
230.538  &610&0.6& \docem\,2-1 &  242 \\ %##spw1 -- 27
231.901  &607&0.6& \htal\,+He30$\alpha$ &  245 \\ %##spw0 -- 25
232.900  &2416&1.3& He30$\alpha$ &  243 \\ %##spw2 -- 29 
\hline                  
\end{tabular}
\tablefoot{Flux calibration uncertainties are $\sim$10\%.}
%\tablefoottext{*}{Not detected}
\end{table}
%
%-------------------------------------------------------------

The observations of M\,2-9 were conducted using the ALMA 12-m array as
part of projects 2016.1.00161.S and 2017.1.00376.S. The observations
were carried out in Band 3 (3\,mm) and Band 6 (1\,mm), with a total of
twelve different spectral windows (SPWs) dedicated to mapping the
emission of various mRRLs (and CO lines) as well as the continuum
(Table\,\ref{tab:spws}). The Band 3 and Band 6 observations were
conducted separately in two different execution blocks, with a
duration of approximately 1.5 and 1.1\,hr, in October and November 2017,
respectively.
%% Band 3: total duration 95.37 minutos ~1.5h
%% Band 6: total duration 69      minutos ~1.0h
%% band-3: 49min (ON SOURCE)
%% band-6: 33min (ON SOURCE) 
The observations utilized 45-50 antennas, with baselines
ranging from 41.4\,m to 16.2\,km for Band 3, and from 113.0\,m to
13.9\,km for Band 6.  The maximum recoverable scale (MRS) of the
observations is $\sim$0\farc8 and $\sim$0\farc7 at 3 and 1\,mm,
respectively.

The total time spent on the science target, M\,2-9, was
about 49\,min in Band 3 and 33\,min in \mbox{Band 6}. A number of sources
(J\,1751+0939, J\,1658$-$0739, J\,1718-1120, and J\,1733$-$1304) were also observed as bandpass, complex gain, pointing, and flux
calibrators. The flux density adopted for J\,1751+0939 is 1.95\,Jy at 108.970\,GHz with a spectral index $-$0.294, and  and 1.38\,Jy at 231.845\,GHz
with the same spectral index.

An initial calibration of the data was performed using the automated
ALMA pipeline of the Common Astronomy Software Applications
(CASA\footnote{\tt https://casa.nrao.edu/}, versions 4.7.2 and 5.1.1).
Calibrated data were used to identify the line-free channels, allowing for the creation of 
initial images of both the lines and continuum in the various 
SPWs. Strong continuum emission was detected in all SPWs, with the noise 
being dominated by secondary lobes triggered by residual
calibration errors. 
%When strong source emission was detected, the noise in those
%channel maps was dominated by secondary lobes triggered by residual
%calibration errors. Therefore,
Given the high signal-to-noise ratio
(S/N$\gsim$200) achieved in the continuum images and the absence of
significant/complex large-scale structure in M\,2-9 in our
observations (see Sect.\,\ref{res_cont}), we self-calibrated our data
using the initial model of the source derived from the standard
calibration to improve the sensitivity and fidelity of the images.
%% SELF-CALIBRATION ...................
Self-calibration as well as image restoration and deconvolution was
done using the GILDAS\footnote{{\tt
    http://www.iram.fr/IRAMFR/GILDAS}.} software MAPPING.

Continuum images were generated for each of the 12 SPWs by utilizing
line-free channels.  Flux measurements of the continuum emission for
each SPW were obtained integrating the surface brightness within a
circular aperture with a radius of 0\farc25 fully enclosing the continuum source
(Table\,\ref{tab:spws}). The absolute calibration uncertainties,
estimated at around 10\%, were determined by comparing the measured
continuum flux at different frequencies for the phase, check, and flux
calibrators (before and after self-calibration) with the 
values adopted by the ALMA calibration pipeline.
%The absolute calibration uncertainties, estimated to be approximately
%10\%, have been determined through meticulous measurements of the
%continuum flux at various frequencies for the phase, check, and flux
%calibrators. These measurements were conducted both before and after
%the self-calibration process. The obtained flux values were then
%compared with the adopted values to assess the calibration
%uncertainties.
Line emission cubes were constructed by subtracting the corresponding
continuum data from each SPW, specifically subtracting the continuum
of the SPW containing the specific line of interest.  The final
self-calibrated line cubes and continuum images presented here were
created using the Hogbom deconvolution method with a
robust weighting scheme\footnote{We used a value of 0.5 for the
threshold of the robust weighting in MAPPING.}  resulting in angular
resolutions of $\sim$40$\times$30\,mas at 1\,mm and
$\sim$90$\times$60\,mas at 3\,mm.  Continuum maps with a circular
  restoring beam of half power beam width (HPBW) of 40\,mas at 1\,mm
  and 3\,mm were also created to facilitate a better comparison between
  the two images.  The typical rms noise level per channel of our
spectral cubes is $\sigma$$\sim$1\,m\jb\ (1\,mm) and
0.80\,m\jb\ (3\,mm) at 3\,\kms\ resolution. The rms noise level range
in the continuum maps is $\sim$25-85\,$\mu$\jb\ and
$\sim$60-100\,$\mu$\jb\ for the individual SPWs within Band 3 and Band
6, respectively.

%% Aqui o arriba, antes de hablar de la self-cal? 
%Flux measurements of the continuum emission for each SPW were obtained
%from the non-selfcalibrated continuum maps (Table\,\ref{tab:spws}). The surface brightness was
%integrated within a circular aperture with a radius of xx\arcsec.

%\section{Results}
%\label{results}

\section{Continuum emission}
\label{res_cont}

\subsection{Surface brightness distribution}
\label{cont_maps}

%% Fig. 1 -- continuum maps ---------------------------------------
\begin{figure*} %%[htbp!]
  \centering
%  \sidecaption
   \includegraphics[width=0.425\hsize]{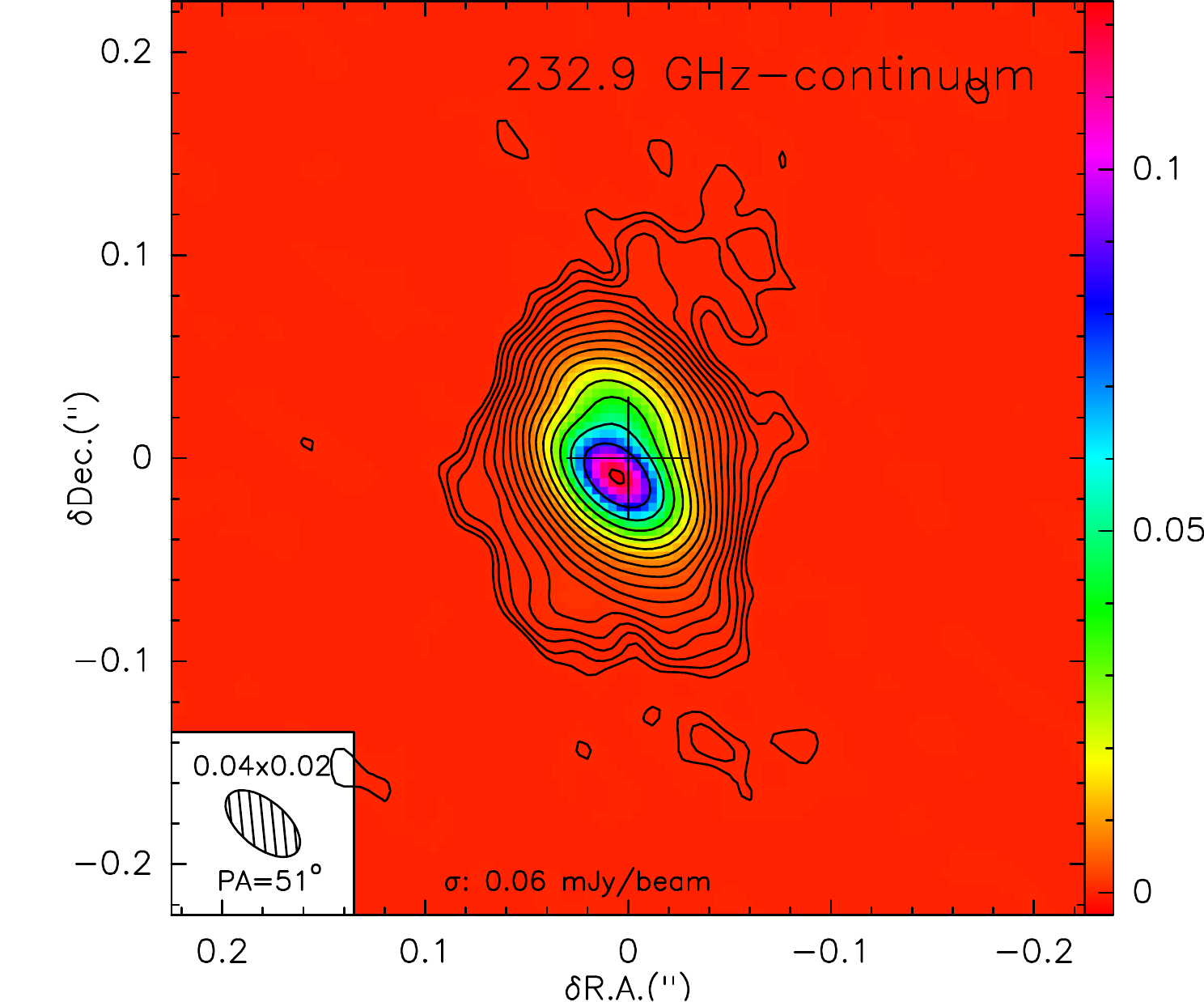}  %% 1mm (spw2=spw29) BW=1.875GHz
   \includegraphics[width=0.425\hsize]{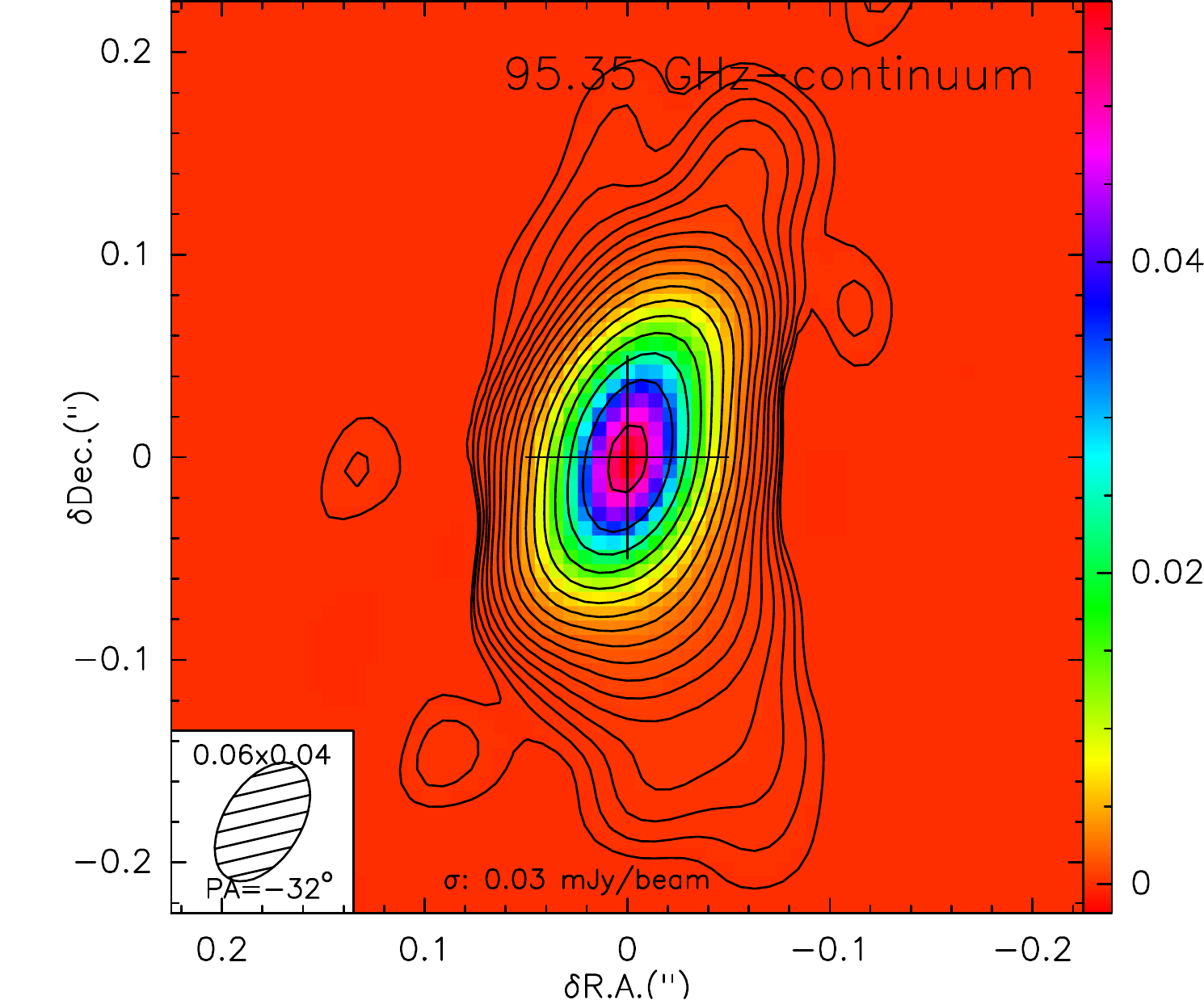}        %% 3mm (spw37, BW=1.875GHz).
   \includegraphics[width=0.425\hsize]{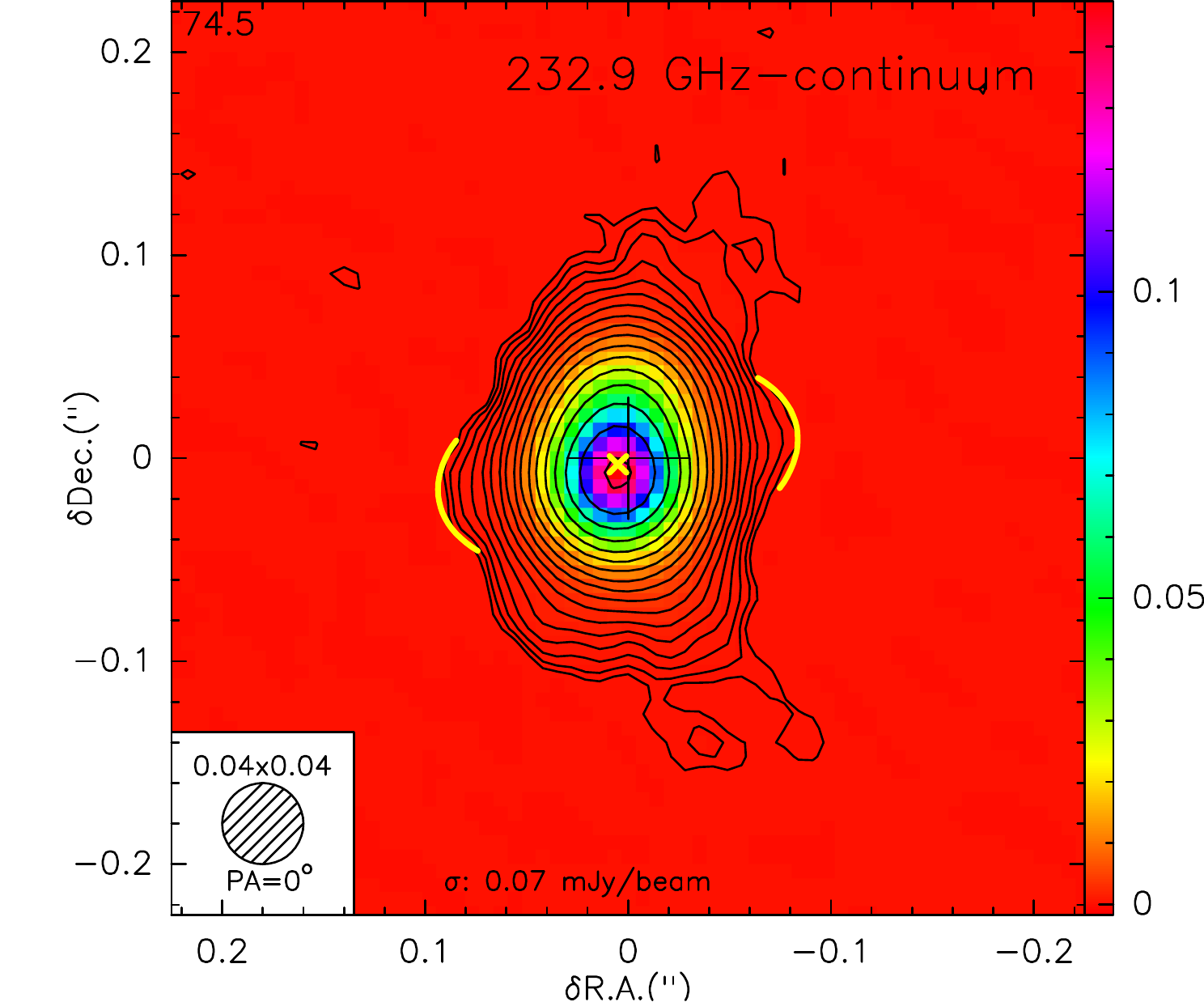}
   \includegraphics[width=0.425\hsize]{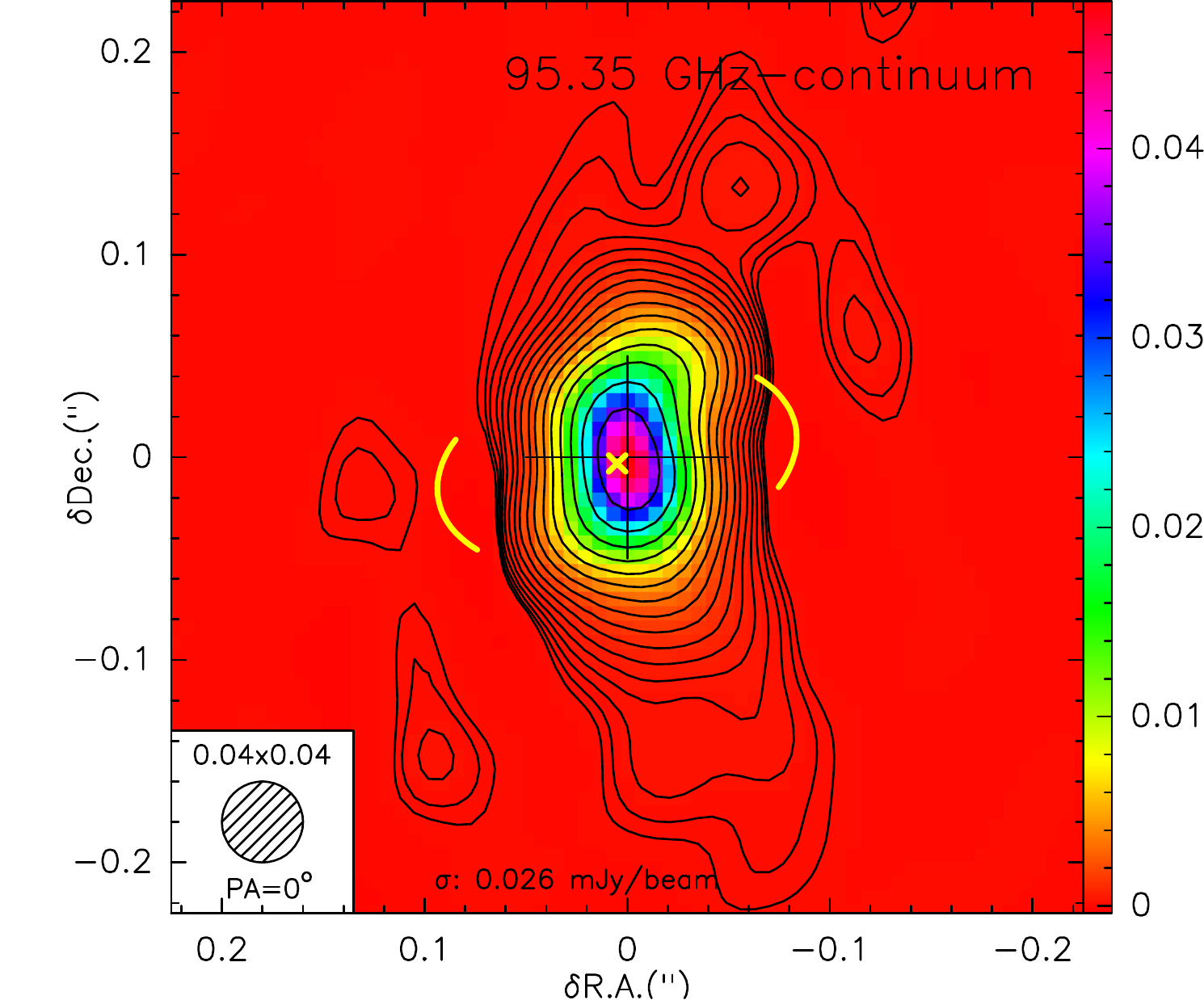}   
   \caption{ALMA continuum emission maps of M\,2-9 at 232.9\,GHz
     (left) and 95.4\,GHz (right). The top row displays the images
     after image restoration using the nominal beam size at each
     frequency (Sect.\,\ref{obs}). The bottom row presents the images
     with a circular restoring beam of 0\farc04 at both frequencies,
     allowing for an optimal comparison between the two. The level
     contours are (3$\sigma$)$\times$1.5$^{(i-1)}$\,Jy\,beam$^{-1}$,
     $i$=1,2,3...  The central cross marks the 3\,mm-continuum surface
     brightness peak at coordinates J2000 R.A.=17\h05\m37\fs9668 and
     Dec.=$-$10\degr08\arcmin32\farc65 (J2000). The yellow arcs, centered at the small yellow cross, 
     represent the broad-waist structure, plausibly a dust disk,
     detected at 1\,mm but not at 3\,mm.
   \label{f-cont}}      
\end{figure*}

   %% @fig_cont.greg --> /pcdisk/jbell3/csanchez/m2-9/selfcal-csc2/
   %%                 %% MAPPING> sic directory "/pcdisk/jbell3/csanchez/m2-9/selfcal-csc2" ; @fig_cont.greg
   %%                 1mm: M2-9_spw2_contNav475-selfcal.uvt
   %%                 3mm:
   %% /pcdisk/jbell3/csanchez/m2-9/selfcal-csc2/mascont/fig_cont.greg 
   %% -------------------%

ALMA continuum emission maps of M\,2-9 at two representative
frequencies at 1\, and 3\,mm are shown in Figs.\,\ref{f-cont} and
  \ref{f-hstcont}. There are no appreciable differences between the
continuum maps for the different SPWs within the same band (band 3 or
6), therefore as representative we have chosen those SPWs with larger
bandwidth (BW=1.875GHz) at the higher and lower ends of the observed
frequency range.
% In addition, none of these SPWs contained emission lines, so that the full bandwidth could be used to generate continuum maps.
The peak surface brightness of the 3\,mm continuum is located at
coordinates R.A.=17\h05\m37\fs9668 and
Dec.=$-$10\degr08\arcmin32\farc65 (J2000)\footnote{Absolute
astrometric uncertainties are of 0\farc003.}.  These coordinates serve
as the reference point ($\delta$x=0\arcsec, $\delta$y=0\arcsec) for
positional offsets in all figures representing image data throughout
the paper.
%Axis 1     A0     17:05:37.96679        Axis 2     D0    -10:08:32.6498

The mm-continuum emission, which is spatially resolved, appears
elongated along the north-south direction, i.e.\, along the symmetry
axis of the nebula, with full ($\sim$3$\sigma$ level) dimensions of
$\sim$0\farc22$\times$0\farc13 and $\sim$0\farc4$\times$0\farc13 at 1
and 3\,mm, respectively.  The elongated shape of the mm-continuum
surface brightness is similar to what is observed at cm wavelengths,
although the mm-continuum emission is more compact 
\citep{kwo85,lim03}. This compactness is expected based on the
inverse relationship between the angular size of the continuum
emission along the nebula axis and the observing frequency observed
before in M\,2-9 at cm wavelengths  \citep[$\approx$$\nu^{-0.84}$,][]{dlF22}, which
reflects the inverse proportionality between the free-free continuum
opacity and the frequency \citep{pana75}.

%% TAMAÑO en el eje menor el tamaño sin embargo casi no ha cambiado...
%% wr.t. kwok85... incluso parece algo mas grande... 

%% sigue la dependencia con la frecuencia esperada? SI 
%% Lim & Kwok 2003??
%% Kwok et al. 1985 
%% MEDIR BIEN LAS DIMENSIONES 3sigma y half-intensity:

The elongated morphology of the mm-continuum emission in M\,2-9
exhibits a slight bending towards the northwest (NW) and southwest
(SW) directions, revealing a subtle C-shaped (mirror symmetric)
curvature.  This curvature is most clearly appreciated at 3\,mm.  The
orientation of the C-shaped curvature in our maps is the same as in
the JVLA maps of the 7\,mm-continuum emission last observed in 2006
\citep{dlF22}.
%% but opposite to that observed in 1999. 

We observed some differences in the brightness surface of the
continuum emission at 1\,mm and 3\,mm, which are worth highlighting.
For example, the major-to-minor axis ratio is larger at 3\,mm.  While the
length of the emission along the nebula's axis is nearly twice as long
at 3\,mm compared to 1\,mm, the average size of the emission in the
equatorial direction (perpendicular to the nebula's axis) is larger in
the 1\,mm continuum maps, which shows a slight widening in the central
equatorial region compared to higher latitudes, i.e.\,a broad-waisted
morphology (indicated by the incomplete yellow ellipse in
Fig.\,\ref{f-cont}). The broad-waist is centered around the
1\,mm-continuum emission peak (at offset $\sim$0\farc007)
and is not perfectly aligned with the
equatorial plane of the large-scale nebula.
The central and
brightest regions of the 1\,mm continuum map exhibit a higher degree
of asymmetry with respect to the equator compared to the 3\,mm map. At
1\,mm, these regions appear to have an egg-shaped morphology, with the
emission from the northern part being weaker than that from the
southern part. Furthermore, the emission peak at 1\,mm is slightly
shifted to the south-east compared to the 3\,mm peak.

These small but noticeable differences between the 1 and 3\,mm
continuum maps (including those restored with the same beam) are
likely due, in part, to the presence of dust in these central regions,
with a larger contribution to the mm-continuum emission at shorter
wavelengths. The presence of a broad-waisted morphology in the 1\,mm
continuum emission suggests the possibility of a dust distribution in
the form of an equatorial disk or belt surrounding the central ionized
region. This interpretation is supported by independent indications in
the CO emission maps that point towards the existence of such an
equatorial structure (Sect.\,\ref{res-co} and Sect.\,\ref{disk_co}). Given
the system's inclination (with the northern lobe moving away from us),
this dust structure likely obscures part of the northern ionized
region, explaining its weaker 1\,mm emission. Furthermore, the
broad-waist feature observed in the 1\,mm-continuum maps does not
appear in the integrated intensity maps of the \htal\ and
\htnal\ lines (see Sect.\,\ref{res-mrrls}), which exclusively trace the
free-free (not the dust) emission.
%We note that the broad-waist is centered around the
%1\,mm-continuum emission peak and it exhibits a tentative orientation
%with a PA$\sim$102\degr. This orientation suggests a slight deviation
%from the equatorial plane of the large-scale nebula, indicating that
%they are not perfectly aligned.
As further discussed in the next subsection, the broad-waist could
represent the outer, cooler regions of the compact warm disk observed
in the middy at the core of M\,2-9 \citep{smg05,lyk11}. 

\subsection{Spectral energy distribution: free-free and dust continuum emission}
\label{sed}

%% Fig. ---------------------------------------
   \begin{figure*}[htbp!]
     \centering
     \includegraphics*[width=0.40\hsize]{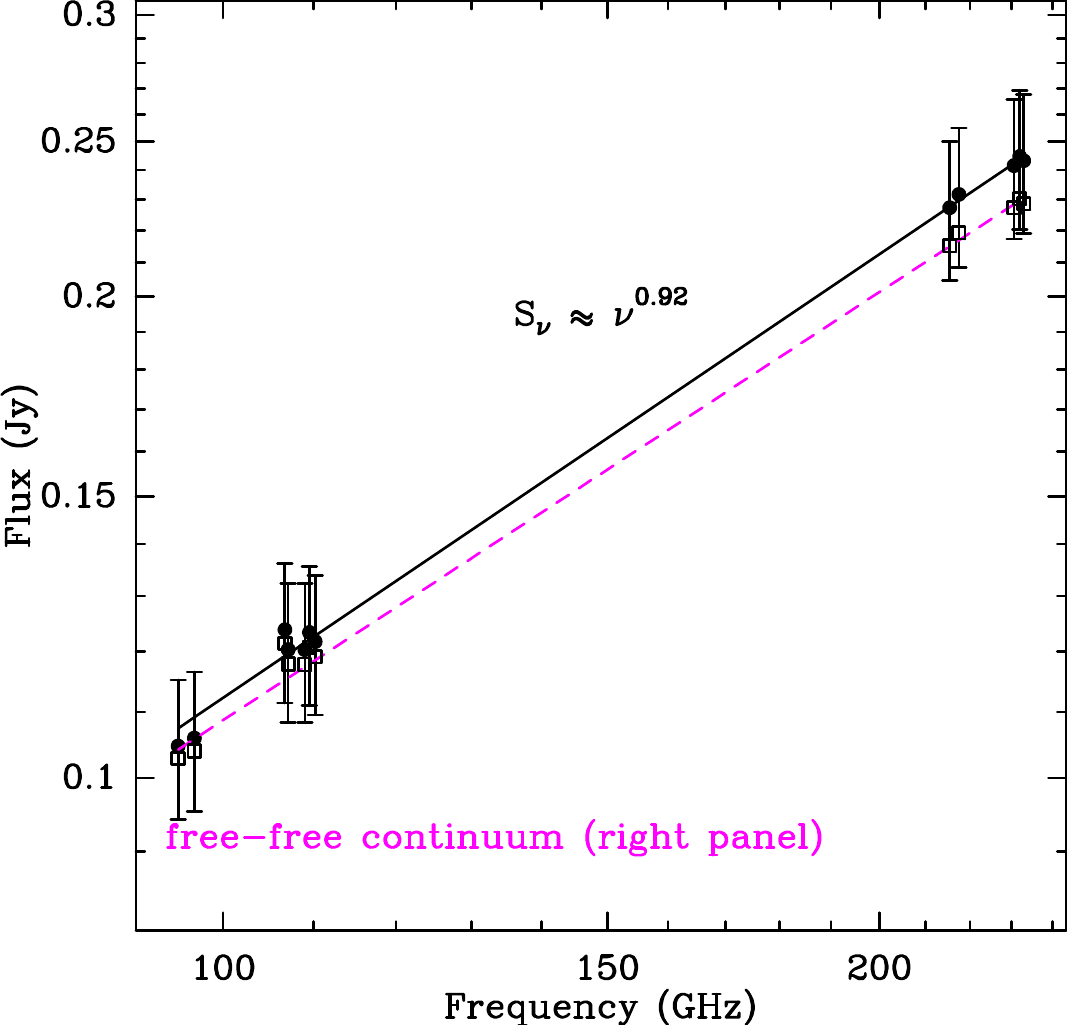}
          \includegraphics*[width=0.55\hsize]{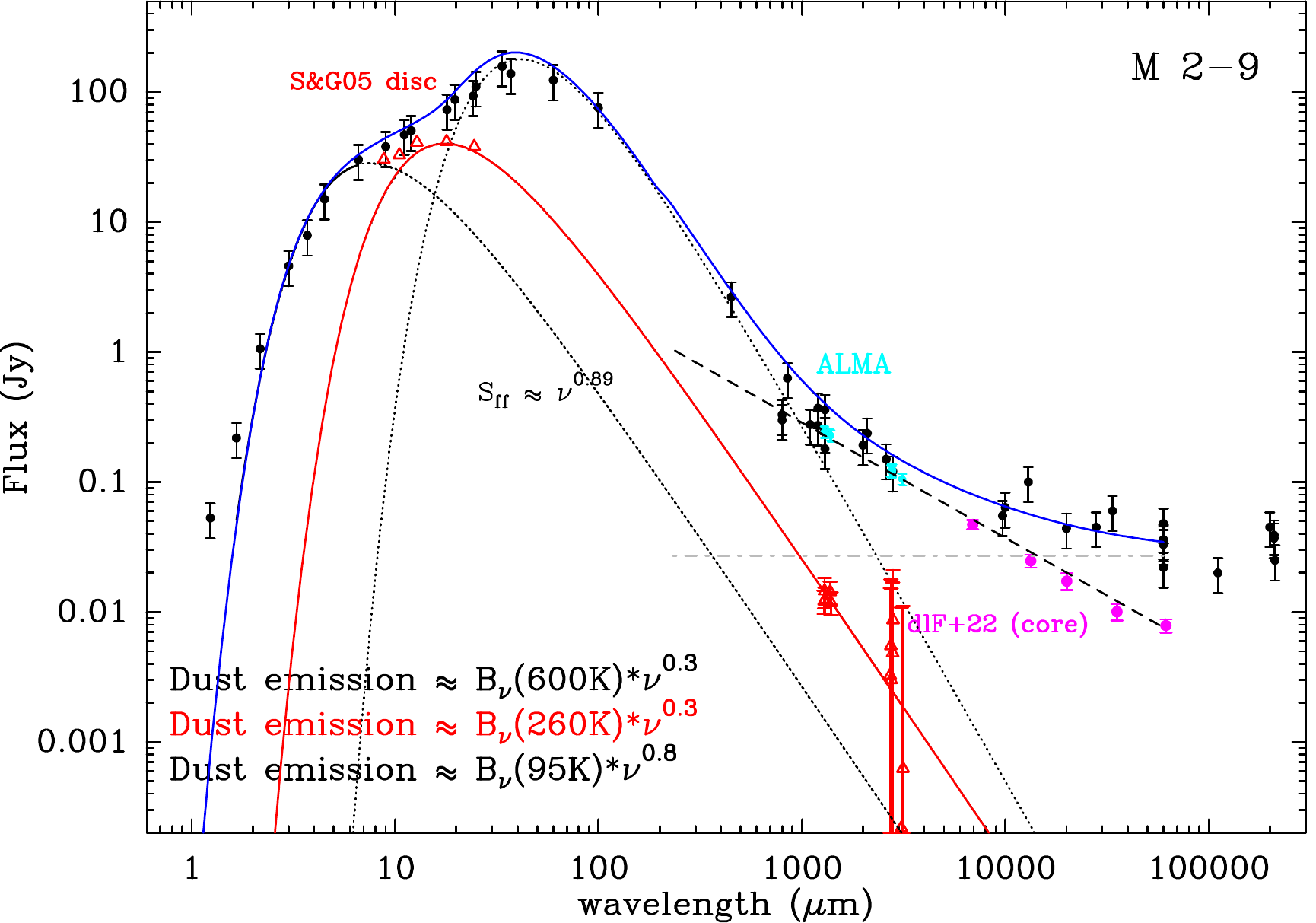}
          \caption{Spectral energy distribution (SED) of M\,2-9. {\bf
              Left)} Total (dust and free-free) mm-continuum fluxes
            measured with ALMA in different SPWs (filled circles,
            Table\,\ref{tab:spws}). The global fit (\snu{0.92\pm0.1})
            is depicted with a solid line. The dashed line represents
            a fit to the free-free emission in the radio continuum,
            using the 4.9-43\,GHz datasets from \cite{dlF22}, which
            also reproduces our mm-continuum fluxes after subtracting
            the contribution from a compact $\sim$260\,K dust
            component (shown in the right panel) -- open squares.
            {\bf Right)} SED of M\,2-9 from the near-IR to the radio
            domain as in \cite{san17} and including our ALMA
            mm-continuum flux measurements (cyan). Radio continuum
            flux measurements at 4.9-43\,GHz from the ionized
            wind/core as reported by \cite{dlF22}
            %and 345\,GHz-continuum flux from the
            %unresolved core reported by \cite{cc17}
            are indicated with pink circles. Mid-IR photometry of the
            compact, warm ($\sim$260\,K) dust disk at the core of
            M\,2-9 is shown with red triangles in the
            8-20\,\micron\ range. Fits to the free-free and dust
            thermal emission are depicted, with the nearly flat
            continuum from the ionized bipolar lobes and the free-free
            emission from the ionized wind/core (\snu{0.89}) shown as
            a dotted-dashed and a dashed line, respectively.  Several
            modified black-body components are also plotted, including
            a $\sim$95\,K component originating from dust located in
            the extended bipolar lobes, and filtered in our ALMA maps,
            (dotted line) and a $\sim$260\,K component arising from
            dust in a compact disk at the center of M\,2-9 (red
            line). For completeness, a hot $\sim$600\,K dust
            component, which is necessary to explain the optical/NIR
            photometry, is also represented. The blue solid line
            represents the combined fits for the free-free and dust
            emission components.  The red triangles at 1\,mm and 3\,mm
            represent the residual dust emission of M\,2-9 after
            subtracting a \snu{0.89} free-free emission component from
            the ionized wind/core.
%% ajusta los puntos de dlF22 y tambien el residuo de nuestros datos despues de quitarles el polvo a 260K. 
            \label{sed-alma}}
   \end{figure*}
   %% @sed_old.greg m2_9 
   %% /pcdisk/jbell3/csanchez/m2-9/continuum/sedm2_9_WARM500K_alpha0p50.pdf
   %% /pcdisk/jbell3/csanchez/m2-9/continuum/sedm2_9_WARM600K_alpha0p30.pdf
   %% /pcdisk/jbell3/csanchez/m2-9/continuum/ datos con calibracion STANDARD. 
   %% @sed_alma_ff-sc_FREQ.greg
   %% ----------------------------------------------------------------

%   The continuum fluxes
%   for the various SPWs observed in this project
%were measured within a circular aperture with a radius of 0\farc25,
%centered on M\,2-9, fully enclosing the mm-continuum emitting source
%(Table\,\ref{tab:spws}).
%These flux values are provided in
%Table\,\ref{tab:spws} and their variation with frequency (wavelength)
%is plotted in the left (right) panel of Fig.\,\ref{sed-alma}.
The overall fit to the continuum fluxes (Table\,\ref{tab:spws}) derived 
from 1\,mm to 3\,mm as a function of frequency yields \snu{0.92\pm0.1}
(Fig.\,\ref{sed-alma}-left), consistent with predominantly free-free
emission from the compact ionized wind/core of M\,2-9 with a secondary
contribution from dust thermal emission, as discussed in
Sect.\,\ref{m29}.
%% REMOVE ----------------------------------------------------------------------
%% QUIZA SE PUEDE DECIR SOBRE LOS DATOS DESPUES DE SUSTRAER EL POLVO. 
%If the continuum at 1\,mm and 3\,mm is fitted separately, it is
%observed that the spectral index is slightly higher at 3\,mm
%(\snu{1.1}) compared to 1\,mm (\snu{0.85}). This could suggest that the
%turnover frequency of the free-free emission from the compact ionized
%core of M\,2-9 could be around $\lsim$1\,mm. At this frequency, the
%energy distribution would start to flatten out, consistent with a
%transition from partially optically thick to optically thin emission
%(see Sect.\,\ref{model}). {\bf MODELO? si no lo corrobora mejor no
%  mencionar y no hacer ajustes por separado...}
%% REMOVE ----------------------------------------------------------------------
%As discussed by \cite{san17} and several other works referenced there,
%in addition to free-free, thermal dust emission also contributes to
%the mm-wavelength continuum flux.
%Although our observations reveal a
%broad waist in the 1\,mm continuum maps (Sect.\,\ref{res_cont}),
%suggestive of dust emission from a compact equatorial disk, accurately
%isolating free-free emission from the dust emission in our ALMA maps
%is difficult.

The spectral energy distribution (SED) of M\,2-9 from the near
infrared (NIR) to the radio domain, including our ALMA continuum flux
measurements, is shown in Fig.\,\ref{sed-alma} (right).  Comparing the
ALMA fluxes with the majority of the mm-continuum fluxes obtained with
single-dish telescopes reveals flux losses of
approximately 20\% in our high-resolution interferometric
measurements. This finding indicates that ALMA, in the extended
configuration used in this work (with MRS$\sim$0\farc7-0\farc8,
Sect.\,\ref{obs}), is filtering and losing the contributions from both
the free-free emission originating from the extended ionized lobes,
which results in a nearly flat continuum flux of $\sim$30\,mJy
\citep{kwo85}, and the dust emission from extended structures.
%% , which exhibit different modified black-body spectra.
These extended
structures include the relatively cool dust present in the large-scale
bipolar lobes \citep[with temperatures ranging from $\sim$5-25\,K up
  to $\sim$100\,K,][]{san98,smg05} as well as the dust in the inner
$\sim$1-2\arcsec-sized (in radius) equatorial disk 
\citep{cc17} and whose temperature remains unconstrained.

Besides the extended dust structures, the mid-IR images of M\,2-9
reported by \cite{smg05} revealed the presence of an unresolved
central core, suggesting the existence of a compact dust component at
the center.  A fitting of the mid-IR photometry of this unresolved
central core (red solid line in Fig.\,\ref{sed-alma}-right) indicates
an approximate temperature of $\sim$260\,K. This finding was further
supported by \cite{lyk11}, who used mid-IR interferometry to confirm
the presence of a compact disk with approximate dimensions of
37$\times$46 mas at 13\micron. Unlike the emission from the extended
dust structures mentioned in the previous paragraph, the emission from
this compact disk is not expected to be affected by interferometric
flux losses in our ALMA maps. Therefore, it is likely that the compact
warm disk, which exhibits its peak emission in the mid-IR, is
contributing to the total observed mm-continuum flux.  Moreover, it is
plausible that this compact disk corresponds to the inner, warmer
regions of the broad-waisted structure detected in our 1\,mm-continuum
maps, as supported by the fact that the dimensions of the broad-waist
are comparable to (although slightly larger than) the size of the disk
observed in the mid-IR (Fig.\,\ref{f-cont}).

Under the assumption that a fraction of the mm-continuum flux observed
with ALMA originates from the warm ($\sim$260\,K) compact disk we can
achieve a good fit of the mm-continuum with a
combination of a $\sim$200-300\,K dust contribution (with emissivity
ranging from $\sim$0.2-0.5) and a free-free component that varies as
\snu{0.89} (Fig.\,\ref{sed-alma}).  The latter component
accurately reproduces both the cm-wavelength continuum (exclusively
from the ionized wind/core as reported by \cite{dlF22}\footnote{The
spectral index derived by us, 0.89$\pm$0.06, and derived by
\cite{dlF22}, 0.85$\pm$0.05, 
are consistent with each other within the uncertainties and considering that
these authors only use radio-continuum data.}) and the
residual mm-wavelength continuum obtained by subtracting the
$\sim$260\,K dust emission from the total mm-continuum flux observed
with ALMA.
%This decomposition of the continuum flux also
%aligns well with the anticipated free-free and dust emission
%components of the unresolved emission component identified by
%\cite{cc17} in their $\sim$0\farc25-resolution continuum maps at 345\,GHz
%obtained with ALMA (highlighted with a pink circle and a red triangle adjacent
%to the label "CC+17 (core)'' in Fig.\,\ref{sed-alma}-right).

It is important to emphasise that the decomposition/fit described
above is subject to uncertainties and should be considered as a rough
estimate, its main purpose being to provide an approximate value for
the purely free-free continuum flux at mm-wavelength to be compared
with the predictions from the model of the central ionized region
presented in Section \ref{model}.

\section{Line emission}
%% recombination, but also CO -- solo para mostrar la absorcion y el anillo? 
\label{res-lines}

We observed several mRRLs (including, \htal, He30$\alpha$, \htnal, and
\hcg) in M\,2-9 using ALMA. The He30$\alpha$ and \hcg\ lines are first
detections, while \htal\ and \htnal\ single-dish spectra were
previously reported by \cite{san17}.  Despite the high sensitivity of
our ALMA data, the weak \hce\ and \hsd\ lines were undetected.  We
have also mapped two rotational transitions of carbon monoxide
(\docem\,2--1 and \trecem\,1--0).  While the \docem\,2--1 emission has
previously been mapped with NOEMA at a resolution of
0\farc8$\times$0\farc4 \citep{cc12}, there are no known reports of
\trecem\,1--0 line maps.
%and the \docem\,3--2
%line has been mapped with ALMA at $\sim$0\farc3-resolution
%\citep{cc17}.

\subsection{Millimeter recombination lines}
\label{res-mrrls}
%% ========================== Fig. H30a
   \begin{figure*}[htbp!]
     \centering
     \includegraphics*[width=0.30\hsize]{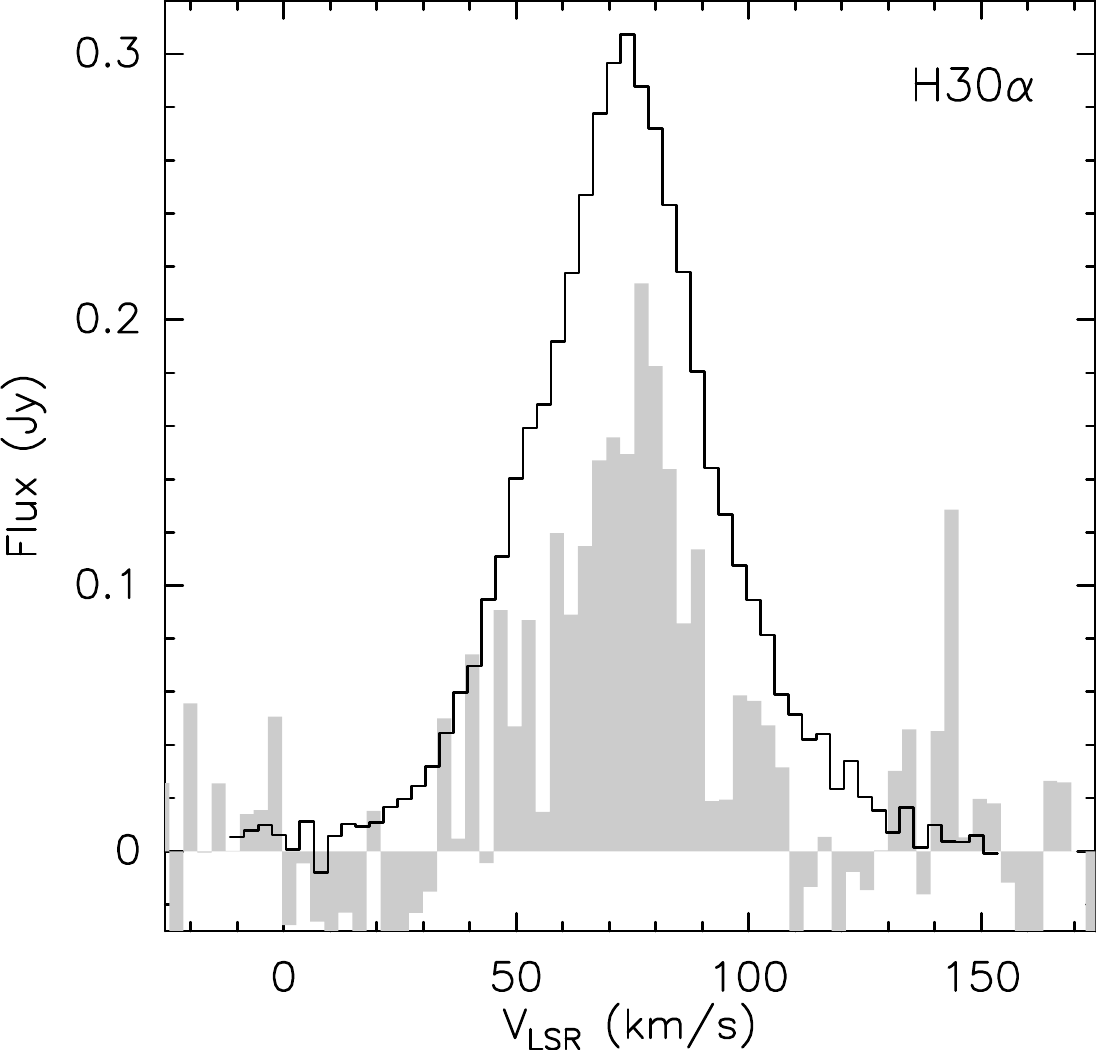}                %1dspec
     \includegraphics*[width=0.33\hsize]{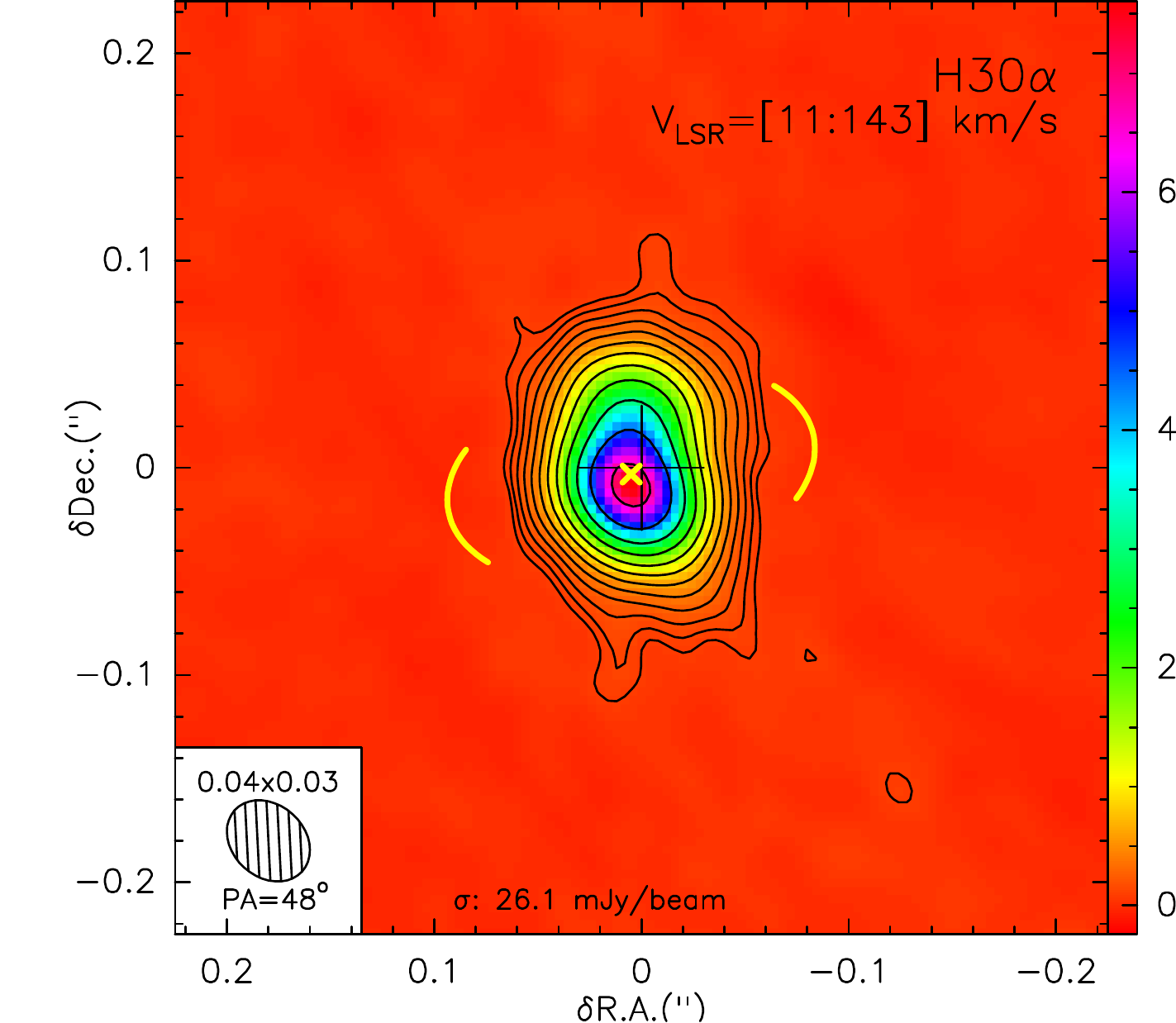}    %area
     \includegraphics*[width=0.34\hsize]{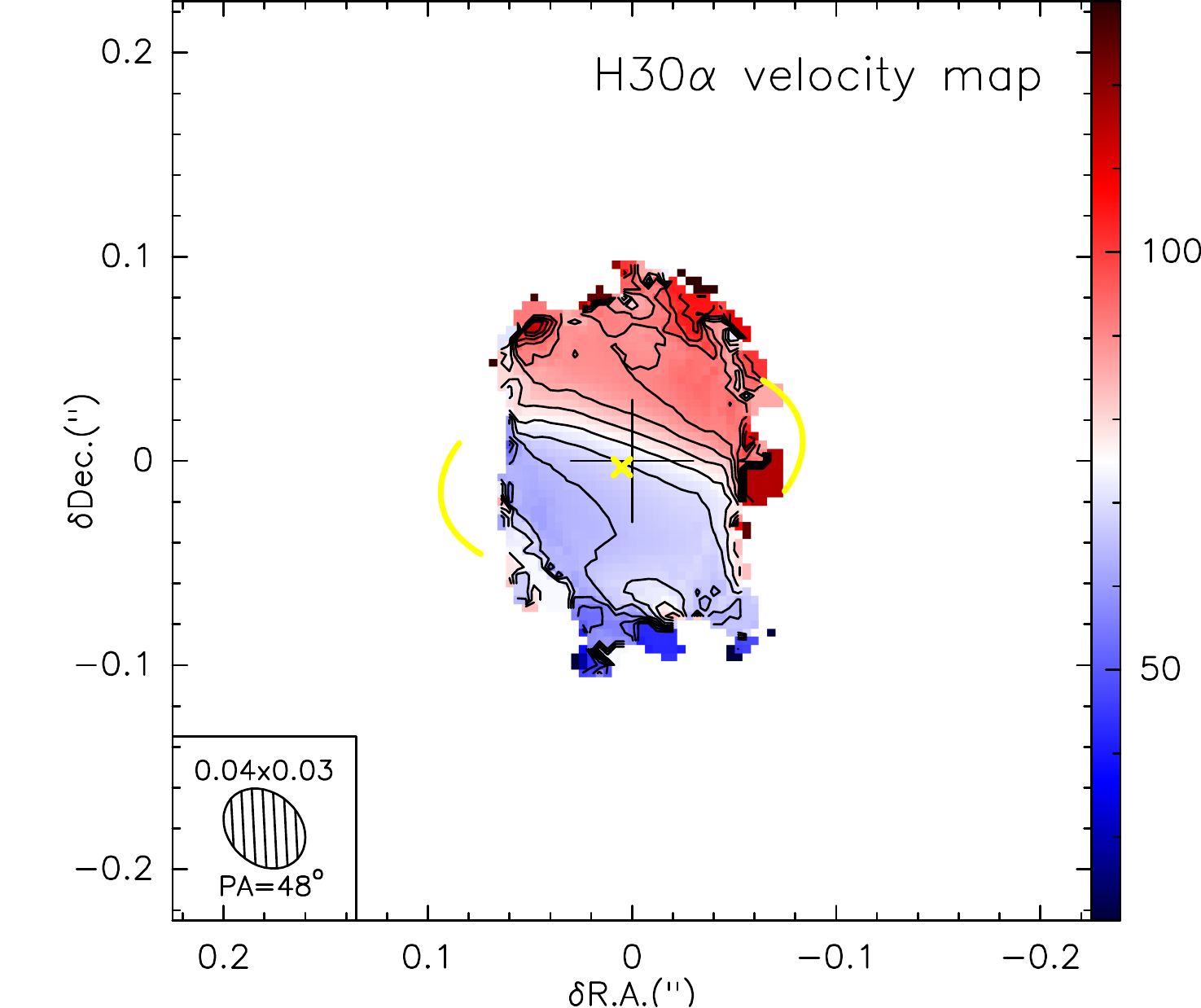}               %velo
     \\
     \includegraphics*[width=0.30\hsize]{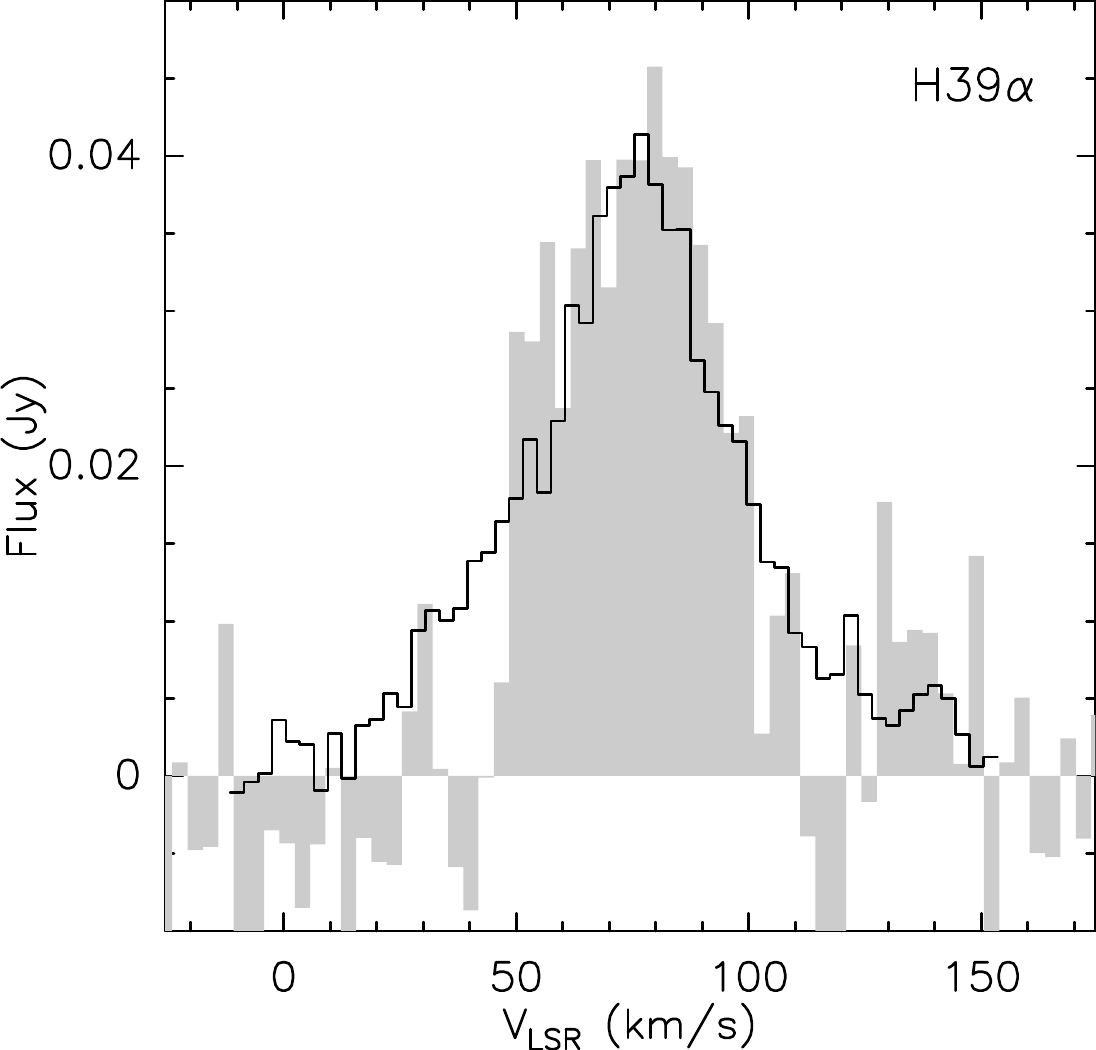}                 %1dspec 
     \includegraphics*[width=0.33\hsize]{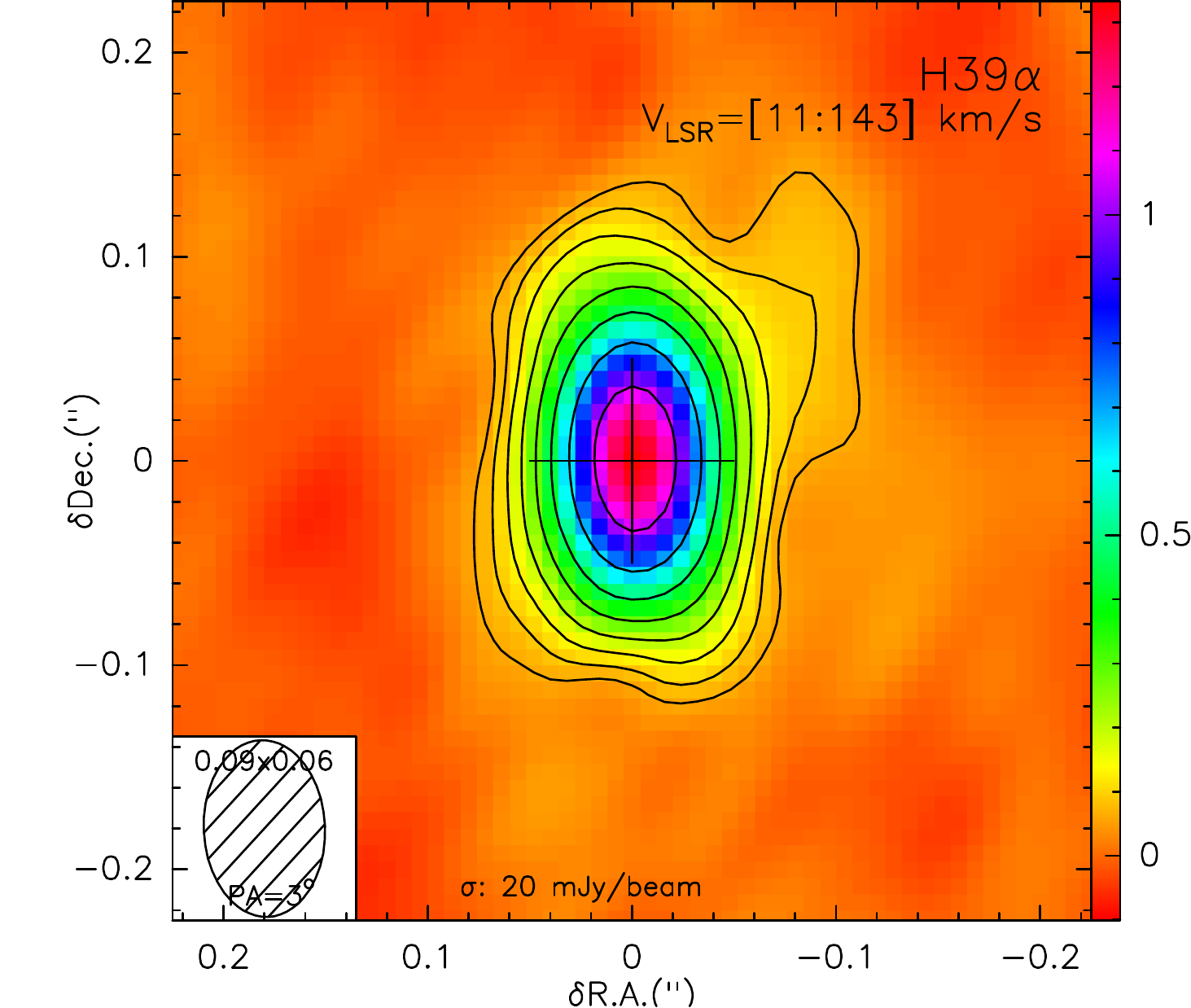}     %area             
     \includegraphics*[width=0.34\hsize]{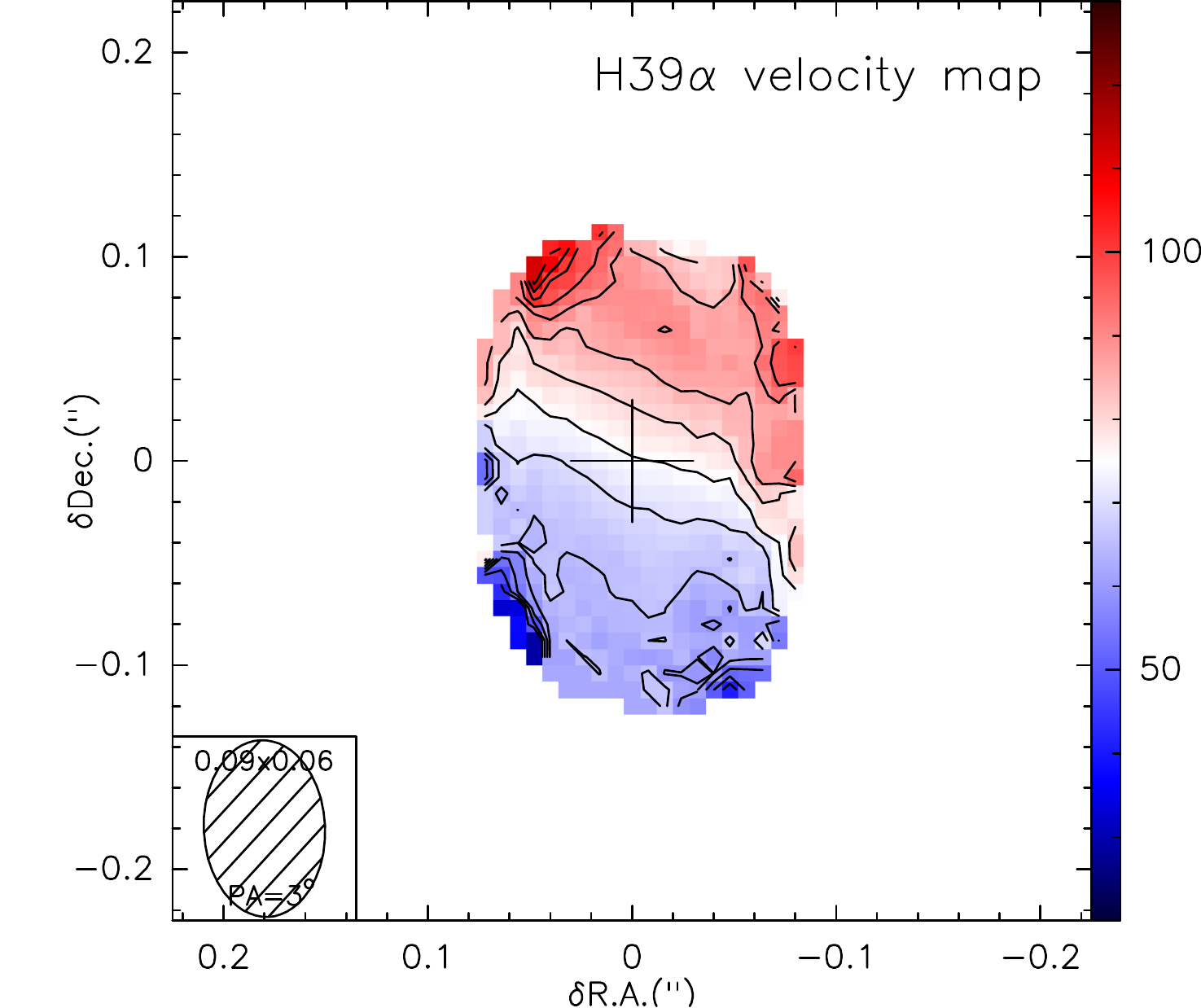}               %velo
         \caption{Summary of ALMA data of the \htal\ (top) and
           \htnal\ (bottom) lines (see also Figs.\,\ref{f-cubeH30a}
           and \ref{f-cubeH39a}). {\bf Left)} Integrated line spectrum
           obtained with ALMA (black lines) and with the
           \iram\ antenna (grey histogram, CSC17). {\bf Middle)} Line
           emission maps integrated over the line profile (LSR
           velocity range [11:143]\,\kms). The level contours are
           (3$\sigma$)$\times$1.5$^{(i-1)}$\,Jy\,beam$^{-1}$,
           $i$=1,2,3... {\bf Right)} First moment map. Contours going
           from \vlsr=45 to 115\,\kms\ by 5\,\kms.  The color bars
           indicate the \vlsr-colour relationship.
           \label{f-h30yh39}}
   \end{figure*}
   %% /pcdisk/jbell3/csanchez/m2-9/selfcal-csc2/:   @1dspec.greg   @H30area.greg 

%% ========================== Fig. H30a
%% ========================== Fig. H30a
   \begin{figure*}[htbp!]
     \centering
     \includegraphics*[width=0.45\hsize]{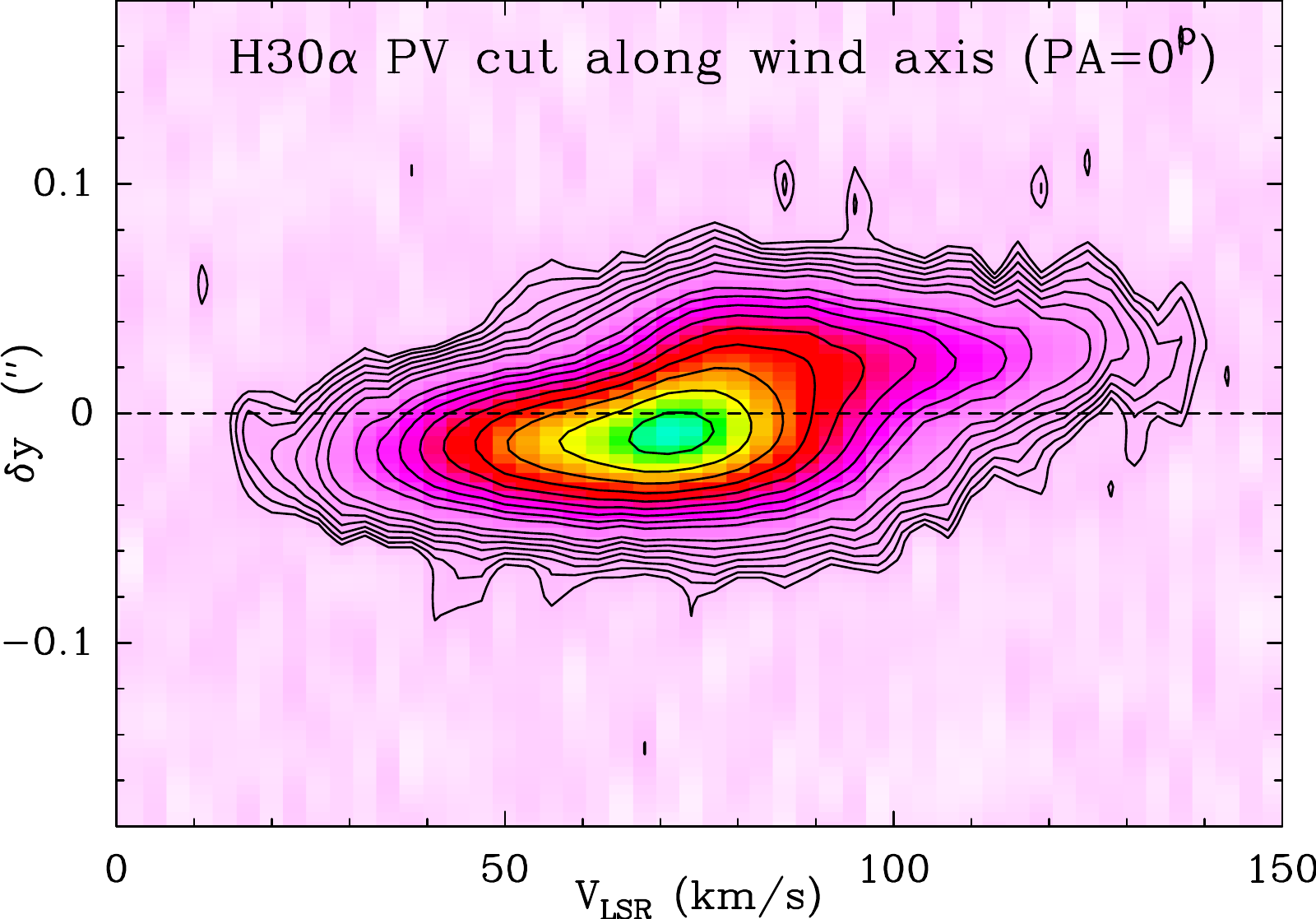} %% H30 PV-axis
     \includegraphics*[width=0.45\hsize]{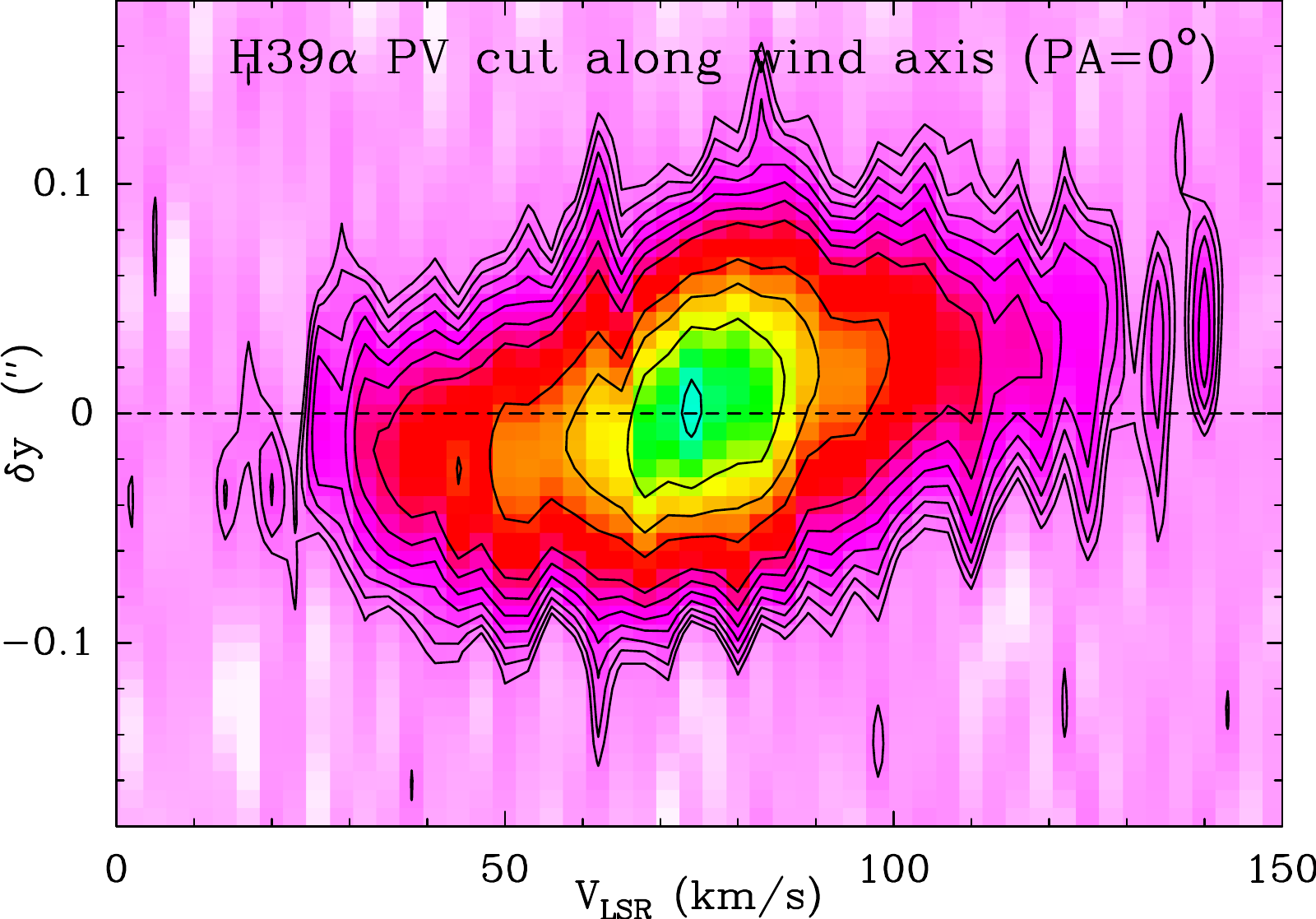} %% H39 PV-axis
     \\
     \includegraphics*[width=0.45\hsize]{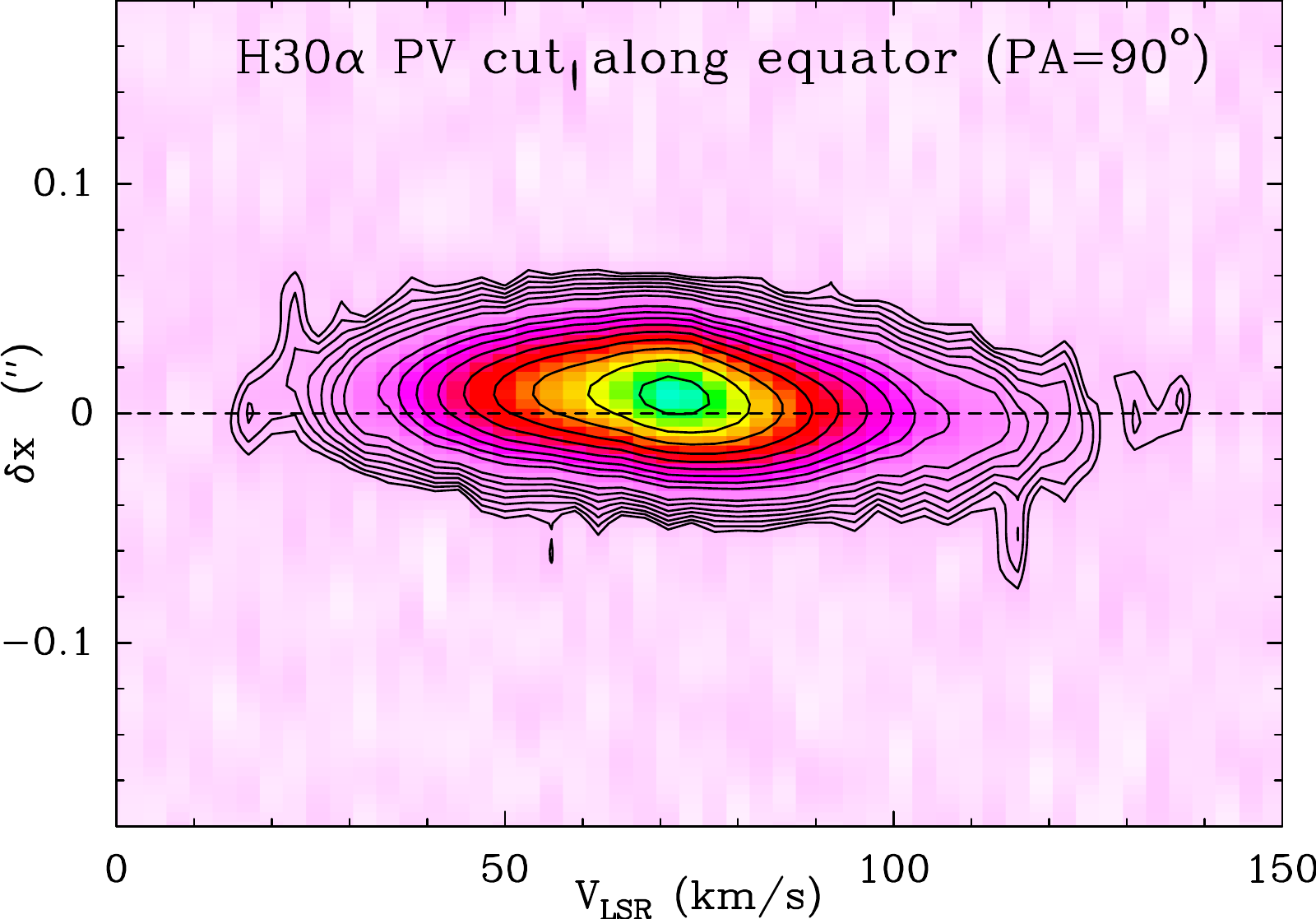} %% H30 PV-equ
     \includegraphics*[width=0.45\hsize]{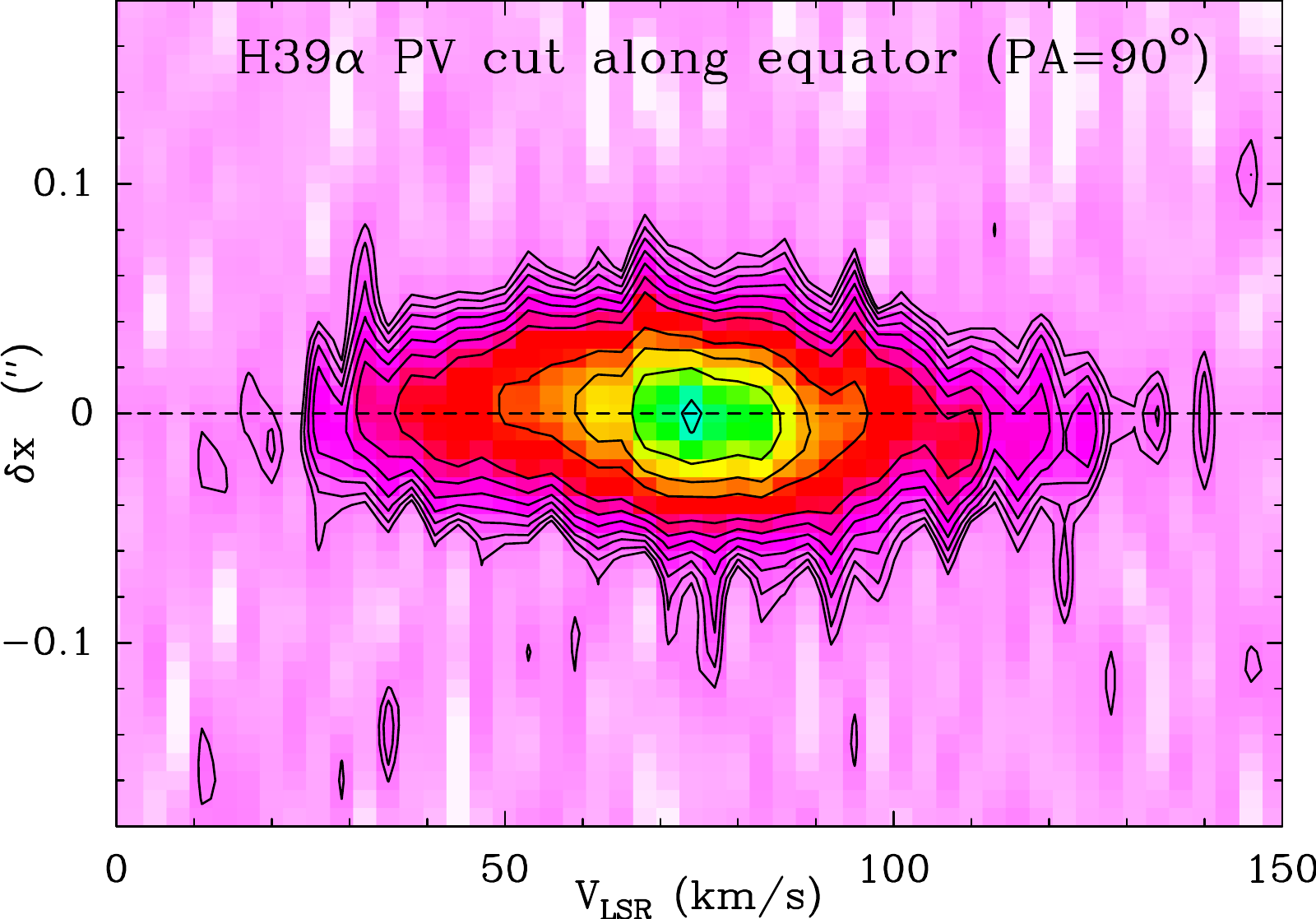} %% H39 PV-equ
         \caption{Position velocity cuts of the
     H30$\alpha$ (left) and \htnal\ (right) lines through the center along the wind
     axis (left, PA=0\degr) and the perpendicular direction
     (right).
     Levels are 2.5$\times$(1.3)$^{(i-1)}$ for \htal\ and 1.5$\times$(1.3)$^{(i-1)}$ for \htnal\ with $i$=1,2,3... 
           \label{f-pv}}
   \end{figure*}
   %% /pcdisk/jbell3/csanchez/m2-9/selfcal-csc2/:   @1dspec.greg   @H30area.greg 

%% ========================== Fig. H30a

%_____________________________________________________________
%%\input{t_rrls.tex}
\begin{table}[htbp]
\small
\caption{Line parameters from Gaussian fitting to the source-integrated profiles of the mRRLs in this study.} % title of Table
\label{t-parms}      % is used to refer this table in the text
\centering                          % used for centering table
\begin{tabular}{l c c c c c }        % centered columns (4 columns)
\hline\hline                  % inserts double horizontal lines
Line & Freq. &   \intil   &   \vlsr  & FWHM   & \il\  \\    % table heading 
     &  (GHz)    &  (Jy\,\kms) &  (\kms)  & (\kms) & (mJy)   \\
\hline                        % inserts single horizontal line
H30$\alpha$ & 231.900 & 13.3$\pm$0.3 & 73.0$\pm$0.4 & 44.9$\pm$0.9 & 280$\pm$5 \\
H39$\alpha$ & 106.737 & 2.07$\pm$0.08 & 75.1$\pm$0.7 & 54.0$\pm$1.7 &  36$\pm$1\\
H55$\gamma$ & 109.536 &  0.17$\pm$0.05 & 78$\pm$4 & 50$\pm$10 & 3.3$\pm$0.6 \\
He30$\alpha$  & 231.995  & 0.96$\pm$0.07 & 71.5$\pm$1.0 & 44$\pm$3 & 20$\pm$1 \\
\hce\ &  215.663 &  \dotfill &   \dotfill &   \dotfill & $<$12\tablefootmark{a} \\ %%rms~4mJy (en el espectro integrado)
\hsd\ & 95.964 &  \dotfill &   \dotfill &   \dotfill &  $<$1.2\tablefootmark{a}  \\ 
\hline                                   %inserts single line
\end{tabular}
\tablefoot{ 
\tablefoottext{a}{Upper limits correspond to three times the formal rms noise of the spectra for a $\Delta \upsilonup$\,$\sim$3\,\kms\ resolution.} \\
}
\end{table}
%
%_____________________________________________________________

The observational results from our ALMA data of the \htal\ and
\htnal\ transitions, which are main targets of this study, are
summarized in Figs.\,\ref{f-h30yh39} and \ref{f-pv}. The complete
velocity-channel maps of these transitions are reported in
Figs.\,\ref{f-cubeH30a} and \ref{f-cubeH39a}.
%In the appendix, we
%include a figure (Fig.\,\ref{f-iram}) that presents a comparison of
%the main line parameters (flux, width, and centroid) obtained from the
%ALMA maps and from the \iram\ data collected two years earlier by \cite{san17}. 
Additionally, we summarize our results for the He30$\alpha$
and \hcg\ transitions in Figures \ref{f-he30a} and \ref{f-h55g}.
The line parameters derived from the source-integrated line profiles of the
mRRLs detected are reported in Table\,\ref{t-parms}.
%% TBD 

%% AREA OF EMISSION: structure, morfology...
{\sl Morphology and dimensions of the mRRL-emitting region.}
Consistent with expectations, the \htal\ and \htnal\ emission
originate from a compact region that shares similar shape and
dimensions with the continuum emitting area at 1 and 3\,mm,
respectively.  The integrated intensity maps of the \htal\ line do not
exhibit any emission corresponding to the broad-waist component that
is observed in the 1\,mm-continuum maps at a similar frequency and
angular resolution (Fig.\,\ref{f-cont}). Since \htal\ exclusively
traces free-free emission, the lack of detection of the broad-waist
component in the \htal\ line supports our earlier conclusion that this
component is primarily associated with dust emission in the equatorial
region.

After deconvolution with the beam, the \htal\ line emission in our
ALMA maps (at a 3$\sigma$ level) extends along the axis to a radial
distance of $\sim$0\farc07 ($\sim$45au at $d$=650pc) from the
center. In the perpendicular direction, it extends to $\sim$0\farc045
($\sim$30\,au). As expected, the optically thicker \htnal\ line, extends
up to slightly larger radial distances of $\sim$0\farc11
($\sim$70\,au) and $\sim$0\farc063 ($\sim$40\,au) along the axis and
equator, respectively (after beam deconvolution and at a 3$\sigma$
level). The major-to-minor axis ratio is larger for the \htnal\ line
($\sim$1.75) compared to \htal\ ($\sim$1.5). This is well aligned with
a previously reported size-to-frequency dependency that differs along
the axis and along the equator \citep{lim03}.

%% KINEMATICS
{\sl Kinematics.} For both the \htal\ and \htnal\ transitions, we observe a clear
velocity gradient along the axis of the nebula. The emission shows a
redshift in the North (receding) lobe and a blueshift in the South (approaching) lobe
indicating a global expansion along the axis. 
%This
%pattern indicates a global expansion along the axis, since
%the North lobe is known to be receding and pointing away from our line of sight
%while the South lobe is approaching and pointing towards us.
%This finding is  consistent with the overall expansive kinematics of the large-scale
%nebula (Sect.\,\ref{intro}).
%% PV
The position-velocity (PV) diagrams shown in Fig.\,\ref{f-pv} offer
additional insights into the velocity structure of the ionized bipolar
wind in M\,2-9. The \htal\ emission exhibits a distinct S-shape
pattern in the axial PV diagram, indicating an abrupt increase of the
range of the line-of-sight velocities 
%% fast/abrupt
%% acceleration of the gas or
in two diametrically opposed, compact regions located at
offsets $\delta$y$\sim$$\pm$0\farc02 along the axis.  While the
S-shape is less pronounced in the \htnal\ transition, a similar
behaviour is observed. 
%%, i.e.\, an abrupt velocity rise from the center to a compact region with the largest velocity spread. 
As shown in Sect.\,\ref{model}, our model-based analysis of the
mRRL emission maps suggests velocities of $\sim$70-90\,\kms\ in these regions, 
which are well described by high-density, high-velocity shell-like structures.
We refer to these compact regions as the high-velocity spots/shells (HVSs).

%Assuming that the wind is flowing axially and
%Considering the inclination of M\,2-9's axis ($i$$\sim$17\degr) and
%the full width at zero intensity FWZI$\sim$130\,\kms\ of the mRRLs, the ionized bipolar wind
%reaches maximum deprojected expansion velocities of $<$220\,\kms.
%% \ at the outer boundary ($\sim$0\farc03-0\farc04).
%The upper limit arises because the wind is not flowing axially but radially, 
%% the expansion is not perfectly directed along the axis,
%as we find from our model-based analysis of the
%mRRL emission maps (Sect.\,\ref{model}).
%This value of the expansion velocity is rather moderate and significantly smaller than that inferred from
%the broad H$\alpha$ emission profile from the central core, which is on the order of a few thousand \kms.
% En estos altos valores (deducidos de Halpha) se ha sustentado la hipotesis de que hay un objeto compacto (WD?) en el core de m2-9.
% A la vista de nuestros datos, esto ya no es necesario. 

\mbox{The average velocity gradient observed in the inner regions} of \mbox{the
ionized wind is
\vgrad$\sim$1150\kms\,arcsec$^{-1}$$\sim$1.8\,\kms\,au$^{-1}$}, which
results in extremely short kinematical ages of less than one year
(after deprojection considering $i$$\sim$17\degr, Sect.\,\ref{m29}).
%% 1.8 km/s au-1 = 1.8e5 cm/s /1.5e13cm = 1.2e-8 s-1 --> t ~1/1.2e-8 ~ 2.6 anos  --> deprojection xtan(17) ~0.8 yr  
This implies that observable changes in the profile of the mRRLs can occur
in less than a year if there are variations on a similar timescale in
the mass-loss rate, structure, or kinematics of the ionized wind.
The PV diagrams along the equator of both \htal\ and \htnal\ reveal
the presence of a subtle velocity gradient perpendicular to the
lobes. Specifically, the east side of the equator
($\delta$x$>$0\arcsec) exhibits blue-shifted velocities, while the
west side ($\delta$x$<$0\arcsec) shows red-shifted velocities. This is
most evident in the velocity (first moment) maps of the lines, where
the isovelocity contours are inclined rather than running parallel to
the equatorial direction (Fig.\,\ref{f-h30yh39}). This behaviour suggests rotation
(counterclockwise) of the ionized bipolar wind.

%% LINE PROFILES 
{\sl Line profiles.} Both the \htal\ and \htnal\ transitions show a
nearly Gaussian source-integrated profile. The \htal\ line has a peak
of emission at \vlsr=73$\pm$0.4\,\kms\ and full width at half maximum
of FWHM$\sim$44.8$\pm$0.9\,\kms. The weaker \htnal\ transition is
centered at a slightly larger velocity \vlsr=75.1$\pm$0.7\,\kms\ and
is somewhat broader, with a FWHM$\sim$54.0$\pm$1.7\,\kms. For both
transitions, weak emission wings are observed with FWZI$\sim$130\,\kms.

The comparison of the source-integrated line profiles as observed with
ALMA and with the \iram\ single-dish reveals apparent differences
(Figs.\,\ref{f-h30yh39}). The most significant changes are observed in
the \htal\ line, which exhibits a notable increase in brightness (by a
factor of $\sim$2.2 in line flux and $\sim$1.6 in line peak intensity)
and broadening of the profile, with the FWHM increasing by
$\sim$11\,\kms.  For \htnal, we do not confirm an increase in
brightness, but the line profile is broader by $\sim$14\,\kms\ in
FWHM.  In both transitions, the presence of broad wings in the ALMA
observations that were not detected in the previous \iram\ data
suggests that these extended features have recently emerged or become
more prominent. These differences are unlikely due to mispointing or
flux calibration uncertainties \cite[within nominal values of
  $\sim$2-3\arcsec\ and $\sim$20\% for the \iram\ data,][]{san17}, as
these factors do not account for the observed variation in profile
width.

The observed temporal variations of the mRRL profiles indicate changes
in the gas dynamics or physical conditions of the central ionized wind
of M\,2-9 occurring over short timescales of $\sim$2\,yr or less.
This is not surprising given the short kinematic age estimated for the
inner regions of the mRRL-emitting regions.  Changes in the profiles
of mRRLs on short ($\sim$2-25\,year-long) timescales have been
reported in the pPNe CRL\,618 \citep{san17} and the post-supergiant
candidate MWC\,922 \citep{san19}.
%% en crl618 son ~25 años. 
%% en MWC922 son ~2años. 
\subsection{CO lines}
\label{res-co}

%% ========================== Fig. CO
   \begin{figure}[htbp!]
     \centering
     \includegraphics*[width=0.90\hsize]{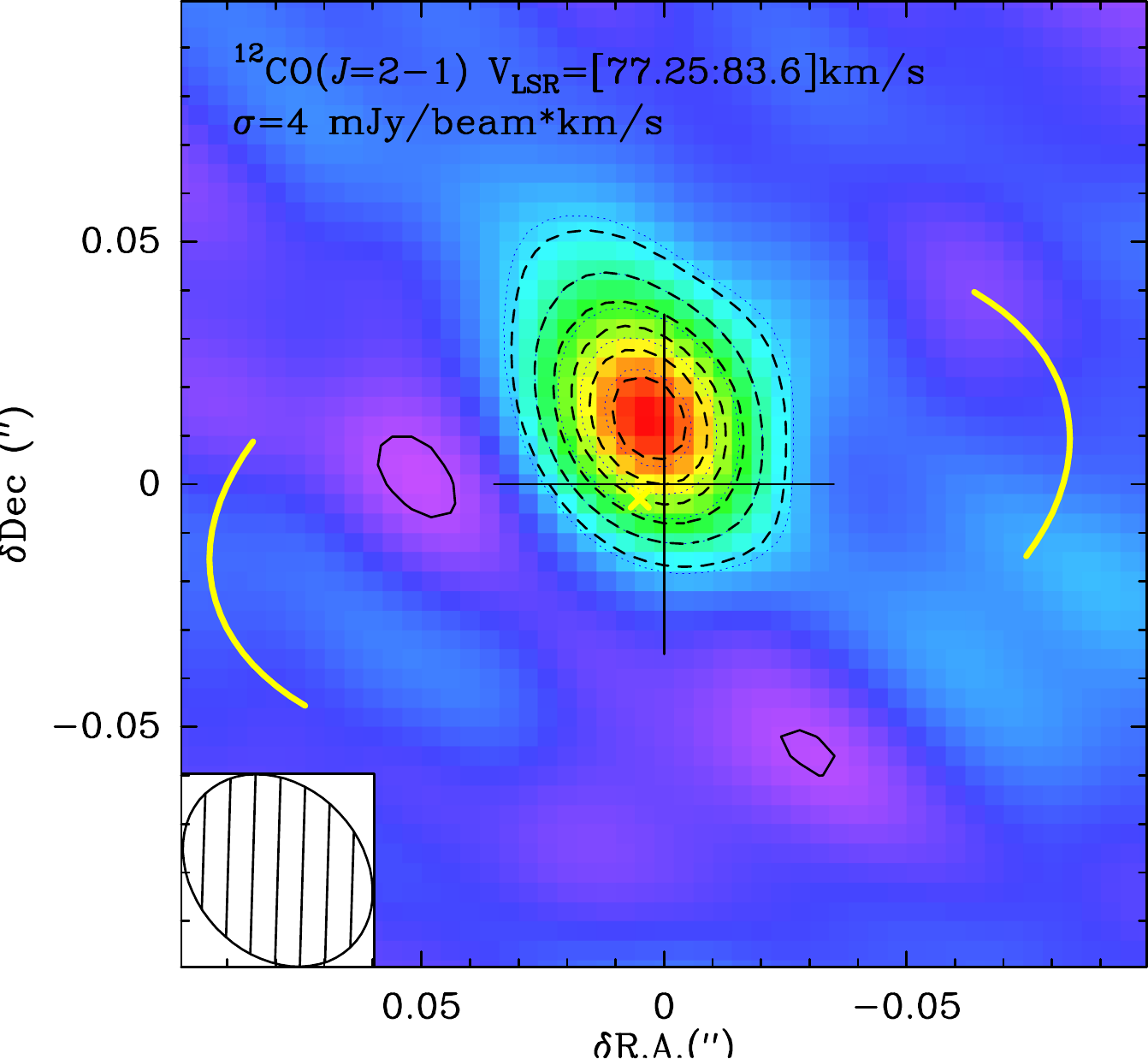}
     \includegraphics*[width=0.90\hsize]{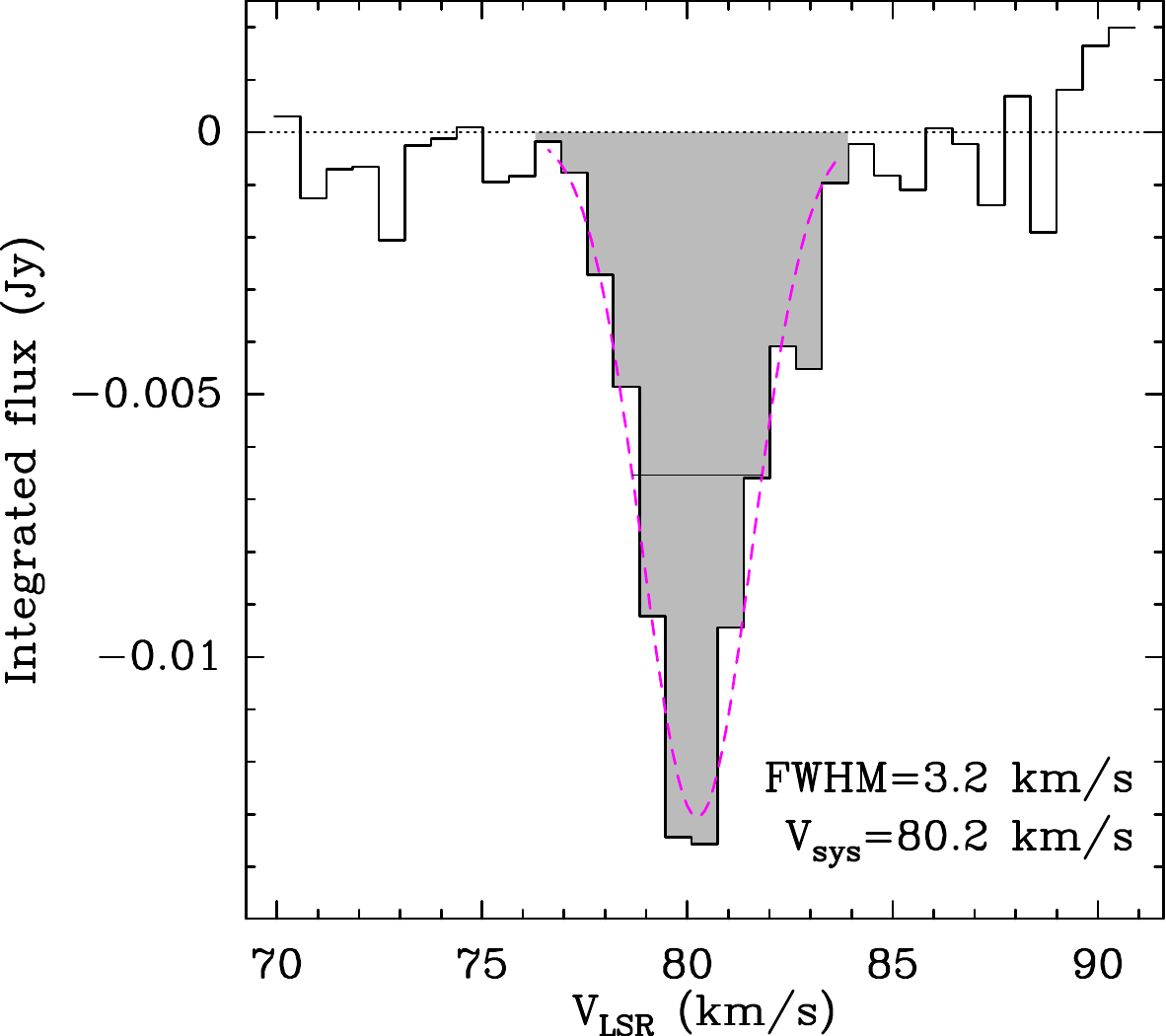}
      \caption{Compact \docem\,($J$=2--1) absorption feature observed
        toward the central regions of M\,2-9. {\bf Top)} Integrated
        intensity map of the CO absorption in the velocity range
        \vlsr=[77.25:83.6]\,\kms. Dashed and solid lines are used for
        negative and positive contours, respectively. Level spacing is
        2$\sigma$ to $-$13$\sigma$ by $-$2$\sigma$
        ($\sigma$=4\,mJy/beam). The yellow arcs represent the outer
        boundary of the broad-waist structure, plausibly a compact dust
        disk, observed in the 1\,mm-continuum maps. {\bf Bottom)} Line
        profile integrated over the area where CO absorption is
        observed. \label{f-co}}
   \end{figure}
   %% /pcdisk/jbell3/csanchez/m2-9/selfcal-csc2/COspec.greg >> 1d-spectrum (integrated over the area where absorption is observed).
   %% /pcdisk/jbell3/csanchez/m2-9/selfcal-csc2/COarea.greg >> First moment map integrated over channels 12:22 [77.25:83.6]km/s
   %%@broad-waist pinta el disco circumbinario

%% ========================== Fig. CO

%   {\bf Mapas del anillo interior (el outer esta muy filtrado)? ACC
%     uvtaper? 13co1-0 nunca se ha publicado y el 12co2-1 tiene mejor
%     resolucion que el de NOEMA, puede tener su intereres ponerlos.}
   
While our study primarily focuses on mRRLs, we simultaneously
mapped with ALMA the \docem\,2--1 and \trecem\,1--0 lines (Table\,\ref{tab:spws}).
As discussed in Sect.\,\ref{intro}, CO (sub)mm-wavelength emission traces
two off-centered, expanding equatorial rings (with radii of $\sim$1\farc1 and $\sim$3-4\arcsec) that have been previously mapped
with the NOEMA and ALMA interferometers and extensively studied by 
\cite{cc12} and \cite{cc17}, respectively. 
As a result, we will not study the CO emission from the rings in this work. 
%as the structure,
%dynamics, and physical properties of these rings have already been
%well characterized in the previous works mentioned.
%Moreover, the CO emission from the rings in our maps is significantly
%affected by interferometric flux losses due to the size of the rings
%(particularly the outer one) and the ALMA configuration of our
%observations.
%Nonetheless, we
%provide the maps of both transitions in the appendix for completeness (Fig.\,XX).

Our only focus will be on a newly discovered absorption component
detected towards the continuum source both in the \docem\ and
\trecem\ observed transitions (Fig.\,\ref{f-co} and Fig.\,\ref{f-13co}).
We observe a narrow (FWHM$\sim$3\,\kms) CO absorption feature
below the continuum level near the center of the nebula at velocity
\vlsr$\sim$80--81\,\kms, i.e., redshifted relative to the centroid of
the mRRLs. The absorption indicates that the excitation temperature in
the CO absorbing layers is lower than that of the continuum background
source, which corresponds to the compact collimated ionized wind.  
Interestingly, the CO absorption feature is not 
centered at the nebula's center but is slightly offset by
$\delta$y$\sim$0\farc012 towards the north. Given that the north lobe is moving
away from the observer (with an inclination $i$$\sim$17\degr), this offset suggests that the
structure responsible for the absorption is likely an equatorial torus
or disk surrounding the collimated ionized wind and positioned
perpendicular to the lobes. In this configuration, the front side of
the disk partially blocks the emission from the base of the receding
north lobe, resulting in the observed CO absorption.

The observed redshift rules out that the absorption is produced
neither in the arcsecond-scale central rings nor in the
hourglass-shaped structure emerging from them reported by
\citep{cc17}, since all these structures are in expansion and would
necessarily produce a blue absorption given their spatio-kinematic
structure, which is very well characterized in the mentioned previous
works.  The redshift of the CO absorption feature by
  $\sim$6.2\,\kms\ (deprojected velocity adopting $i$=17\degr) with respect to
the velocity centroid of the mRRLs (Table\,\ref{t-parms}) ,
%% background collimated ionized wind
likely signifies gas infall movements from the disk toward the central source. This is
further discussed in Sect.\,\ref{disk_co}.
%, where the physical conditions
%and kinematics of the compact equatorial disk are constrained.

\section{Analysis}
\label{analysis}

\subsection{Non-LTE radiative transfer modeling of mRRL and free-free continuum emission}
\label{model}

%In order to constrain the structure, phyical conditions, and
%kinematics of the collimmated ionized wind emerging from M\,2-9,
We have modelled the free-free continuum and mRRL emission in M\,2-9
using the non-LTE three-dimensional radiative transfer code \coral\ (Code
for 3D Computing Continuum and Recombination Lines). The code and our modeling approach are 
described in detail in Appendix \ref{ap-co3ral} and \ref{ap:approach}, respectively.

%\( \vec{V}(r) \)
%\(\overrightarrow{V}(r)\)

Our modeling of the compact ionized wind in M\,2-9 with \coral\ began
by employing the identical physical model utilized by CSC17 (see
their Table 4).  This model has been previously shown to accurately
replicate the single-dish free-free continuum flux measurements and
mRRL profiles observed in 2015 by these authors.  As expected, this
original model (which generates synthetic single-dish data closely resembling
those from MORELI\footnote{MORELI is the code used in the analysis by
CSC17 \citep{bae13}.}) fails to capture some of the characteristics
of the ionized wind now revealed by our ALMA
$\sim$0\farc03-0\farc06-resolution maps (Appendix \,\ref{ap:csc17}, Fig.\,\ref{f:csc17}).

%%%% FIGURES: Input MODEL
% Fig.---------------------------------------
   \begin{figure*}[ht!]
   \centering 
   \includegraphics*[width=0.45\hsize]{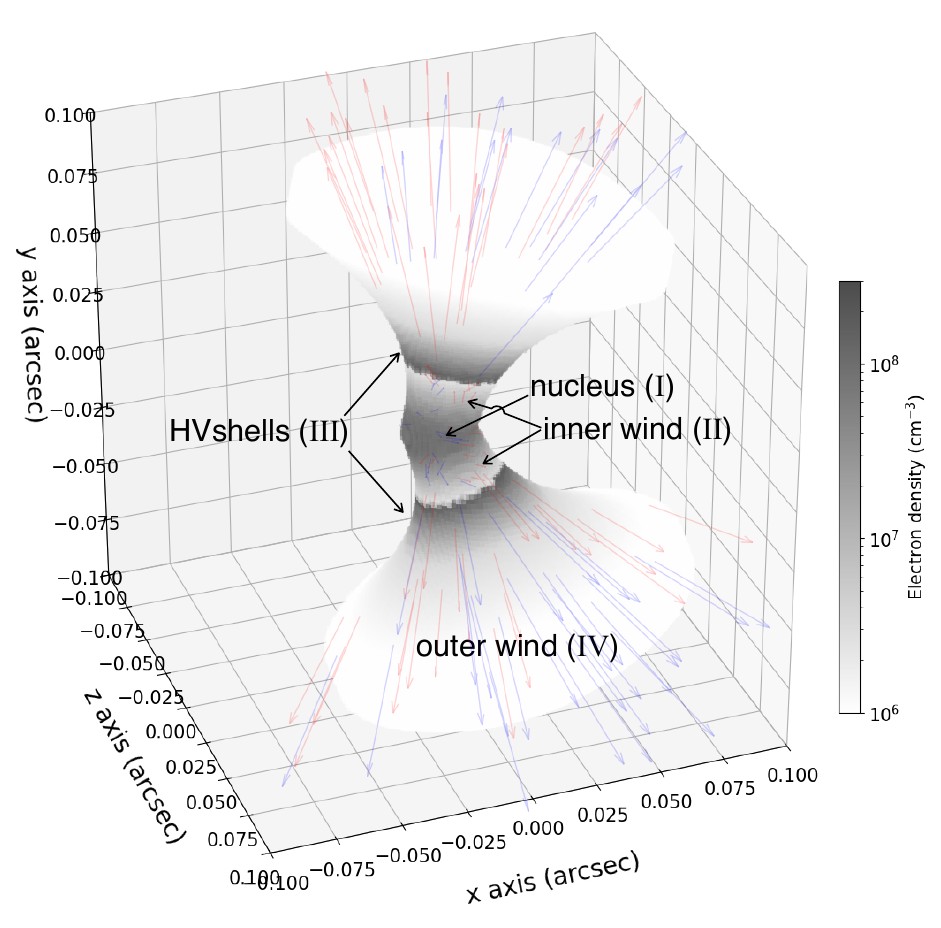}
   \includegraphics[width=0.475\hsize]{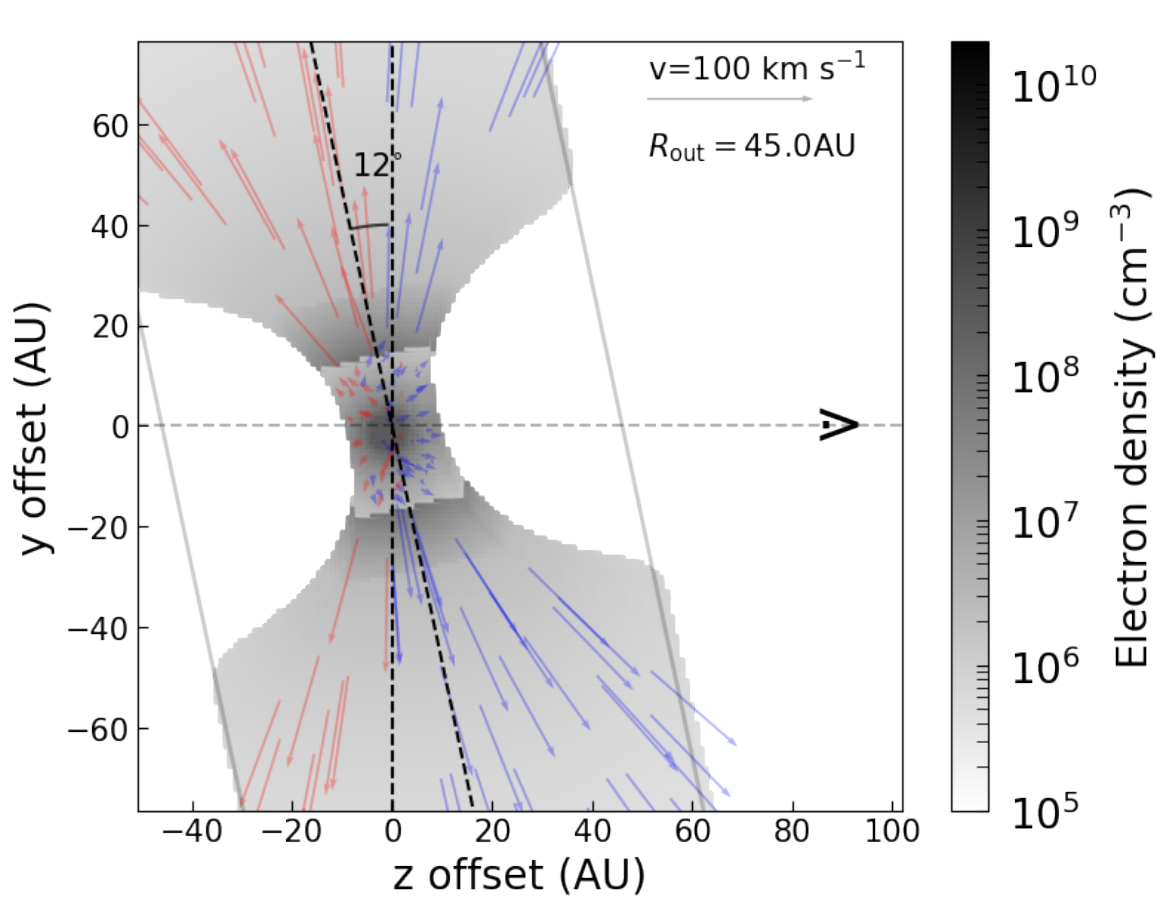}
   \includegraphics[width=0.475\hsize]{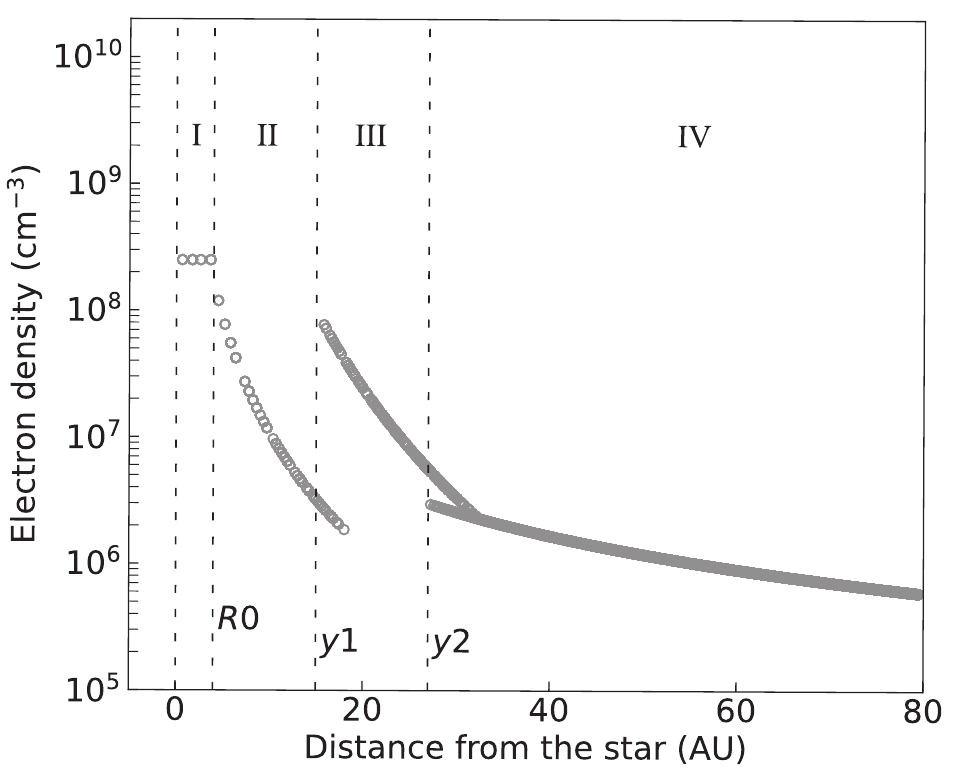}
   \includegraphics[width=0.475\hsize]{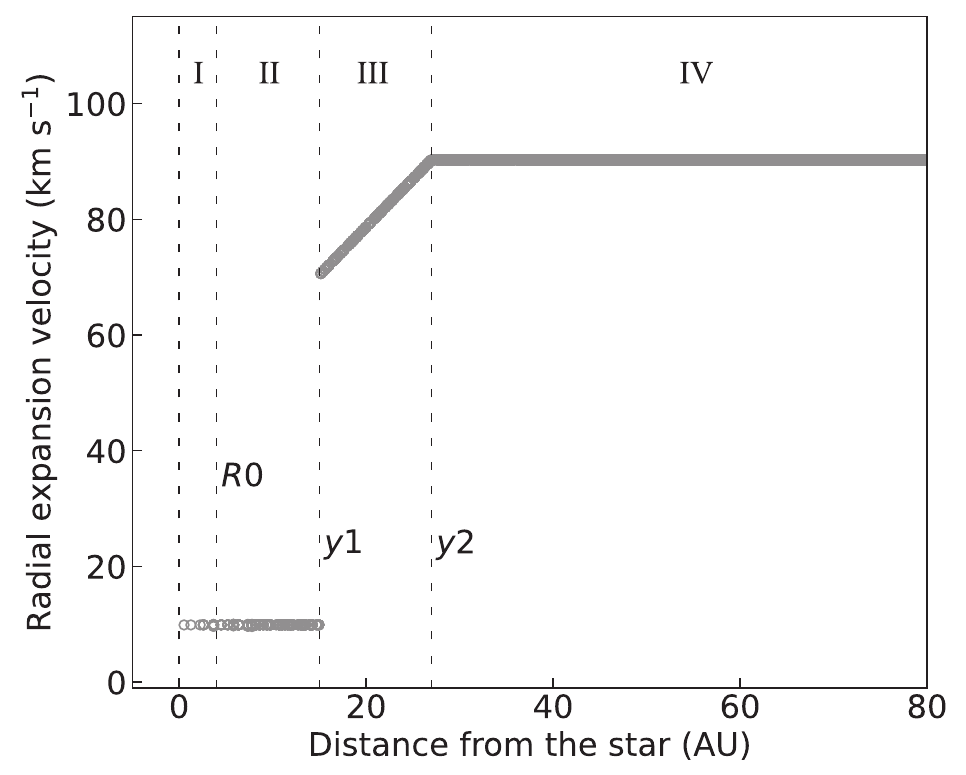}
   \caption{Schematic view of the geometry and main parameters of our
     model of the ionized core of M\,2-9 (Sect.\,\ref{model} and Table\,\ref{t-model}). Top-left)
     3D view indicating the four major structural components in our
     model, namely: nucleus (I), inner wind (II), high-velocity
     spots/shells (HVS, III), outer wind (IV). Colour scale represents
     density as indicated in the right wedge; arrows represent the
     velocity field (colour indicate blue or red Doppler shift with
     respect to the line of sight). The line of sight runs along the
     z-axis. Top-right) 2D view through the x=0 offset. Bottom panels
     show the radial distribution of the electron density (left) and
     radial expansion velocity modulus (right) adopted in our model. Vertical dashed lines indicate the boundaries of regions I-IV.
          \label{f-model}
        }
   \end{figure*}
   %% Fig.------------------------------------------------------------------------------------

   %% Table with input parameters of the model:
\begin{table}
\caption{Properties of the central ionized core of M\,2-9 derived from our radiative transfer modeling (Sect.\,\ref{model} and Fig.\,\ref{f-model}).}
   \small
\label{t-model}
\centering 
\begin{tabular}{l c}        % 
  \hline\hline                      % inserts double horizontal lines
%Parameter & M\,2-9 \\ 
%\hline
\multicolumn{2}{c}{\it Input parameters} \\ 
Distance ($d$) & 650\,pc\tablefootmark{*} \\ 
LSR Systemic velocity (\vsys) & $+$78\,\kms \\ 
Geometry & {\it C-shaped horn}\tablefootmark{\dag} \\
Inclination ($i$) & 12\degr \\ %& \cite{sto95}, b31=b30\\  
%Opening angle ($\theta_{\rm a}$) & 45\degr \\
%Opening angle exponent ($\alpha$) & 2.5\degr \\
Equatorial radius ($R_{\rm eq}$) & 8\,au \\
Outer radius (\rout) & 40\,au \\ 
%Cavity inner radius (\rin)  & 1\,au  \\
Nucleus outer radius ($R_0$) & 4\,au \\
HVSs inner boundary ($y_1$) & 15\,au \\
HVSs outer boundary ($y_2$) & 28\,au \\

Electron temperature (\te) & 15000\,K \\
Electron density ($n_{\rm e}$): & \\
\hspace{0.5cm} nucleus (I) & 2.5\ex{8}\,cm3\ \\
\hspace{0.5cm} inner wind (II) & 1.7\ex{8}($\frac{r}{4{\rm au}}$)$^{-3.0}$\,\cm3\  \\
\hspace{0.5cm} HVSs (III) & 1.0\ex{8}($\frac{r}{15{\rm au}}$)$^{-4.7}$\,\cm3\  \\
\hspace{0.5cm} outer wind (IV) & 2.5\ex{6}($\frac{r}{28{\rm au}}$)$^{-1.4}$\,\cm3\  \\ 
Kinematics & {\it radial expansion + rotation} \\ 
Expansion velocity (\vexp):  & \\ 
\hspace{0.5cm} nucleus (I) & 10\,\kms\ \\
\hspace{0.5cm} inner wind (II) & 10\,\kms\ \\
\hspace{0.5cm} HVSs (III) & 70+1.5$\times$($r$-15\,au)\,\kms\\
\hspace{0.5cm} outer wind (IV) & 90\,\kms\ \\ 
Rotation velocity (\vrot) & 7-10\,\kms\ \\
%\\
%\hline
\\
\multicolumn{2}{c}{\it Implied physical properties of the jet} \\ 
Ionized mass ($M_{\rm ion}$) & 4.3\ex{-6}\msun\ \\   
Mass-loss rate (\mloss) &  7.2\ex{-7}\,\my\ \\ 
Kinematic age (\tkin) & 1.0-10.3\,yr ($\sim$6\,yr) \\ %% rango, desde las partes mas internas a las mas externas...
Scalar momentum ($P$)\tablefootmark{\ddag} & 6.4\ex{34}\,g\,cm/s \\  %% L/c=7.72e25
%% g cm/s --> can be compared with the “momentum” carried by radiation
%% pressure per unit time, calculated as L/c. Sale un factor 30 mayor,
%% no es overluminous.
Total kinetic energy ($E_{\rm k}$) & 2.8\ex{41}\,erg \\
Mechanical luminosity (\mloss$\times$\vexp) & 1.5\ex{33}\,erg/s\\ 
\hline 
\end{tabular} \\
\tablefoot{In all cases, $r$ is the radial distance to the center. Departure coefficients $b_n$ are from \citep{sto95}. Regions I-IV within the wind are shown in Fig.\,\ref{f-model}. The model presented has 333$^3$ cubic cells of 0\farc0015$\times$0\farc0015$\times$0\farc0015 ($\sim$1\,au$\times$1\,au$\times$1\,au).
\tablefoottext{*}{Adopted (see Sect.\,\ref{intro}).} 
\tablefoottext{\dag}{See Fig.\,\ref{f-model} for details.} 
\tablefoottext{\ddag}{Linear momentum defined as a ``scalar'' magnitude computed as the sum of the momentum moduli for every cell
  as defined in \cite{buj01}.}
}
\end{table}
%%%%%%%%%%%%%%%%%%%%%%%%%%%%%%%%%%%%%

%\begin{equation}   (Req+(Rout-Req)*(tan(theta|180)*(yp|Rout)))^exp.
%\[ \text{rcil} < \left( \text{Rout\_eq} + (\text{Rout} - \text{Rout\_eq}) \cdot \left( \tan\left(\frac{\text{op\_angle} \cdot \pi}{180}\right) \cdot \frac{|\text{yp}|}{\text{Rout}} \right)^{\text{op\_ang\_exp}} \right) \] 
%\end{equation}
%______________________________________________________________________________________________

\begin{figure}[ht!]
   \centering 
   \includegraphics[width=0.95\hsize]{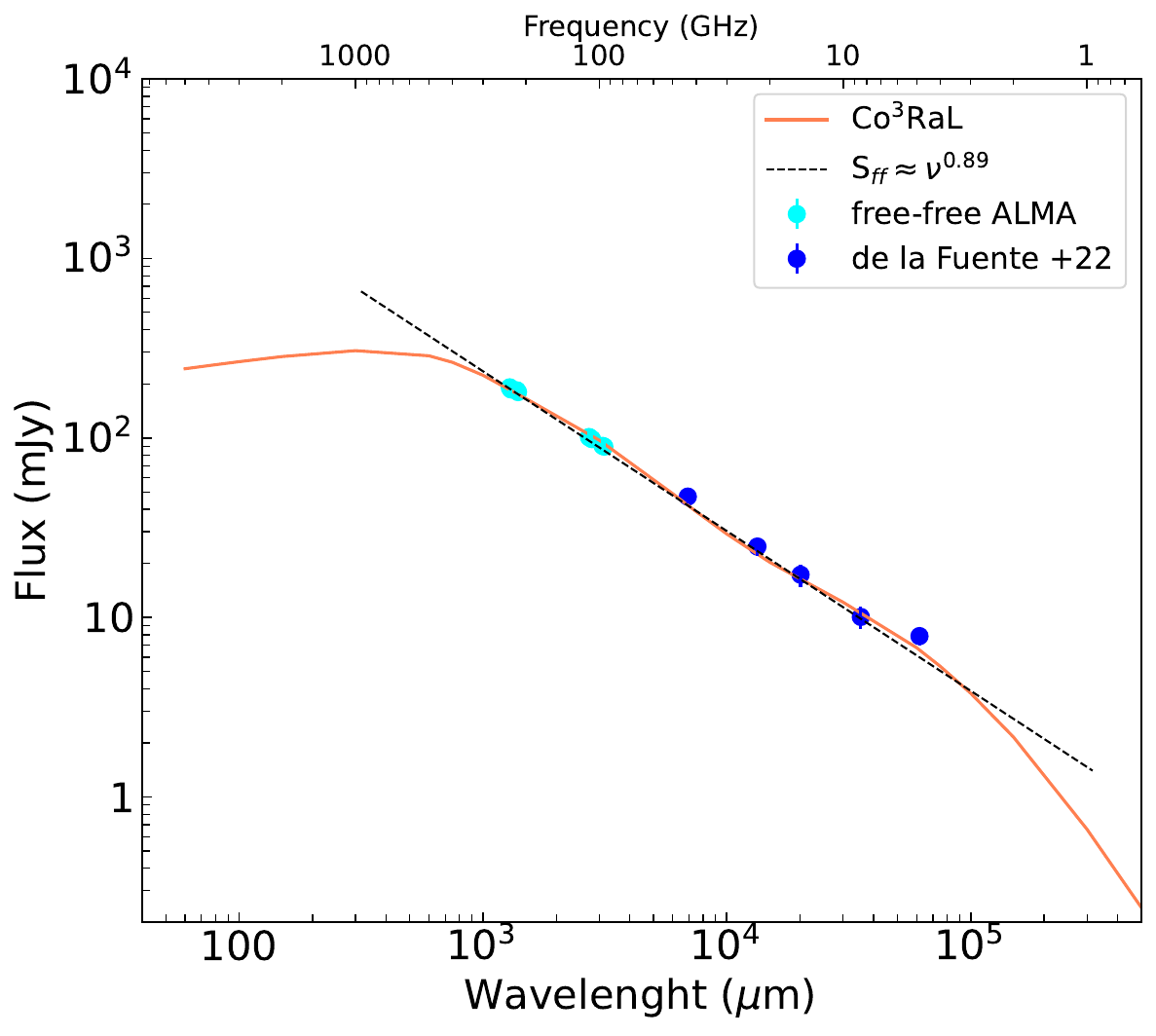}  %% es igual a la 297 (no esta puesta)
   \includegraphics[width=0.99\hsize]{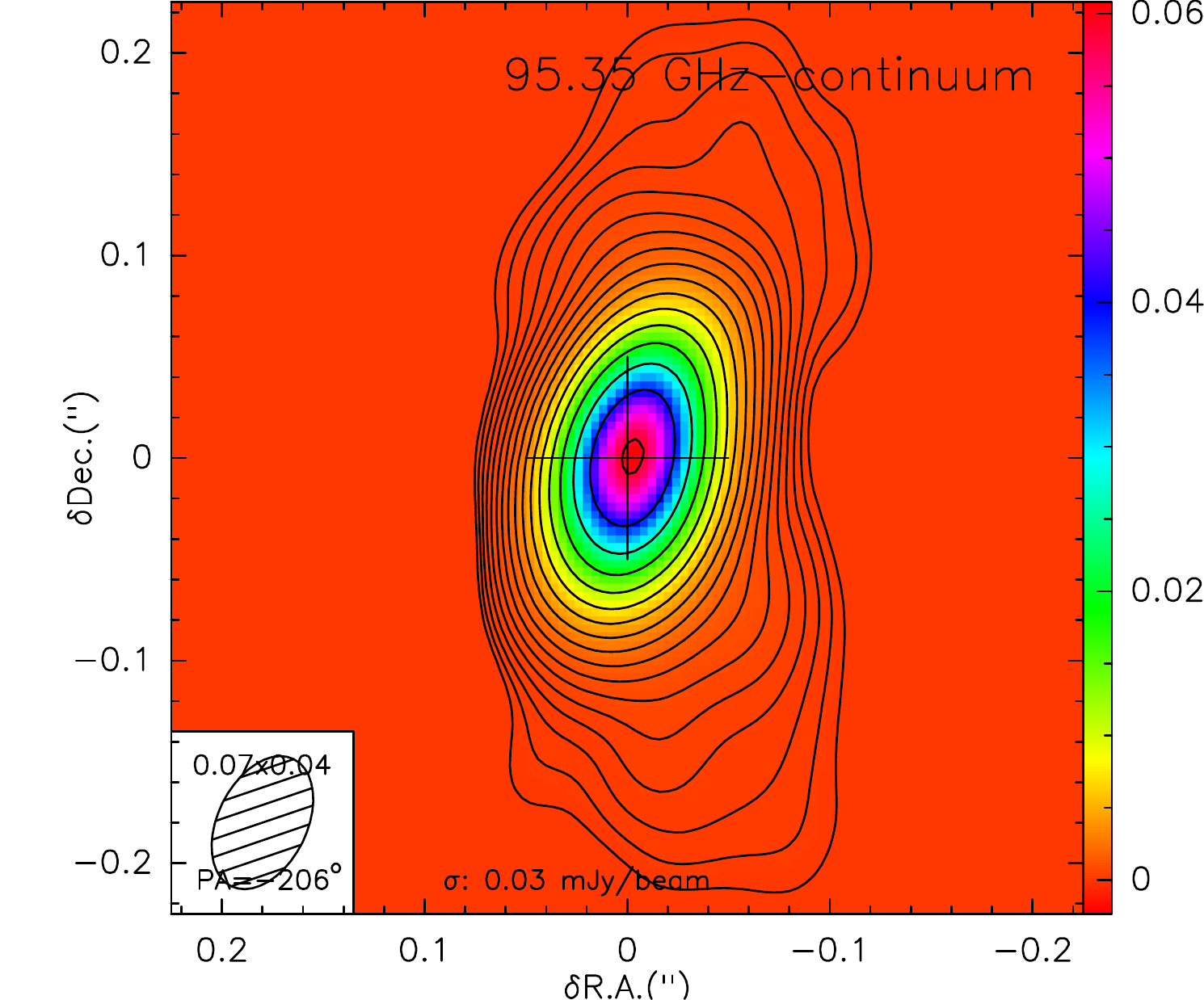} %% 297 final; 299 = 297 con resol 1au  

   \caption{Free-free continuum model predictions. Top) Synthetic SED
     (red) along with empirical values of, and fit to, the mm-to-cm
     continuum flux as in Fig.\,\ref{sed-alma}-right. Bottom)
     Synthetic 3mm-continuum map as in Fig.\,\ref{f-cont} top-right.
          \label{f-contmod}
        }
   \end{figure}

\begin{figure*}[htbp!]
     \centering
     \includegraphics*[width=0.30\hsize]{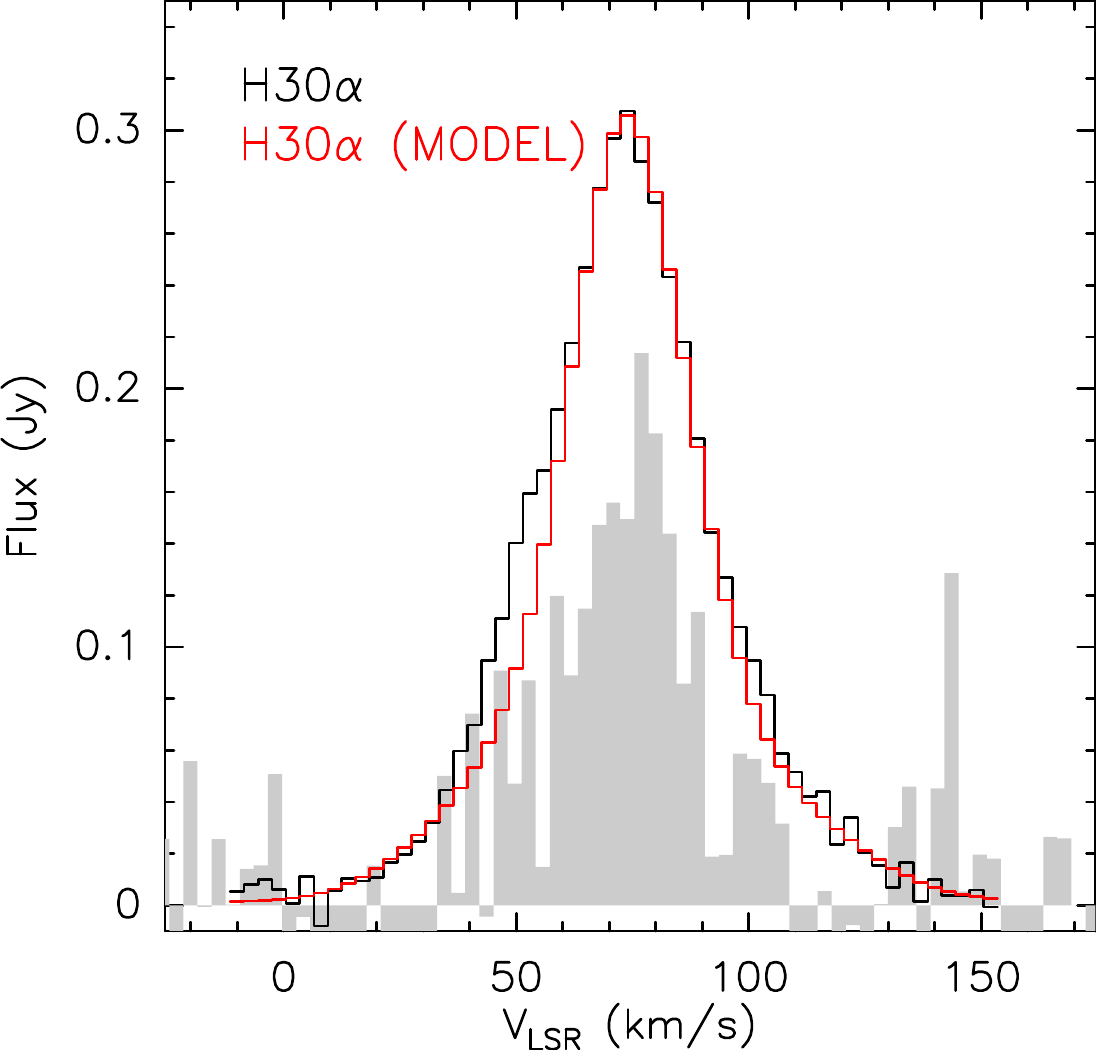} 
     \includegraphics*[width=0.33\hsize]{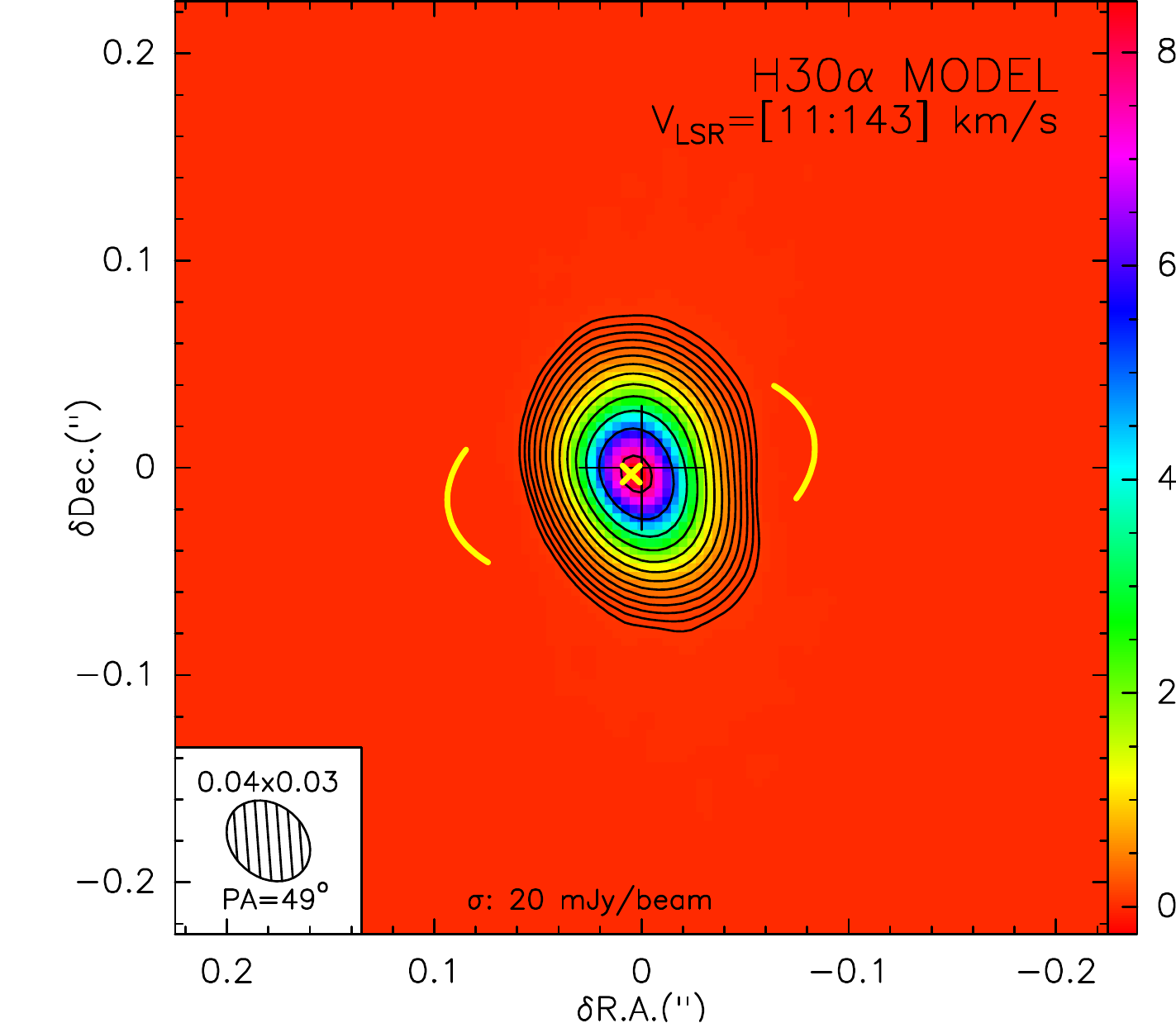} %% area MODEL
     \includegraphics*[width=0.34\hsize]{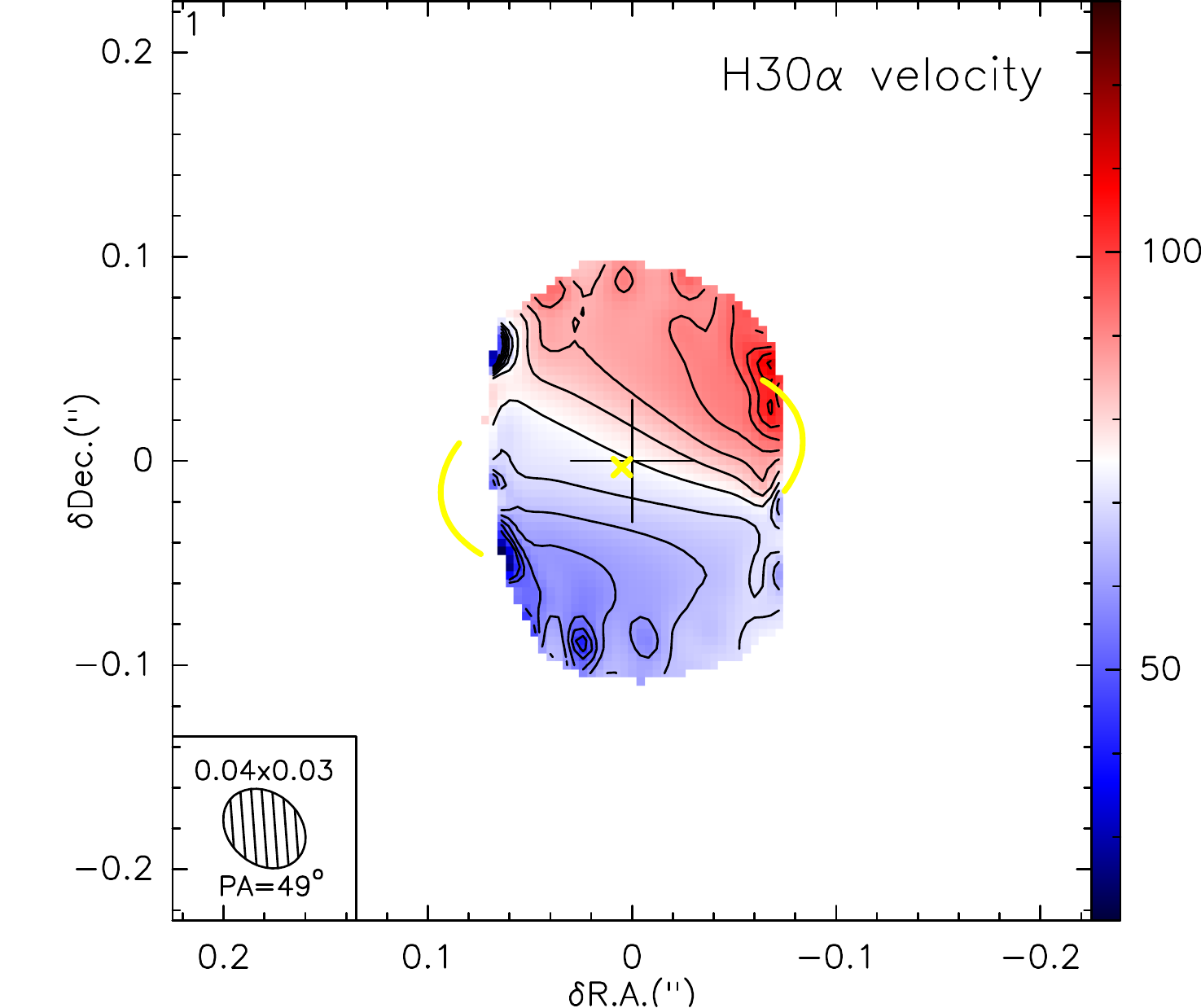} %velo MODEL 
     \\
     \includegraphics*[width=0.30\hsize]{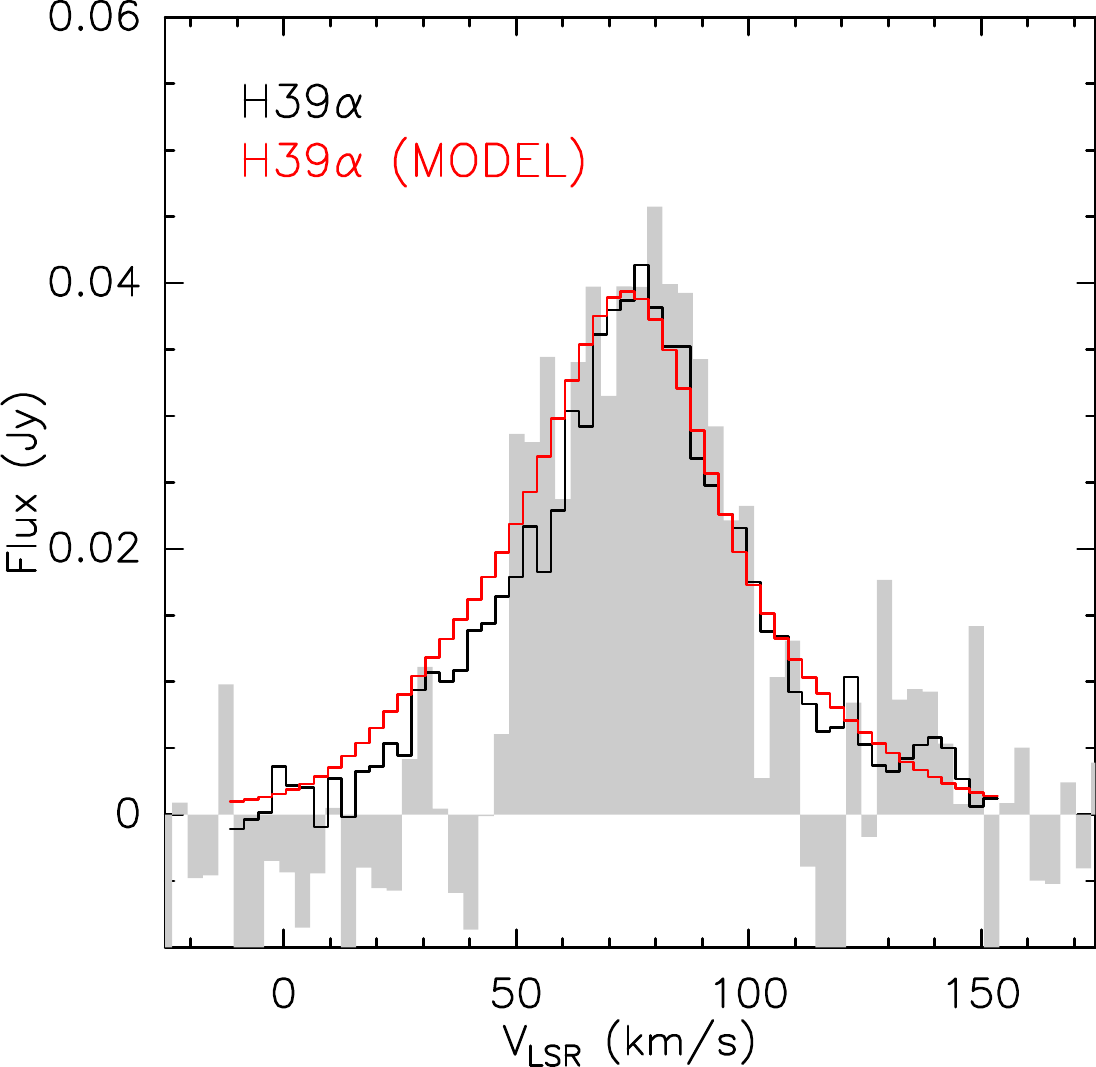} 
     \includegraphics*[width=0.33\hsize]{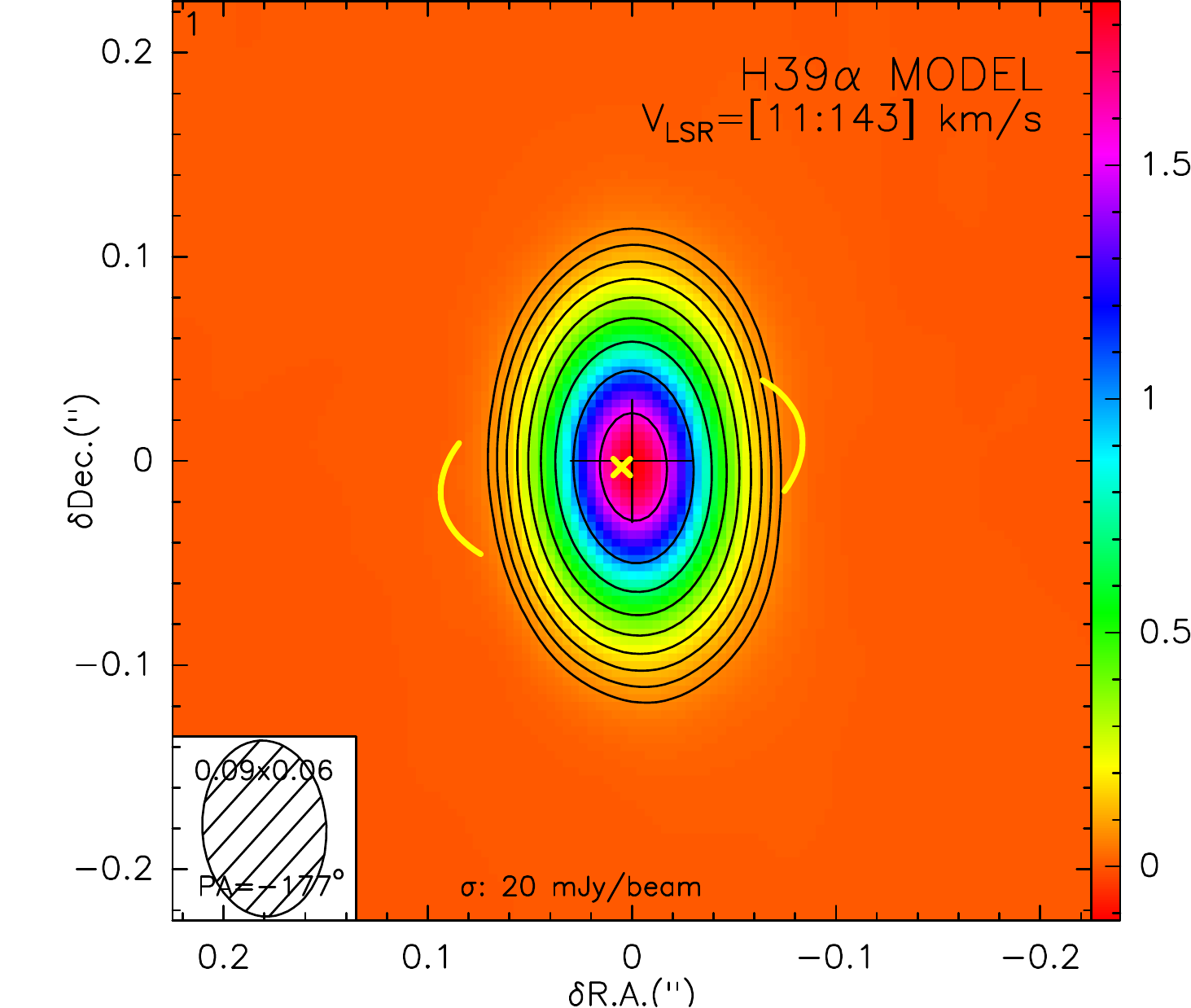} %% area MODEL
     \includegraphics*[width=0.34\hsize]{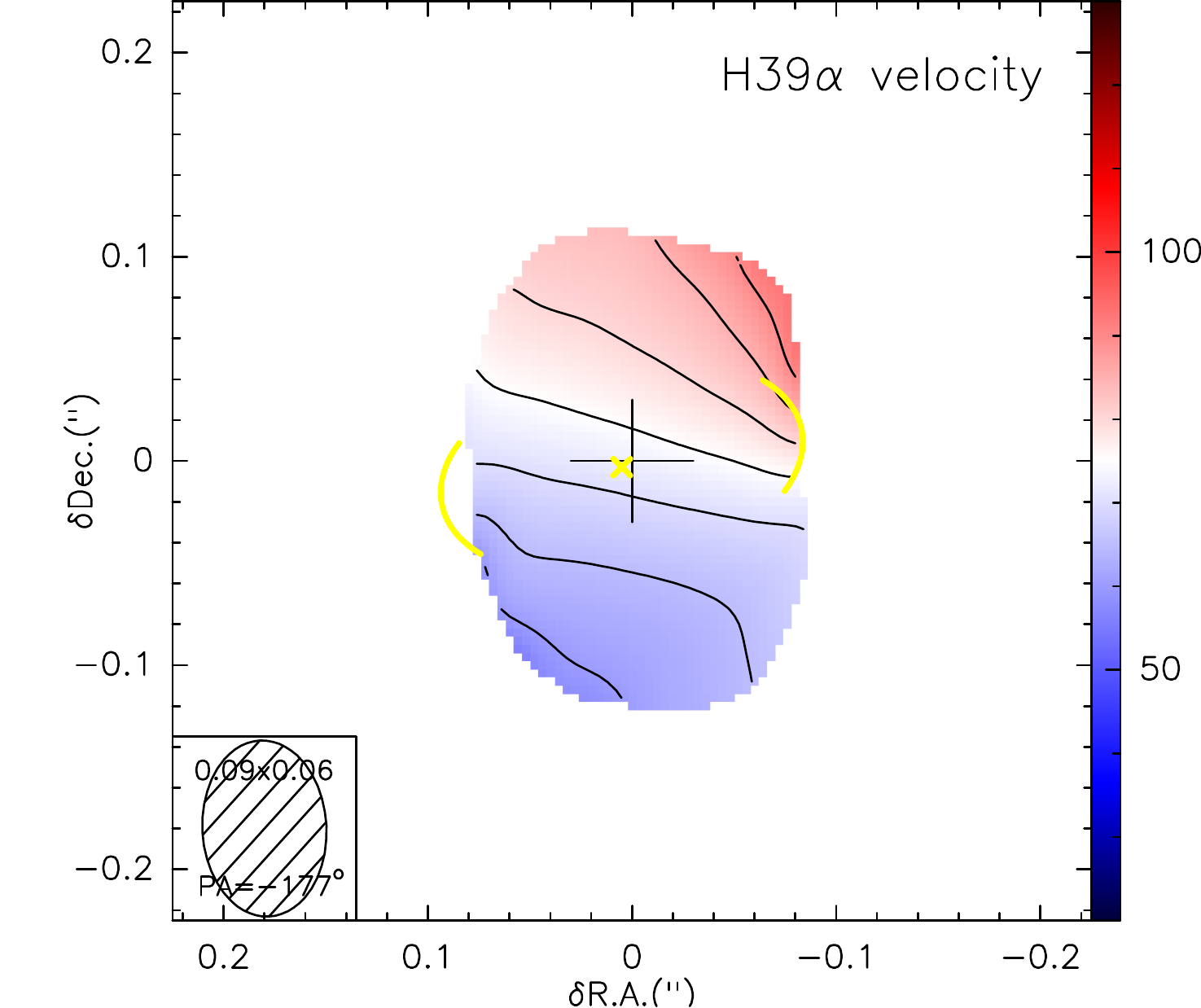} %velo MODEL 
         \caption{Summary of model predictions of the \htal\ and \htnal\ lines as in Fig.\,\ref{f-h30yh39}  
           \label{f-modfig3}}
   \end{figure*}

\begin{figure*}[htbp!]
     \centering
     \includegraphics*[width=0.45\hsize]{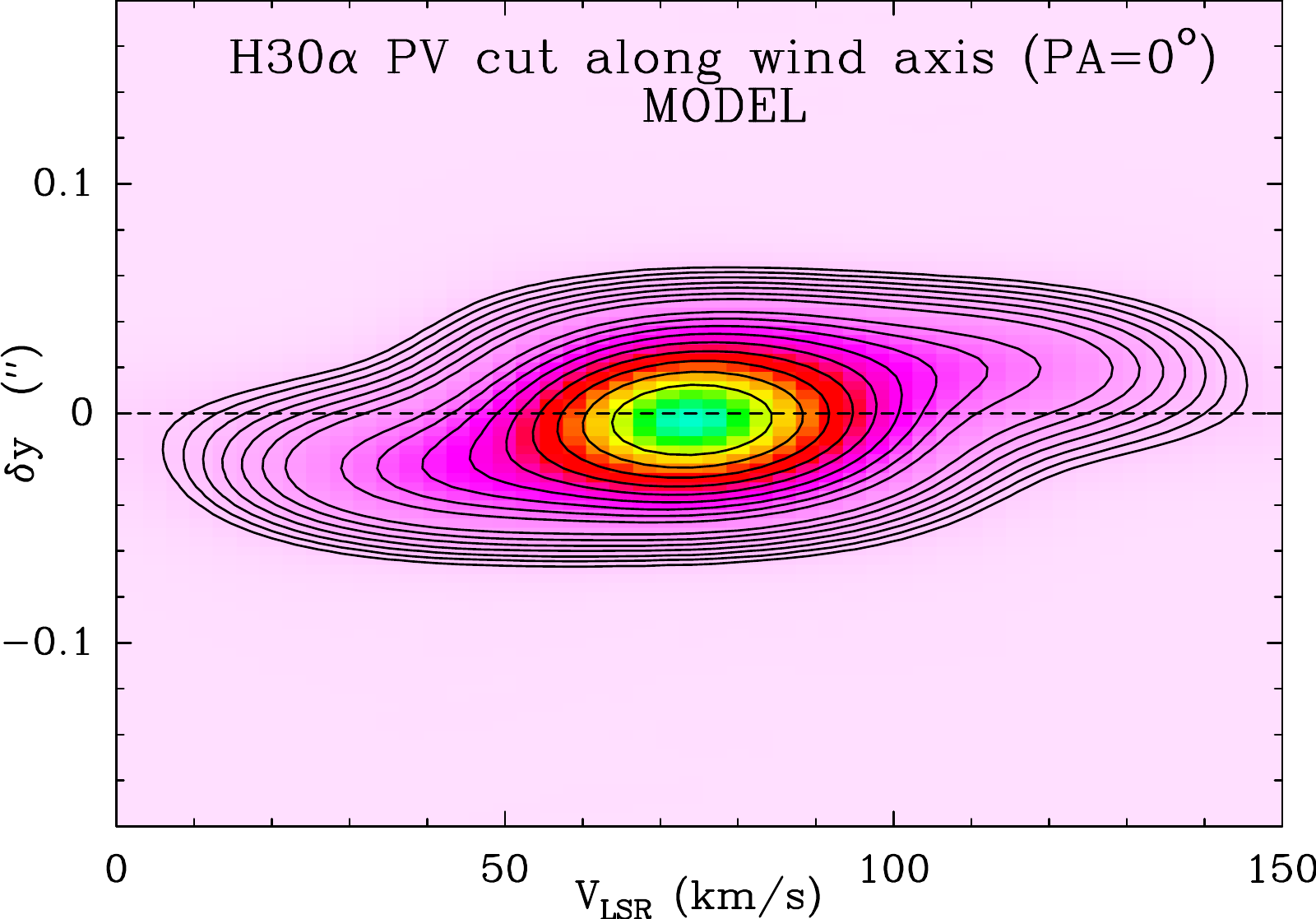} %% MODEL
           \includegraphics*[width=0.45\hsize]{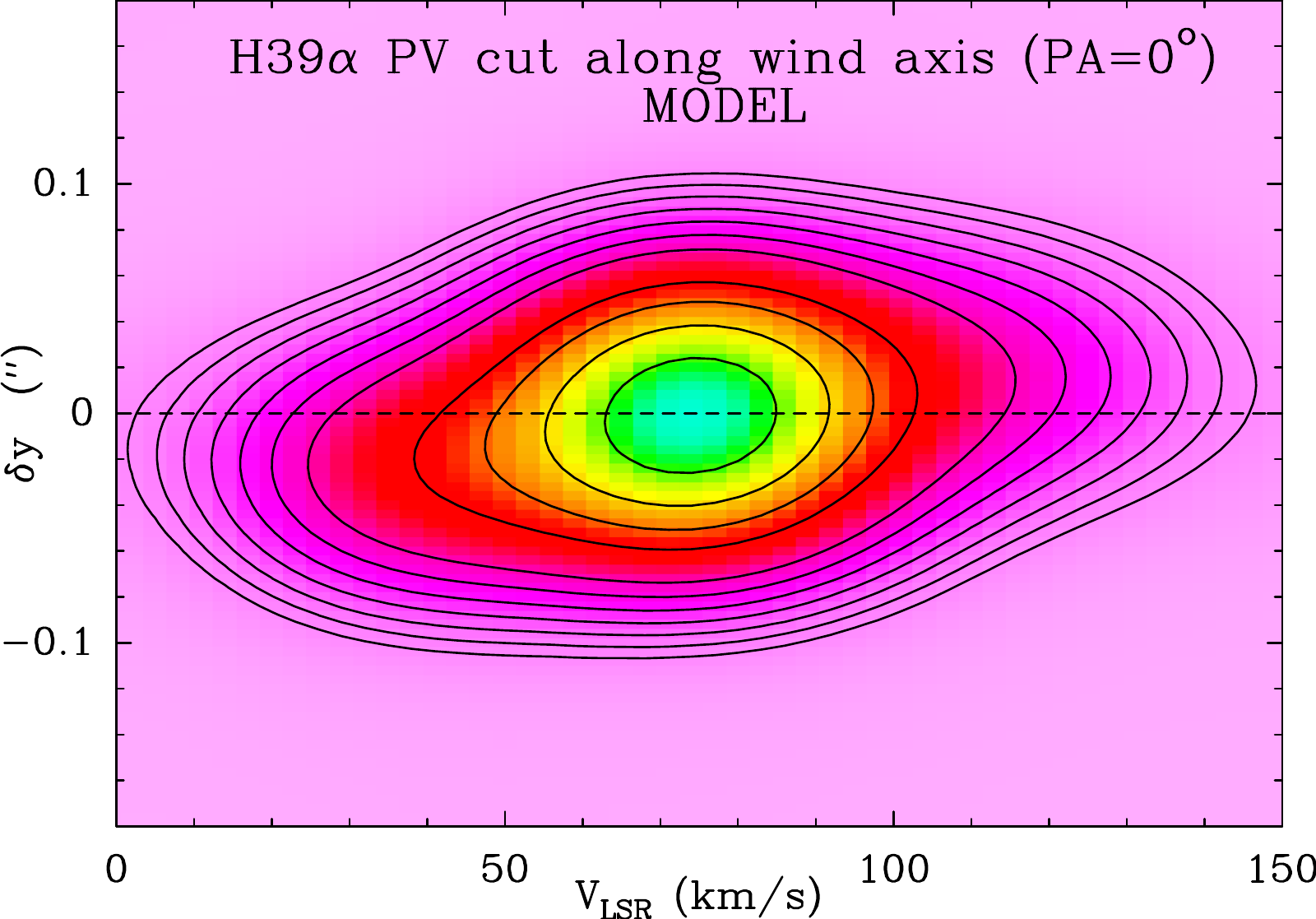} %% MODEL
     \\
      \includegraphics*[width=0.45\hsize]{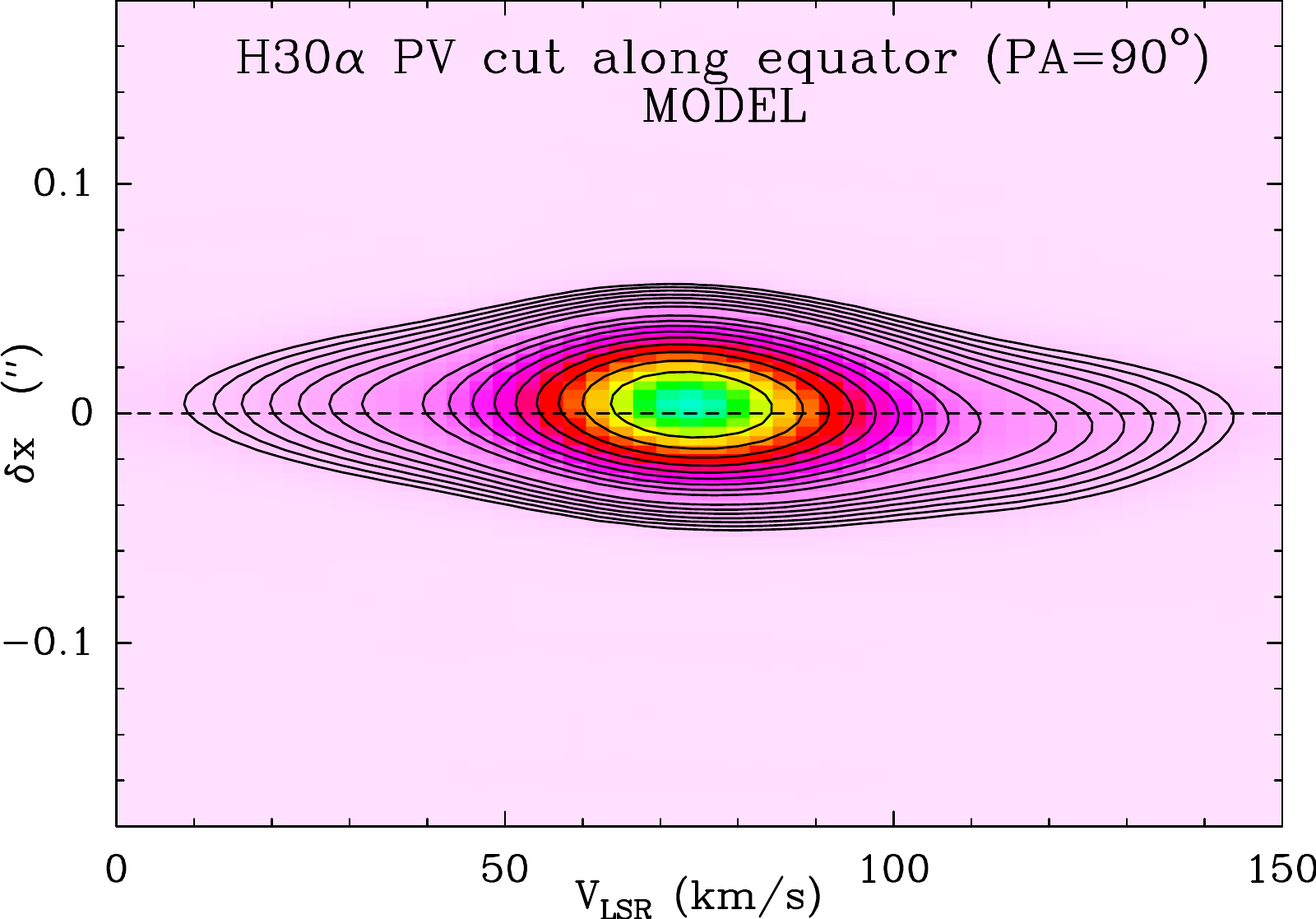} %% MODEL
\includegraphics*[width=0.45\hsize]{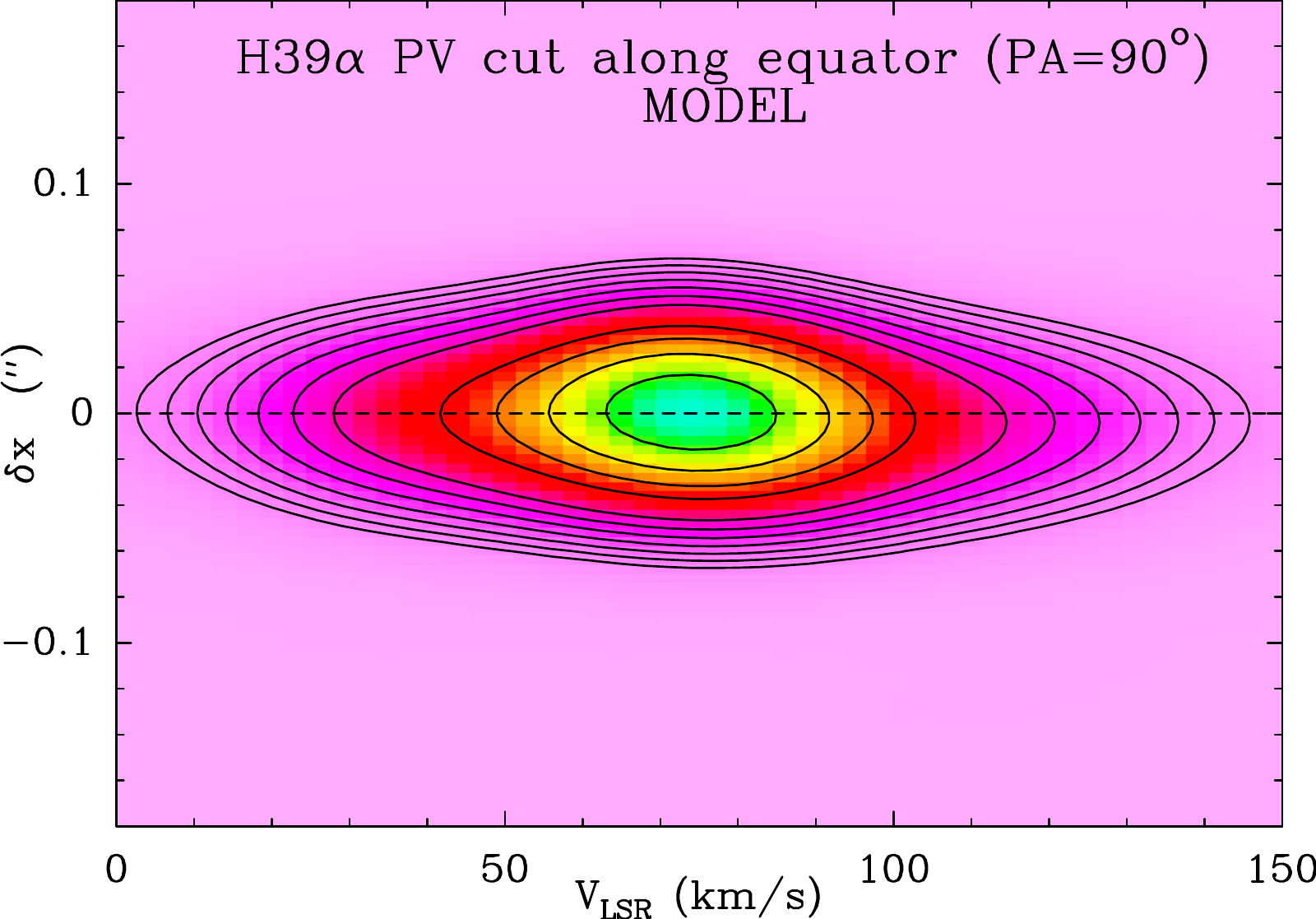} %% MODEL
      \caption{Model position velocity cuts through the center of along the wind
     axis (left, PA=0\degr) and the perpendicular direction
     (right) as in Fig.\,4. \label{f-modfig4}}
   \end{figure*}

To obtain an improved model that fits these new details, we
progressively adjusted the parameters of the input model (exploring a
wide range of physical conditions and over $\sim$300 models) until a reasonable match was
achieved between the model predictions and the data.
The ALMA-based improved M\,2-9 model presented here, contrary to the
one in \cite{san17}, has not been run under the LTE approximation but
in non-LTE.  This is because the new ALMA data show a frequency
dependence of the integrated flux of the Hn$\alpha$ lines steeper than
observed in 2015 with the \iram\ radiotelescope and deviating from the expectations in LTE. 
%% se puede mencionar que el mismo modelo CSC17's de m2-9 (Tabla 4,
%% san17) cuando se ejecuta en no-ETL produce un perfil en la linea
%% h30a que se desvia bastante del observado en esa epoca con el 30m,
%% en partocular, muestra un prominente pico de emision desviado de
%% Vsys (al azul, creo recordar) que indica que hay una no
%% despreciable contribucion de emision estimulada. La unica forma de reproducir un perfil como el observado, sin ese pico,
%% es subir la Te de 7000K (modelo original) a >15,000K!! esto es lo que constriñe mas fuertemente la Te. 
The best-fit model parameters are summarized in Table\,\ref{t-model}
and Fig.\,\ref{f-model}, and the predicted free-free continuum and
mRRL emission maps are shown in
Figs.\,\ref{f-contmod}-\ref{f-modfig4}. Throughout the modeling
process, we kept the number of model parameters as low as possible,
aiming at simplicity.

%%%% FIGURES: MODEL continuum: 
% Fig.---------------------------------------------------------------------------------------------------------------------
   %% hole282 (vrot=10km/s) = hole297 (vrot=7km/s)
%% Fig.-------------------------------------------------------------------------------------------------------------------

With regard to the wind's structure, although the cylindrical geometry
remains a reasonable approximation, the new observations necessitate
two significant changes: $i$) a much narrower bipolar structure
(particularly an outer radius value of \rout$\sim$40\,au), 
to reproduce
the elongation of the maps and avoid generating excessively extended
emission in the transverse direction compared to the data, and $ii$) a
narrow waist at the base of the wind (\req$\sim$8\,au) to prevent
excessive emission from the central regions. Additionally, we
introduced a C-shaped curvature to the bipolar outflow to closely
match the analogous shape observed in the 3mm-continuum maps.  While
this paper does not address centimeter-wavelength properties of the
free-free continuum, which originates well beyond the mm-wavelength
emission's spatial range, we observed that gradually widening the wind
beyond axial distances of 15-20\,au (i.e.\,emulating a trumpet-like or hyperbolic horn
shape) yields also a better match of the cm-continuum fluxes and
best reproduces the faint, extended emission observed in the
3mm-continuum maps beyond $\delta$y$\sim$0\farc12.
%Similar to the
%model in CSC17, a central cavity is required to account for the
%turnover frequency of the free-free continuum to be at $\nu_{\rm
% t}$$\gsim$230\,GHz.
%as deduced from the stable spectral slope over $\sim$10-230\,GHz range.
%% --- realmente no, si se ajusta la densidad del nucleo 
%For simplicity, the cavity is assumed to be
%spherical with an inner radius of $\sim$1\,au, which is within the
%range derived by CSC17.
Various inclinations for the line-of-sight
axis of the wind around $i$=17\degr\ were examined, with the most
successful models having an inclination of $i$=12\degr.

%% ------------------- 

%% PRESENTAR Las subcomponentes de I a IV. 

Regarding the wind's density structure, a notable difference compared
to the previous model is the necessity to include well-defined regions
with different density profiles. These regions are defined as the
nucleus (I), the inner wind (II), the HVSs (III), and the outer wind
(IV) -- see Fig.\,\ref{f-model}.  Specifically, to reproduce the
relatively bright emission from the HVSs it becomes necessary to
introduce a relatively dense, shell-like structure at axial distances
of $\sim$15--30\,au from the center within the wind.
%(Later in this section, we will
%see that this shell is associated with a sudden increase in velocity.)
Without these dense shells (i.e.\,with a unique, smooth density
power-law) is not possible to reproduce the S-shape of the axial
position-velocity diagrams observed for \htal\ and \htnal.
Immediately below and above the HVSs are the inner and outer winds,
respectively. The density in the outer wind primarily determines the
cm-continuum flux and the surface brightness of the 3mm-continuum
immediately beyond the HVSs. Meanwhile, the density of the inner wind
is constrained by the 1mm-continuum maps and the line-to-wings intensity
ratio of the mRRLs. The maximum electron density in the central
regions interior to the HVSs is constrained to \dense$<$3\ex{8}\,\cm3
by the full width of the \htal\ line: higher electron densities would
produce excessively broad emission wings from the core (offset
$\delta$y=0\arcsec) due to electron pressure broadening (see also
model caveats in Appendix \ref{ap:caveats}). Since the density of the inner wind cannot 
be larger than this maximum \dense\ value, a nuclear compact region
(assumed spherical and with constant \dense=2.5\ex{8}\,\cm3, for simplicity) is then
necessary to reproduce the mm-continuum flux level, which otherwise
would fall below the observed values.
%% --> MOVER A model caveats, uncertainties, etc ... 
%For the adopted maximum \dense, the radius of the
%nucleus is $\sim$4\,au, however, a slightly lower density with a
%larger radius could also be possible.
%This would reduce the pressure broadened wings of the sinthetic mRRLs, notably of \htnal, however,
%% it is impossible to reproduce the SED with densities as low as few\ex{7}\,cm3, which are the values needed
%% for the preassure broadened wings of H39a to reduce significantly. 

%%
%The density in the mm-wavelength emitting region
%is in the range \dense$\approx$10$^8$-10$^6$\cm3.

%% FIGURAS DATA_MODEL comparison
%% ========================== Fig.3
   %% /pcdisk/jbell3/csanchez/m2-9/selfcal-csc2/:   @1dspec.greg   @H30area.greg
   %% los macros de estas figuras para el MODELO son: 
   %% /pcdisk/jbell3/csanchez/m2-9/MODELO/hole270/: H30area-mod.greg H30spec-mod.greg H30velo-mod.greg pl_pv-mod.greg  (ver notas en tomboy: M2-9:MODELO-RRLs (hole270)
   %%
   %% hole270 - m29_csc.pdf  1er modelo optimo
   %% hole282 -- nuevo optimo! aunque mejor bajar algo vrot~8.5km/s ... 
   %% ========================== Fig.3

   %% ========================== Fig.4
   %% /pcdisk/jbell3/csanchez/m2-9/selfcal-csc2/:   @1dspec-mod.greg   @H30area-mod.greg 
   %% los macros de estas figuras para el MODELO son: 
   %% /pcdisk/jbell3/csanchez/m2-9/MODELO/hole270/: H30area-mod.greg H30spec-mod.greg H30velo-mod.greg pl_pv-mod.greg  (ver notas en tomboy: M2-9:MODELO-RRLs (hole270)
   %%
   %% ========================== Fig.4

To simplify, we maintain a constant value for the electron temperature
throughout the various regions of the wind. It is worth noting that
while we cannot entirely rule out the possibility of temperature
fluctuations within the wind, this factor alone cannot provide a
satisfactory fit of the observed brightness distribution in the mRRL
emission maps\footnote{As discussed by CSC17 a temperature
stratification has a rather limited effect on the free-free continuum
flux.}.
%Additionally, finding a plausible physical explanation
%for significant temperature variations across the ionized mm-emitting
%layers would be quite challenging.
The value deduced \te$\sim$15000\,K, which should be taken as a representative average
value within the ionized wind regions under study, is higher than than derived by CSC17 based on observations
obtained 2 years earlier (\te$\sim$7500\,K).
%% La Te=7500K en No-ETL no reproduce el perfil H30a tampoco el del 2015, (aunque en ETL si). 
%% En no-ETL la Te>15000K para producir una line-to-continuum ratio consistent with the ALMA (and also the IRAM 30m) obs. 
While there are uncertainties of up to
$\Delta$\te$\sim$15--20\%, the observed increase in the average
electron temperature in the ionized wind might indeed reflect real
variations in the wind's properties over time, for example, an
increase of the electron temperature may reflect an increase of
collisional ionization as a result of recent propagation/development
of shocks.

%% ===================
%Overall, the electron temperature of approximately 10,000 K is a
%result of the balance between various heating and cooling processes
%within HII regions, driven primarily by the intense UV (ionizing)
%radiation from the central star.
%% Collisionally ionized gasses can have signiﬁcantly higher electron
%% temperatures (than 10,000K) Prozesky & Smits 2020!  EL aumento de
%% temp. podria indicar ionization colisional al menos en parte, lo
%% cual seria otro signo de la presencia de choques ... que como se
%% sabe pueden injectar energia adicional en el gas y/o romper el
%% equilibrio termico del viento ionizado.

In terms of wind kinematics, the distinctive S-shape observed in the
position-velocity diagrams along the axis of the mRRLs can only be
explained by incorporating a steep velocity rise at the HVSs, which
has been approximated by a step function.  This step function
represents a low velocity of $\sim$10\,\kms\ within distances less
than $\sim$15\,au; beyond this point, i.e.\,within the HVSs, the
velocity rises from $\sim$70 to $\sim$90\,\kms.  Beyond the HVSs, in
the outer wind, the velocity remains unconstrained as the line
emission from these tenuous regions is below our detection limit.
%% REALMENTE ??? 
%% DECIMOS?? 
%This velocity profile resambles that predicted by numerical
%simulations of the interaction of a slow, desde wind and a
%jet blown by an orbiting companion by \citep[][see their Figs. 3, 5, 7
%  top-left panels]{gar04}, which primarily reflects the acceleration
%of part of the slow wind due to its interaction with the jet.
Within the HVSs, we explored alternative velocity functions that
varied in steepness (including constant velocity), but these produced
a poorer match to the data.
%Within the HVSs, we explored alternative velocity functions that were less steep but
%found them to consistently predict broader mRRL line profiles at the
%center, which did not match the observed data.
Our best-fit model also includes rotation, which is needed to
reproduce the observed velocity gradient perpendicular to the lobes
(Sect.\,\ref{res-mrrls}).  In the Appendix\,\ref{ap-figures} we show a
model with no rotation for comparison.  Since the angular resolution
of the data does not allow us to distinguish between Keplerian or
momentum conservation velocity profiles, we have used a constant
rotation velocity of 7\,\kms for simplicity. This value (at a
representative radius of \req=8\,au) implies a lower limit to the
central mass of $\gsim$0.4\,\msun.

%hole270: 
%#number of particles=3.92e+51, ionised_mass=8.53e+27 g = 4.26e-06 M_sun
%#total scalar momentum=6.40e+34 g*cm/sec, = 3.20e+01 M_sun*cm/sec = 3.20e-04 M_sun*km/sec
%#total kinetic energy=2.76e+41 ergs
%#average mechanical luminosity=1.53e+33 ergs/s = 0.398765 L_sun
%#average kinematical time scale=6 yr
%#minimum time scale=1.0 yr; maximum time scale=10.3 yr
%#average M_dot=1.52e+13 g/sec, = 7.48e-07 M_sun/yr; n_cell=84384

The total ionized mass within the model region that produces the
observed (sub)mm-to-cm free-free continuum and mRRL emission is
$M_{\rm ion}$$\sim$4\ex{-6}\,\msun, similar to that obtained by CSC17.
Considering the mean kinematical age of the wind, <\tdyn>$\sim$6\,yr,
the average mass-loss rate implied from the model is
\mloss$\sim$7\ex{-7}\,\my, which is also in good agreement with the
value obtained by CSC17. This agreement is expected because, as
discussed in CSC17, the mass/mass-loss rate is almost exclusively
constrained by the flux of the free-free continuum, which has not
shown discernible variations considering flux uncertainties
($\sim$5--10\% in our ALMA data and $\sim$20--30\% in the \iram\ data
analyzed by CSC17).  The determination of the mass/mass-loss rate is
subject to some uncertainty due to factors such as the uncertain
distance to the source, the simplified model used for the structure
and kinematics of the ionized wind, and the moderate range of
acceptable values for different model parameters. This likely results
in a mass/mass-loss rate uncertainty of a factor $\lsim$2--3.  We
caution that given the relatively complex structure and density
stratification in the ionized central regions of M\,2-9, interpreting
the meaning of the mass-loss rate is not straightforward (see
Sect.\,\ref{dis-mloss}). The total scalar momentum, kinetic energy, and
the average mechanical luminosity of the ionized wind have been
computed and are given in Table\,\ref{t-model}.  Model caveats
  and other final modeling remarks are discussed in appendix
  \,\ref{ap:caveats}.

\subsection{CO absorption and 1\,mm continuum: Evidence of a Compact Equatorial Disk}
\label{disk_co}

%% a continuacion, presentamos una serie de evindencias que indican la presencia
%% de un disco ecuatorial de gas y polvo que rodea la region hii central y es perpendicular
%% a los lobulos. 
%% 

In this section, we provide compelling evidence from our ALMA data
supporting the existence of an equatorial disk composed of gas and
dust surrounding the collimated ionized wind emerging from 
M\,2-9 and oriented perpendicular to the lobes. This equatorial disk
is probably the counterpart of the compact ($\sim$37$\times$46\,mas)
flattened dust structure observed along the equatorial plane of the nebula using VLTI/MIDI
at 8-13\micron\ \citep{lyk11}.

First, as shown in Fig.\,\ref{f-co} and briefly discussed in
Sect.\,\ref{res-co}, the structure responsible for the CO absorption
observed against the compact continuum source at the center of M\,2-9
is likely an equatorial torus or disk surrounding the collimated
ionized wind and positioned perpendicular to the lobes. In this
configuration, the front side of the disk partially blocks the
emission from the base of the receding north lobe, resulting in the
observed CO absorption. Furthermore, differences in the brightness
distribution of the continuum at 1 and 3\,mm can also be attributed to
the presence of an equatorial dust disk, which is expected to be the
counterpart of the gas disk inferred from the CO absorption. In
particular, at 1\,mm, where dust-related effects are more prominent
due to higher dust opacity and a larger contribution of dust thermal
emission to the total emission (Fig.\,\ref{sed-alma}), the lower
brightness of the northern lobe compared to the southern lobe can be
partially explained by absorption of the free-free continuum from the
collimated ionized wind that is behind the front side of the disk.

Additionally, the broad-waisted structure observed in the 1\,mm
continuum map (but absent at 3\,mm and in the \htal\ maps) can be
attributed to dust thermal emission from regions outside the narrower
background continuum source, indicative of the presence of a dusty
disk surrounding the ionized wind.  Indeed, in contrast to the
expected behaviour for free-free continuum emission, we observe that
the size of the waist is larger at 1\,mm (reaching a maximum radius,
at 3$\sigma$ level, of $\sim$0\farc09 at 232.9\,GHz --
Fig.\,\ref{f-cont}).
%% Rout~56AU... parece que la forma eliptica sugiere i~30degr, un PA=102\degr
%% y su centro podria estar desplazado 0.005arcsec (3.25au) al ESTE. 
However, in ionization-bounded or density-bounded \ion{H}{ii} regions, the
free-free continuum emission decreases with increasing frequency or
remains constant, as it is indeed observed in the mm-to-cm continuum maps
of M\,2-9 in the axial direction and at lower frequencies in both the
equatorial and axial directions \citep{lim03,dlF22}.
Therefore, the larger waist observed at higher frequencies is
consistent with the contribution of dust thermal emission from the
disk.
%This further supports the interpretation of an equatorial disk
%composed of gas and dust surrounding the collimated ionized wind in M\,2-9.
%% {\bf podemos llamar jet al viento ionizado? dlF+22 lo hacen ... }
%% depende de la velocidad maxima que deduzcamos.

%% si esto es asi el radio del disco es 0.035" @ spw1
%% si esto es asi el radio del disco es 0.045" @ spw2 -- highest frecuency.  

%% NO SE SI DECIRLO O NO... es tentativo y muy secundario // quiza en la discusion 
%The 1mm maps of M,2-9 reveal that the equatorial plane of the disk is
%oriented at PA$\sim$102\degr. This orientation is different from the
%equatorial plane of the large-scale nebula and CO rings
%(PA$\sim$90\degr).  The tentative 12\degr\ difference in the
%orientation of the equatorial planes suggests that the compact disk is
%tilted with respect to the larger-scale structure of the nebula.

In the following, we attempt to establish some constraints on the physical conditions,
including temperature and CO line opacity, as well as the dimensions
(radius and vertical thickness) of the disk and its kinematics
using the information contained in the CO absorption profile.

\subsubsection{Disk's physical conditions}
\label{disk_phys}

%% colum density, temperature
As shown in detail, for example, by \cite{san22} in their analysis of
the compact NaCl absorption feature observed at the central rotating
disk of the post-AGB nebula \oh, the presence of absorption enables
constraining the excitation temperature and opacity of the absorption line observed. This is done by comparing the brightness temperature
of the absorption line (after continuum subtraction), \tlon, the
brightness temperature outside the continuum source, 
%(where the
%dominant background emission source is the 2.7\,K cosmic microwave
%radiation),
\tloff, and the continuum brightness peak in the same
spectral window as the line is observed, \tc.  Taking into account
the values of the absorption and the upper limit to the emission
measured in the CO maps of M\,2-9 along with the continuum peak in the same
spectral window\footnote{For a proper line-to-continuum comparison, we
use the continuum maps of the same spectral window where the line was
observed cleaned and restored exactly in the same manner and, thus,
imaged with the same beam.}  ($T_{l}^{\rm on}$$\sim$$-$308\,K,
$T_{l}^{\rm off}$$\lsim$50\,K, and $T_{\rm c}$$\sim$2350\,K), we
deduce \tex$\lsim$330\,K and $\tau$$\sim$0.16. 
%We note that, despite the moderate opacity, the CO absorption is relatively intense due to the large difference between \tc\ and \tex, since \tlon$\sim$(\tex-\tc)$\times(1-e^{-\tau})$. 
This value of the CO line opacity and \tex\ imply a CO column density
of \nco$\lsim$6\ex{17}\,\cm2, for a line width of $\sim$3\,\kms\ (in
this calculation we used the on-line version of the non-LTE line
radiative transfer code RADEX\footnote{\tt
http://var.sron.nl/radex/radex.php} by \cite{radex}).  Adopting a
CO-to-H$_2$ relative abundance of $X_{\rm CO}$=2\ex{-4}, which is
typical of O-rich sources \citep[e.g.][]{ram18}, we derive a total
column density of \nh$\sim$3\ex{21}\cm2 across the absorbing layers
(along the line of sight) of the disk.
%% Consistente con la extincion total hacia el centro. Av~2-5mag estimada por Calvet & Cohen y de Hora & Latter 1994...
%% sale menor y es normal porque nuestro dato solo da cuenta de la abs. por el disco compacto y no por el resto de material
%% de la nebulosa de gran escala que estan en la linea de mirada entre la estrella central y nosotros.
However, this value represents a lower
limit for the column density because the CO
abundance used in the calculation is probably significantly overestimated
due to strong photodissociation caused by intense ionizing radiation
or shocks. 

%% The nebula is O-rich with C/O < 0.5 (Liu et al. 2001), 

The outer radius of the disk can be estimated by considering the
angular size of the broad-waist observed in the 232.9\,GHz continuum
map. From this analysis, an approximate value of \rout$\sim$0\farc075
(beam-deconvolved; equivalent to $\sim$50\,au at a distance of
650\,pc) is obtained, as discussed earlier in this section. This
relatively large radius, along with the likely binary orbital
  separation of $\gsim$15\,au \citep{lyk11,cc12,san17}, suggests
the disk is most likely circumbinary rather than an accretion
  disk, which typically has a size comparable to or slightly larger
  than the radius of the accreting compact object.  A lower limit to
the disk radius can also be obtained from the offset to the north
where the CO absorption is observed ($r_{\rm
  abs}$$\sim$0\farc012$\sim$8\,au). This lower limit arises from
geometric considerations, as the CO absorption corresponds to the
radius within the disk where the column density reaches its maximum,
which is not expected to occur at the very edge of the disk.
%{\bf no se si se entiende pero no se como explicarlo sin un dibujo
%(?)}
Assuming a line-of-sight inclination of $i$=12-17\degr\ for the disk
equator, we can estimate a lower limit of \rout$\gsim$27-38\,au.
%% es dificil explicar por que es un limite inferior: porque la absorcion solo
%marca el rdio dentro del disco donde la densidad de columna es maxima, aunque el disco podria extenderse algo mas en radio pero dada 

%% No sabemos si esta hueco o es continuo...
% radius and thickness ... --> dimensions suggest circumbinary disk 

Based on the disk's dimensions and the lower limit to the derived
column density, we estimate the total mass to be $M_{\rm
  disk}$$>$2\ex{-6}\,\msun.  Additionally, the disk could be larger
than what is seen in our ALMA maps, as the emission is probably
limited by sensitivity, which would result in an even larger value for the disk gas mass. 

The radius and the temperature of the equatorial disk in M\,2-9 are
similar to those of the circumbinary disk recently discovered at the
center of \oh, a well-known massive bipolar outflow associated with an
AGB+main-sequence binary system \citep[with a radius of $\sim$40\,au
  and temperature $\sim$450\,K,][]{san22}.  Assuming the average
density of both disks is similar, \dens$\approx$10$^8$\,\cm3, we
estimate the line-of-sight path length for CO absorption to be $r_{\rm
  t}$$>$\nh/\dens$\sim$2\,au. The lower limit arises from the upper
limit to the CO/H$_2$ fractional abundance linked to a possibly
enhanced CO photodissociation.  This implies a vertical thickness of
the M\,2-9 disk, after deprojection, of $h$$>$0.6\,au, resulting in an
aspect ratio $h/r$ larger (or much larger if strong CO photodissociation) than 0.01. 
%, within the generally
%considered stable range of $\sim$0.1-0.01 \citep{fran02}.

%% %% is consistent with the minor-to-major axis ratio of the
%% %% elliptical distribution of the equtorial emission in the 1mm
%% %% continuum maps.

%% La vel de
%% expansion cerca al centro parece bastante baja --- se puede
%% considerar esto una indicacion de la presencia de una estrella AGB?

%% Disco de LYKOU??? 

%% KINEMATICS: 

\subsubsection{Disk's kinematics}
\label{disk_kin}

%The redshift of the CO absorption feature by approximately
%6\,\kms/cos(17\degr)$\sim$6.3\,\kms\ with respect to the systemic
%velocity of the background collimated ionized wind allows for two
%possible interpretations: infall or rotation (Sect.\,\ref{res-co}). Let's
%explore each of these scenarios further.

The (deprojected) redshift of the CO absorption feature by
$\sim$6.2\,\kms\ with respect to the
systemic velocity of the background collimated ionized wind can be
plausibly explained by gas infall from the disk toward the central
region.  Gas infall motions from the inner regions of circumbinary
disks, formed e.g.\ in Wind Roche Lobe OverFlow (WRLOF) binary systems,
is theoretically expected \citep[e.g.][]{mm99,moh07,chen17} and has
indeed been observed in a number of post-AGB objects with interacting
binaries \citep[e.g.\,R\,Aqr,][]{buj21}. An infall velocity of
$\sim$6.2\,\kms\ observed at an average radial distance of $r_{\rm
  abs}$$\sim$27--38\,au from the center (Sect.\,\ref{disk_phys}) would
imply a central mass of $\sim$0.7\,\msun, assuming that the gas
started falling with (near) zero velocity from a point much far beyond
the observed infall radius. This represents a lower limit for the
central system's mass since the closer we assume the fall started, the
larger the estimated mass of the central system becomes. If we
consider that the gas began falling in from the radius of the
broad-waisted structure observed in the 1\,mm continuum maps
($\sim$50\,au) then the central mass would be $\sim$2.0\,\msun.  This
should be regarded as an upper limit for two reasons. First, the
radius of the disk in our maps might only represent a lower limit to
the actual disk outer radius, which could extend beyond our sensitivity
limit. Second, the infall velocity may be smaller than $\sim$6.2\,\kms\ if the centroid of
the mRRLs is slightly blue-shifted relative to the systemic velocity
of the source (as suggested by our model, Sect.\,\ref{model} and Table\,\ref{t-model}). 
%% NO veo razon por la cual si es disco tiene un radio dado la caida
%%no empieze desde ese mismo radio... MODELOS??  50au quiza es un
%%limite inferior al radio del disco... cuyo tamaño esta limitado por
%%sensibilidad de las obs.  e.g. a 90au ---> Mc~0.77Msun. Lo cual
%%cuadra bien con la estimacion de la masa de crl618, ~0.6 Msun, mas
%%la de una compañera de 0.1-0.2 msun, como decia CC+12.  si la
%%m2~0.4msun y la m1~0.6Msun -- mt~1Msun -- distancia es ~60au
%%... donde empezo a caer...
The mass range derived from these crude estimates is consistent with
the previous assessments of the central binary's mass by \cite{cc12},
who proposed a low-mass companion of
$m_2$$\lsim$0.1–0.2\,\msun\ orbiting around a
$m_1$$\sim$0.5-1\,\msun\ mass-losing star.

%Unfortunately, the CO absorption feature originates
%from a very compact region that remains spatially unresolved,
%preventing the identification of velocity gradients that could provide
%a more robust characterization of the disk kinematics and the mass of
%the central system.

In principle, rotation of the disk could also produce a redshifted CO absorption feature
if (and only if) the center of rotation and the center of the ionized
region do not coincide, i.e., if both are off-centered. Only then will
the rotational velocity have a non-zero component in the line of sight
direction. In the context of a circumbinary disk, where the disk
rotates around the center of mass of the system rather than around one
of the binary components, the offset of the disk relative to the
centroid of the ionized wind is expected.  However, to observe a
significant line-of-sight rotational velocity 
it is essential that the separation between the disk's centroid and
the ionized wind (denoted as $r_{\rm dw}$) and the radial
distance where most of the absorption occurs (denoted as $r_{\rm abs}$) are
comparable. Our data allows us to estimate 
$r_{\rm dw}$ to be around $\sim$0\farc007$\sim$5\,au (Sect.\,\ref{cont_maps})
and $r_{\rm abs}$ to be $\sim$38--27\,au (Sect.\,\ref{disk_phys}). Therefore, 
to achieve a line-of-sight rotational velocity of $\sim$6.2\,\kms, the actual rotational
velocity at $\sim$27--38\,au should be $\sim$6.2$\times$$\frac{(27-38)}{5}$$\sim$33--47\,\kms\ (from basic geometric calculations). 
Such a high velocity would imply an unreasonably massive central object, exceeding 30\,\msun. 

\section{Discussion}
\label{discussion}

Both the nature of the central binary system residing in the core of
M\,2-9 and which of the stars is launching and ionizing the bipolar
wind observed in our mm-continuum and mRRL emission maps are two
fundamental unknowns. In this section, we will analyze/discuss some of
the results of our work with the aim of making
progress in this regard.

\subsection{Ionizing central source}
\label{nlyc}
As is well known \citep[e.g.\,][]{con16}, the number of Lyman
continuum photons emitted per second (\nlyc) needed to reproduce the
%free-free and mRRL
emission from the ionized wind at the core of M\,2-9 can be computed
using the measured value of the free-free continuum flux (S$_\nu$) at a given
frequency ($\nu$) as:

\begin{equation*}
\nlyc = 7 \times 10^{46} \left( \frac{\te}{10^4\text{K}} \right)^{-0.45} \left( \frac{S_\nu}{\text{mJy}} \right) \left( \frac{d}{\text{kpc}} \right)^2 \left( \frac{\nu}{\text{GHz}} \right)^{0.1}
  %  \nlyc=7\ex{46}(\te/1e4K)$^{-0.45}$(S$_\nu$/mJy)(d/kpc)$^2$($\nu$/GHz)$^{0.1}$
\end{equation*}

For an electron temperature of \te$\sim$15,000\,K (Sect.\,\ref{model}) 
and a free-free continuum of 102.80\,mJy at 95.3\,GHz (after
subtracting the 260\,K dust emission component from the total
continuum fluxes in Table\,\ref{tab:spws} --- see also
Fig.\,\ref{sed-alma}), we calculate a Lyman continuum photon rate of
log\nlyc$\sim$43.8\,s$^{-1}$.  This value has been compared with the
Lyman continuum output rate from the grid of model atmospheres of
early B-type stars by \cite{lan07} to investigate the central ionizing
source. We find that a star with an effective temperature of around
\teff$\sim$25\,kK (with \rs$\sim$1.3\rsun, adopting
\lstar$\sim$600\,\ls\ at 650\,pc, and with a surface gravity
log $g$[cgs]$\sim$4, adopting a stellar mass of $\sim$0.5\,\msun\ 
\cite{san17}) can account for the observed ionization. A similar
effective temperature for the central ionizing source (early B-type or
late O-type) was obtained by \cite{cal78} from analogous
considerations (and using continuum flux measurements at
cm-wavelengths) but also independently by  \cite{all72} and \cite{swi79} 
from the analysis of the optical line emission spectra.  This result
implies that there is not need to invoke the presence of a hotter
companion star to explain the observed nebular ionization in M\,2-9.

If, as previously speculated, a white dwarf (WD) companion exists at the
core of M\,2-9 and is responsible for the observed nebular ionization,
we can establish a lower limit to its effective temperature. By
comparing the value of \nlyc\ calculated from the mm-continuum flux
with the Lyman continuum output rate for WD stars expected
from model atmosphere models by, e.g., \cite{wel13}, we find that the
effective temperature of the WD must be larger than
50,000\,K, implying a spectral type O\,5 or earlier.

%However, this high temperature contradicts the spectral
%classification of the central star, which has been inferred as early-B
%or late-O (25-35kK) based on the emission line spectrum in the
%optical \citep{all72,swi79,fei84}. If the central star had indeed \teff$>$50kK its spectral
%classification would have been early-O instead, strongly dominated by
%lines from highly ionized species such as He II, C IV, N V, O VI, and
%Ne VIII.
%and lacking low-ionization lines commonly associated with
%late-O or early-B stars.
%% SE PUEDE DESCARTAR DEL TODO?? NO**
%% =============================================
%% OJO: Feibelman: IUE spectrum --> The present IUE data do not yield an
%accurate determination of the central star temperature, but the
%presence of the N v 21240, N iv 21719, and N iv 21487 emis-
%sions, and the slightly rising level of the continuum from 1400
%to 1200 À require a temperature in excess of 60,000 K, unless
%one invokes unconventional processes, such as shocks, for the
%excitation."
%% =============================================

The luminosity and mass of the progenitor star of M\,2-9 remain
uncertain due to the lack of precise distance measurements
(Sect.\,\ref{intro}).
%In this study, we adopt a distance of 650\,pc, the
%lowest value reported in the literature consistent with proper motions
%of the optical knots and of the CO equatorial rings.
%% Schwarz et al. 1997, Castro-Carrizo et al. 2012
%% los mov. prop. fueron recomputadis por Corradi y sacan d=1300pc.
% Sin embargo en CC12 descartan una distancia mayor que 1000pc... 
At a distance of $d$=650\,pc, adopted in this study, the 
luminosity of M\,2-9 is below the expected value of 
$\sim$2500\,\ls\ for the least massive stars (around 0.8\,\msun) that are
expected to evolve off the main sequence within a {\sl Hubble} time
\citep{mil16}.
% QUITAMOS LA MENCION EXPLICITA A LA PRESENCIA DE LA B_MAIN-SEQUENCE PORQUE NO ENCAJA. 
%The relatively low luminosity led \cite{sch97} to
%propose that if there were a B-type star inside M\,2-9, it should be a
%$\sim$B5\,V main sequence star (with a luminosity of a
%few\ex{2}\,\ls). However, as the effective temperature of a B5\,V star
%(around 15,000\,K) is insufficient to account for the observed
%ionization of the core, these authors speculated about the presence of
%a hotter, compact companion (likely a white dwarf) to explain the high
%ionization degree of the gas in the vicinity of the star.
%As we show later in Sect.\,\ref{dis-mloss}, the high mass-loss rate of
%the compact ionized wind deduced from our data, argues against the
%donor star being a main-sequence star.
A possible explanation for the low luminosity of the central star of
M\,2-9 is that it could be a post Red Giant Branch (post-RGB) object,
meaning that it has prematurely left the RGB phase of stellar
evolution. Recent studies have highlighted the existence of post-RGB
stars with many characteristics similar to post-AGB objects, and it
has been speculated that their peculiar evolution, deviating from
standard models, may be influenced by strong interactions with close
binary companions \citep{oud22,kam16,hri20,olo19}.
%This scenario
%opens up new avenues for understanding the central star of M\,2-9 and
%highlights the importance of considering binary interactions and
%non-standard evolution in the study of post-RGB stars and their
%associated planetary nebulae.
%Further investigations are required to
%better constrain the enigmatic nature and evolutionary history of the
%central star in M\,2-9.

%    The B\,5V plus white dwarf companion hypothesis
%    has some important drawbacks, which we will discuss in Sect.\,\ref{xx}.
%% 1) Si B5V main sequence, la progenitora debio de ser mas masiva que
%% 3-5msun.  Evolucion mas rapida, pero hace bastante tiempo, y no se
%% explica que aun conserve eyecciones de masa de ~1200yr. COMPROBAR.
%% 2) el espectro optico NO es consistente con una Teff>50kK, que
%% corresponderia a una early-O type, sino con una early-B or late-O,
%% como dicen Calvet & Cohen, 1973.
% 2) The typical mass-loss rate of the wind from a B5 main-sequence star
% is relatively low compared to more massive stars. It is on the order
% of about 10^(-10) to 10^(-9) solar masses per year (M☉/yr).
%% Es muy bajo y no explicaria la tasa de acrecion necesaria para
%% explicar el jet que estaria lanzando presumiblemente la companera,
%% que se deduce a partir de nuestras observaciones.  3) No hay signos
%% en el presente de eyecciones de alta velocidad que delaten la
%% presencia de un objeto compacto acretando material de la
%% mass-lossing star y relanzandolo convertido en un jet de alta
%% velocidad.

\subsection{Implications from the mass-loss rate} 
\label{dis-mloss}

  The average mass-loss rate of the ionized bipolar wind at the core
  of M\,2-9 is deduced to be of \mloss$_{\rm  w}$$\approx$10$^{-7}$\,\my\ (Sect.\,\ref{model}).
  %%Without knowing which of the stars is launching the wind, interpreting the meaning of the mass-loss rate is not straightforward. 
  Interpreting the meaning of the mass-loss rate is not straightforward when we are uncertain about which of the stars is responsible for launching the wind.
  
  If the ionized wind emerges from the mass-losing star, then, the
  measured mass-loss rate of the present-day wind is significantly
  lower than the rates observed $\sim$1300 and $\sim$900 years ago, $\approx$10$^{\rm -5}$\,\my, which were responsible for forming the
  large-scale nebula, including the massive CO equatorial rings
  (Sect.\,\ref{intro}). In this scenario, the substantial reduction in
  the mass-loss rate with respect to previous epochs can be
  interpreted as evidence that the primary star has transitioned to a post-AGB or
  post-RGB phase.
Alternatively, the star could currently be in the AGB phase with a
mass-loss rate of \mloss$_{\rm AGB}$$\approx$10$^{\rm -7}$\,\my, which is more
in line with what is expected for a low-mass progenitor star, but the
ejection of the large-scale nebula, including the CO rings,
could have occurred at two specific
moments with "abnormally" high rates.
%for the AGB stage.
%The exact cause of these violent and abnormally massive mass ejections 
%could have be related to sudden and significant alterations of the
%binary system's properties.
%(possible scenarios that could contribute to these
%abrupt changes include common envelope evolution, mass transfer
%instabilities, magnetic interactions, etc).
%% 
%% se ha espeulado que dos CE sucesivos en los que el resultado es una disminucion progresiva de la orbita.
%% Aun a dia de hoy no habria un merger... aunque la estimacion que hay de a~20-30au no parece
%% apoyar esta idea. 
  %% ----------------------------------------------------------------------------------------------------------------

  If the ionized wind does not directly originate from the mass-losing
  primary star but, instead, from its companion after mass transfer
  through disc-mediated accretion, it is expected that this wind would
  interact with the slow wind of the donor star. Such an interaction
  would give rise to a complex bipolar structure characterized by
  intricate density and velocity patterns, featuring multiple shock
  formations and a dense equatorial region as shown by analytical
  studies \citep[e.g.][]{sok00,liv01} and numerical simulations
  \citep{gar04}. According to these investigations, if this scenario
  applies to M\,2-9, the region responsible for emitting mRRLs would
  encompass the ionized portion of this complex structure\footnote{The
  specific part of this intricate structure that becomes ionized
  depends on its exposure to the central ionizing source, which may be
  selectively obstructed in specific directions.}, including both the
  primary's slow wind and the companion's fast wind, both of which
  would exhibit significant structural and velocity alterations due to
  their interaction.  In this context, the mass-loss rate estimated in
  our study is likely to fall between the mass-loss rates of the
  mass-losing star and that of the companion.

%Unravelling the nature of M\,2-9's central binary system and 
%the intricate interplay between the primary star and its companion are
%essential for comprehending the mechanisms governing the observed mass
%loss and wind dynamics in this system.

 \subsection{Wind structure and kinematics} 
\label{dis-kin}

Our ALMA observations show the presence of a collimated, ionized wind
emanating from the center of M\,2-9. As observed at mm-wavelengths,
this narrow-waisted wind extends along the nebular axis direction up to radial distances of $\sim$75\,au and
$\sim$40\,au in the perpendicular direction.
%%and opening angle of xx\degr\ at its base,  
%%has a radius of $\sim$40-60\,au and
Based on radio-continuum
emission maps, the ionized wind extends up to least 230\,au along the axis
\citep{lim03,dlF22}, resulting in an aspect ratio of $>$4.5.
%% this corresponds to the part of the wind that is ionized
This wind displays high-velocity motions, reaching speeds of
$\sim$80\,\kms, thereby categorizing it as a jet. One of the
significant findings from our data and modeling is the presence of a
non-uniform density structure within the jet: we find alternating
regions with noticeable density variations, characterized by regions
of high density and areas of low density.
%(Some fluctuations on the
% electron temperature can also be present.)
This density pattern may
reflect the presence of shock-compressed regions and/or fluctuations in the mass-loss
rate on short timescales.

The jet, as seen in the 3\,mm-continuum emission maps
(Fig.\,\ref{f-cont}), displays a distinctive C-shaped curvature,
consistent with that observed in the JVLA 7\,mm-continuum emission
maps $\sim$11 years earlier \citep{dlF22}. As suggested by
\cite{dlF22}, evoking the old model by \cite{liv01}, the mirror
curvature of the jet could be caused by the influence of the dense
primary star’s wind on a jet hypothetically launched by the companion. 
%the one
%star that is not launching the jet, as a consequence of the
%interaction between their winds.
We believe that the presence of poloidal magnetic fields could also be
considered as an alternative explanation, potentially exerting a
bending effect on the ionized wind along the field lines. By
  equating the magnetic and mechanical energy densities, we estimate
  that a magnetic field strength of $\sim$50\,mG could induce the
  observed curvature in the ionized jet, based on typical densities
  ($\sim$2\ex{6}\,\cm3) and expansion velocities ($\sim$80\,\kms) in
  the outer regions of the jet ($\gsim$30\,au), where the curvature is
  observed. This estimate is in good agreement with existing, albeit
  limited, observations of magnetic fields in (post-)AGB envelopes at
  similar or even larger radial distances from the center
  \citep{bla09,vlem18}. 

In both cases, the orientation of the curvature depends
on the relative position of the stars, which changes over time as the
stars move in their orbit.  Since the orientation of the curvature has
remained unchanged since 2006, the star producing the bending effect
is still situated on the same side
%%, either east or west,
relative to
the star launching the jet. This observation is consistent with the
small time lapse between the ALMA and JVLA observations, approximately
one ninth of the orbital period (\porb$\sim$90\,yr).
Our ALMA data have revealed the kinematics of the jet, showing 
a non-uniform velocity field with low-expansion velocities ($\sim$10\,\kms) at the center and
maximum line-of-sight projected velocity widths of $\sim$80\,\kms\ at 
around $\sim$15--30\,au along the axis (HVSs).
%% Beyond this point ?? the emission is too weak: beyond the HVspot the velocity remains constant?? 
The abrupt change in velocity observed at/within the compact HVSs,
relative to the center, can be interpreted in at least two ways. One
possibility is that the wind acceleration occurs at these 15--30\,au
scales through some unknown mechanism, possibly a combination of
line-radiation pressure and the magneto-centrifugal mechanism.
%In this scenario, the HVSs would be regions where the wind is rapidly
%accelerated.
Alternatively, HVSs may indicate a wider range of line-of-sight
velocities arising from shocks within the ionized wind. Analytical
bow-shock models \citep{har87} and numerical simulations \citep{den08}
predict substantial velocity spreads in compact, shocked regions, which are 
indeed observed in various pPNe \citep{den08,rie06,sah06}.
Internal shocks can arise due to variations in the wind expansion
velocity, which aligns with the time-varying mRRL profiles of M\,2-9 
(Sect.\,\ref{res-mrrls}). Shocks could also result from
the interaction of the winds of the two central stars, assuming that
both stars possess winds \citep[e.g.][]{sok00,liv01,gar04}.

\section{Summary and conclusions}
\label{summary}

Using ALMA, we conducted detailed mapping in bands 6 and 3 of the
inner layers (within $r$$\sim$0\farc2$\sim$130\,au at $d$=650\,pc) of
the ionized core of M\,2-9. The mm-continuum and the \htal\ and
\htnal\ line emission were imaged with spatial resolutions reaching
down to $\sim$0\farc03 and $\sim$0\farc06, respectively. Our data
provides detailed evidence for the presence of a compact, bent jet
expanding with velocities of up to $\sim$80\,\kms\ at the core of this
remarkable object.  Using the non-LTE radiative transfer code
\coral\ (Sect.\,\ref{model} and Appendix
\ref{ap-co3ral}-\ref{ap:caveats}), we performed data modeling to
derive valuable insights into the morphology, kinematics, and physical
conditions of the jet. Our findings suggest a potential interaction
between a tenuous companion-launched jet and the dense primary star's
wind. Here is a summary of our key findings:

%% ROTATION?? 

\begin{itemize}

\item[-] The mm-continuum emitting region is elongated along the main
  symmetry axis of the nebula, with dimensions of
  0\farc4$\times$0\farc13 (260$\times$85\,au at $d$=650\,pc) at 3\,mm
  and slightly smaller at 1\,mm due to the lower opacity of the
  free-free continuum at shorter wavelengths.
     
\item[-] The spectral index of the continuum ($\sim$0.9)
  suggests predominantly free-free emission from an ionized wind, with
  a minor contribution from dust possibly originating from a warm
  compact disk.
  
\item[-] The ionized wind is collimated and displays a C-shaped
  curvature at 3\,mm consistent with previous observations taken
  11\,yr earlier at 7\,mm.
  
  \item[-] The 1\,mm-continuum map reveals a broad-waisted morphology
    oriented perpendicularly to the wind lobes, which is plausibly the
    counterpart of the compact $\sim$260\,K dust disk observed in the
    mid-IR.
%    This dust disk partially
%    obscures the emission from the north (receding) side of the
%    ionized wind, resulting in a relatively weaker appearance compared
%    to the south side.

  \item[-] CO and \trecem\ absorption features further support the
    presence of an equatorial disk with a radius of $\sim$50 au
    (likely circumbinary), with gas infall towards the central source.
    The analysis of the CO absorption feature suggests an excitation
    temperature of \tex$\lsim$330\,K and a CO column density of
    \nco$\gsim$6\ex{17}\,\cm2. A lower limit to the mass of the disk of
    $>$2\ex{-6}\,\msun\ is deduced.

  \item[-] Emission from \htal, \htnal, He30$\alpha$, and \hcg\ is
    detected, with line emission extending to radial distances of up
    to $\sim$75\,au along the direction of the nebular axis and $\sim$40\,au in the perpendicular direction.
    %, which
    %is similar but slightly smaller than the extent of the continuum
    %emission.

\item[-] Changes in the \htal\ and \htnal\ line profiles over two
  years indicate variations in the wind's kinematics and physical
  conditions. The wind has become faster during this time period.

%% KINEMATICS: AQUI  
\item[-] Both the \htal\ and \htnal\ transitions exhibit a distinct
  velocity gradient along the axis of the nebula, implying an overall
  expansion along the axis of the bipolar outflow.
  %%which is well aligned with the overall expansive kinematics of the large-scale nebula.
     
%\item[-] An abrupt increase in velocity is evident in the inner
%  regions of the ionized bipolar wind, occurring within $\pm$0\farc02
%  from the center. In this compact region, the line-of-sight
%  velocities reach a maximum value of $\sim$70\,\kms.
  %suggesting expansion velocities of up to 240\,\kms\ (assuming an
  %inclination angle $i$ of 17\degr).

%\item[-]  The observed velocity gradient
%  corresponds to an average (projected) velocity gradient of at least
%  1150\kms\,arcsec$^{-1}$, equivalent to approximately
%  1.8\,\kms\,au$^{-1}$. Such rapid changes in velocity imply extremely
%  short kinematical ages, likely less than one year.

\item[-] Low expansion velocities, of $\sim$10\,\kms, are observed at
  the center.  Rapid wind acceleration or shocks are suggested by peak
  velocities of $\sim$70--90\,\kms\ in two compact regions, situated at a
  radial distance of $\sim$$\pm$15--30\,au along the axis (HVSs).

%  There is a velocity gradient within the HVSs. 
  %% or HV-regions or HV-hotspots ??  MODELO: podrian ser regiones mas
  %% calientes?? No podemos distinguir entre estos casos porque no
  %% estan resueltos espacialmente.
%This intriguing behavior points to localized regions in the compact
%wind where the acceleration processes are particularly intense or
%where internal shocks arise due to ejections with time-variable
%velocity.

\item[-] The ionized wind has a short kinematical age, including
  regions of less than a year old, explaining changes in the mRRL profiles
  on similar year-long timescales.
  
  %% ROTATION?? 
\item[-] The emission maps of \htal, \htnal, and He30$\alpha$ lines
  reveal a subtle velocity gradient perpendicular to the lobes
  suggestive of rotation (\vrot$\sim$7-10\,\kms).

  \item[-] Radiative transfer modeling indicates an average electron
    temperature of $\sim$15000\,K and reveals a non-uniform density
    structure within the ionized jet, with electron densities ranging
    from \dense$\approx$10$^6$ to $\lsim$10$^8$\,\cm3. The mass and
    average mass-loss rate of the ionized wind is deduced to be of
    $M_{\rm ion}$$\sim$4\ex{-6}\,\msun\ and
    \mloss$\approx$10$^{-7}$\,\my, respectively. These results
    potentially reflect a complex bipolar flow pattern resulting from
    the interaction of a tenuous companion-launched jet and the dense
    primary star's wind.
%  If the
%  ionized wind emerges from the mass-lossing star then the measured%
%  mass-loss rate of the ionized present-day wind is significantly
%  lower than the rates observed $\sim$1300 and $\sim$900 years ago,
%  which were responsible for forming the large-scale nebula
%  (Sect.\,\ref{intro}). In this scenrio, the substantial reduction in the mass-loss rate
%  would suggest that the primary star has transitioned to a
  %  post-AGB or post-RGB phase.
%  If the ionized wind represents a complex bipolar structure resulting 
%  from the interaction of a jet launched by the companion and
%  the dense, slow wind of the primary, then the estimated mass-loss
%  rate is expected to lie between the mass-loss rate of the primary
    %  star and that of the companion.

  \item[-] The mass of the central system is uncertain, but it is
    likely in the range of 0.4 to 2\,\msun, based on the rotational
    and infall motions observed in the jet and the circumbinary disks,
    respectively, described in this study.

\item[-] The required number of Lyman continuum photons per second is
  consistent with a central star with an effective temperature of
  $\sim$25\,kK and a luminosity of $\sim$600\,\ls, perhaps a post-RGB
  star. We do not observe empirical evidence of fast
  ($\gsim$1000\,\kms) ejections indicative of a compact companion.
  While a WD companion is not ruled out, our findings do not provide
  empirical evidence for it.

  \item[-] The nature of the companion remains highly
  uncertain. Unravelling the nature of M\,2-9's central binary system
  and the intricate interplay between the primary star and its
  companion are essential for comprehending the mechanisms governing
  the observed mass loss and wind dynamics in this system.

%\item[-] The nature of the companion
%remains highly uncertain. Our findings indicate that 
%a compact object (e.g., a WD) is not strictly necessary
%for two reasons: 1) we do not observe empirical evidence of fast
%($\gsim$1000\,\kms) ejections, and 2) the free-free continuum and mRRLs
%emission can be adequately explained by the 25kK, 600\,\ls\ central
%star. While our results do not rule out the possibility of a WD 
%companion, they do not provide evidence in support of it.
  
  \end{itemize}

%% Data avilability?? upload to ZENODO. 

\begin{acknowledgements}
We would like to thank the referee, Joel Kastner, for his insightful
comments and suggestions, which helped improve the clarity of this
work. This paper makes use of the following ALMA data:
ADS/JAO\-.ALMA\#2016.1.00161.S and ADS/JAO.ALMA\#2017.1.00376.S.  ALMA
is a partnership of ESO (representing its member states), NSF (USA)
and NINS (Japan), together with NRC (Canada), MOST and ASIAA (Taiwan),
and KASI (Republic of Korea), in cooperation with the Republic of
Chile. The Joint ALMA Observatory is operated by ESO, AUI/NRAO and
NAOJ. The data here presented have been reduced and analyzed using
CASA (ALMA default calibration software; {\tt https://casa.nrao.edu})
and the GILDAS software ({\tt http://www.iram.fr/IRAMFR/GILDAS)}.
This work is part of the I+D+i projects PID2019-105203GB-C22 and
PID2019-105203GB-C21 funded by Spanish MCIN/AEI/10.13039/501100011033.
This research has made use of the SIMBAD database operated at CDS
(Strasbourg, France) and the NASA’s Astrophysics Data System.

% Daniel? 
\end{acknowledgements}


\begin{thebibliography}{}


  
\bibitem[Allen \& Swings(1972)]{all72} Allen, D.~A. \& Swings,
  J.~P.\ 1972, \apj, 174, 583. %doi:10.1086/151520 %B-type

\bibitem[Arrieta \& Torres-Peimbert(2003)]{arr03} Arrieta, A. \&
  Torres-Peimbert, S.\ 2003, \apjs, 147, 97. %doi:10.1086/374922

\bibitem[B{\'a}ez-Rubio et al.(2013)]{bae13} B{\'a}ez-Rubio, A.,
  Mart{\'{\i}}n-Pintado, J., Thum, C., \& Planesas, P.\ 2013, \aap,
  553, A45

\bibitem[B{\'a}ez-Rubio et al.(2014)]{bae14} B{\'a}ez-Rubio, A.,
  Mart{\'\i}n-Pintado, J., Thum, C., et al.\ 2014, \aap, 571,
  L4. %doi:10.1051/0004-6361/201424389

\bibitem[Bujarrabal et al.(2001)]{buj01} Bujarrabal, V.,
  Castro-Carrizo, A., Alcolea, J., et al.\ 2001, \aap, 377,
  868. %doi:10.1051/0004-6361:20011090
  
\bibitem[Bujarrabal et al.(2005)]{buj05} Bujarrabal, V., Castro-Carrizo, A., Alcolea, J., \& Ner
i, R.\ 2005, \aap, 441, 1031 %% Red Rect. turb<0.5km/s 

  
\bibitem[Balick \& Frank(2002)]{bal02} Balick, B., \& Frank, A.\ 2002,
  \araa, 40, 439

\bibitem[Balick et al.(2018)]{bal18} Balick, B., Frank, A., Liu, B.,
  et al.\ 2018, \apj, 853, 168. %doi:10.3847/1538-4357/aaa772

\bibitem[Blackman(2009)]{bla09} Blackman, E.~G.\ 2009, Cosmic Magnetic Fields: From Planets, to Stars and Galaxies, 259, 35. %doi:10.1017/S174392130903004X

\bibitem[Bloecker(1995)]{blo95} Bloecker, T.\ 1995, \aap, 299, 755 %%
  evolutionary tracks

\bibitem[Brussaard \& van de Hulst(1962)]{Brussaard1962} Brussaard,
  P.~J. \& van de Hulst, H.~C.\ 1962, Reviews of Modern Physics, 34,
  507. %doi:10.1103/RevModPhys.34.507

\bibitem[Bujarrabal et al.(2021)]{buj21} Bujarrabal, V., Ag{\'u}ndez,
  M., G{\'o}mez-Garrido, M., et al.\ 2021, \aap, 651,
  A4. %doi:10.1051/0004-6361/202141002

\bibitem[Calvet \& Cohen(1978)]{cal78} Calvet, N., \& Cohen, M.\ 1978,
  \mnras, 182, 687 %%B1 V
  
\bibitem[Castro-Carrizo et al.(2012)]{cc12} Castro-Carrizo, A., Neri,
  R., Bujarrabal, V., et al.\ 2012, \aap, 545, A1

\bibitem[Castro-Carrizo et al.(2017)]{cc17} Castro-Carrizo, A.,
  Bujarrabal, V., Neri, R., et al.\ 2017, \aap, 600, A4. %doi:10.1051/0
  004-6361/201630101

\bibitem[Clyne et al.(2015)]{cly15} Clyne, N., Akras, S., Steffen, W.,
  et al.\ 2015, \aap, 582, A60. %doi:10.1051/0004-6361/201526585

\bibitem[Condon \& Ransom(2016)]{con16} Condon, J.~J. \& Ransom, S.~M.\ 2016, Essential Radio Astronomy, by James J. Condon and Scott M. Ransom. ISBN: 978-0-691-13779-7. Princeton, NJ: Princeton University Press, 2016.

  
\bibitem[Corradi et al.(2011)]{cor11} Corradi, R.~L.~M., Balick, B.,
  \& Santander-Garc{\'\i}a, M.\ 2011, \aap, 529,
  A43. %doi:10.1051/0004-6361/201016361

\bibitem[Chen et al.(2017)]{chen17} Chen, Z., Frank, A., Blackman,
  E.~G., et al.\ 2017, \mnras, 468, 4465. %doi:10.1093/mnras/stx680

\bibitem[Decin et al.(2010)]{dec10} Decin, L., De Beck, E., Br{\"u}nken, S., et al.\ 2010, \aap,
 516, A69 %%turbulence IKtau.

\bibitem[Dennis et al.(2008)]{den08} Dennis, T.~J., Cunningham, A.~J.,
  Frank, A., et al.\ 2008, \apj, 679, 1327. %doi:10.1086/587730

\bibitem[Doyle et al.(2000)]{doy00} Doyle, S., Balick, B., Corradi,
  R.~L.~M., et al.\ 2000, \aj, 119, 1339. %doi:10.1086/301267

%\bibitem[Dyson \& Williams(1997)]{dys97} Dyson, J.~E. \& Williams,
%  D.~A.\ 1997, The physics of the interstellar medium.  Edition: 2nd
%  ed. Publisher: Bristol: Institute of Physics Publishing,
%  1997. Edited by J. E. Dyson and D. A. Williams. Series: The graduate
%  series in astronomy. ISBN: 0750303069. %doi:10.1201/9780585368115

  %\bibitem[Feibelman(1984)]{fei84} Feibelman, W.~A.\ 1984, \apj, 287, 353. %doi:10.1086/162695  %% late-O IUE spec
%% absence of HeII lambda1640A


%\bibitem[Frank et al.(2002)]{fran02} Frank, J., King, A., \& Raine,
%  D.~J.\ 2002, Accretion Power in Astrophysics, by Juhan Frank and
%  Andrew King and Derek Raine, pp. 398. ISBN 0521620538. Cambridge,
%  UK: Cambridge University Press, February 2002., 398

  
\bibitem[Garc{\'\i}a-Arredondo \& Frank(2004)]{gar04}
  Garc{\'\i}a-Arredondo, F. \& Frank, A.\ 2004, \apj, 600,
  992. %doi:10.1086/379821

\bibitem[Goldwire(1968)]{Goldwire1968} Goldwire, H.~C.\ 1968, ApJS,
  17, 445. %doi:10.1086/190180

\bibitem[Hartigan et al.(1987)]{har87} Hartigan, P., Raymond, J., \&
  Hartmann, L.\ 1987, \apj, 316, 323. %doi:10.1086/165204

\bibitem[Herman et al.(1986)]{her86} Herman, J., Burger, J.~H., and
  Penninx, W.~H.\ 1986, \aap, 167, 247

\bibitem[H{\"o}fner \& Olofsson(2018)]{hof18} H{\"o}fner, S. \&
  Olofsson, H.\ 2018, \aapr, 26, 1. %doi:10.1007/s00159-017-0106-5


\bibitem[Hrivnak et al.(2020)]{hri20} Hrivnak, B.~J., Henson, G.,
  Hillwig, T.~C., et al.\ 2020, \apj, 901,
  9. %doi:10.3847/1538-4357/abad8c

%\bibitem[$Hubble$ Legacy Archive (2013)]{hst} Hubble Legacy Archive
%  2013, NASA, ESA – Processing: Judy Schmidt, Astronomical Picture of
%  the Day
  
\bibitem[de la Fuente et al.(2022)]{dlF22} de la Fuente, E., Trinidad,
  M.~A., Tafoya, D., et al.\ 2022, \pasj, 74, 594. %doi:10.1093/
  pasj/psac020

\bibitem[Kamath et al.(2016)]{kam16} Kamath, D., Wood, P.~R., Van
  Winckel, H., et al.\ 2016, \aap, 586,
  L5. %doi:10.1051/0004-6361/201526892

\bibitem[Kardashev(1959)]{Kardashev1959} Kardashev, N.~S.\ 1959,
  Soviet Astronomy, 3, 813

\bibitem[Kielkopf(1973)]{Kielkopf1973} Kielkopf, J.~F.\ 1973, Journal
  of the Optical Society of America (1917-1983), 63, 987

\bibitem[Kohoutek \& Surdej(1980)]{koh80} Kohoutek, L. \& Surdej,
  J.\ 1980, \aap, 85, 161

\bibitem[Kwok et al.(1985)]{kwo85} Kwok, S., Purton, C.~R., Matthews,
  H.~E., \& Spoelstra , T.~A.~T.\ 1985, \aap, 144, 321

\bibitem[Lanz \& Hubeny(2007)]{lan07} Lanz, T., \& Hubeny, I.\ 2007,
  \apjs, 169, 83.

\bibitem[Lee et al.(2001)]{lee01} Lee, H.-W., Kang, Y.-W., \& Byun,
  Y.-I.\ 2001, \apjl, 551, L121. %doi:10.1086/319830

\bibitem[Lim \& Kwok(2000)]{lim00} Lim, J., \& Kwok, S.\ 2000,
  Asymmetrical Planetary Nebulae II: From Origins to Microstructures,
  1 99, 259

\bibitem[Lim \& Kwok(2003)]{lim03} Lim, J., \& Kwok, S.\ 2003,
  Symbiotic Stars Probing Stellar Evolution, 303, 437

\bibitem[Livio \& Soker(2001)]{liv01} Livio, M. \& Soker, N.\ 2001,
  \apj, 552, 685. %doi:10.1086/320567

\bibitem[Lykou et al.(2011)]{lyk11} Lykou, F., Chesneau, O., Zijlstra,
  A.~A., et al.\ 2011, \aap, 527,
  A105. %doi:10.1051/0004-6361/200913845

\bibitem[Martin \& Maass(2022)]{Martin2022} Martin, P. \& Maass,
  F.\ 2022, Results in Physics, 35,
  105283. %doi:10.1016/j.rinp.2022.105283

\bibitem[Mart{\'\i}nez-Henares et al.(2023)]{mar23}
  Mart{\'\i}nez-Henares, A., Jim{\'e}nez-Serra, I.,
  Mart{\'\i}n-Pintado, J., et al.\ 2023, \apj, 955,
  119. %doi:10.3847/1538-4357/acebcd
  
\bibitem[Mastrodemos \& Morris(1999)]{mm99} Mastrodemos, N. \& Morris,
  M.\ 1999, \apj, 523, 357. %doi:10.1086/307717

\bibitem[Menzel(1937)]{Menzel1937} Menzel, D.~H.\ 1937, ApJ, 85,
  330. %doi:10.1086/143827

\bibitem[Mezger \& Hoglund(1967)]{Mezger1967} Mezger, P.~G. \&
  Hoglund, B.\ 1967, ApJ, 147, 490. %doi:10.1086/149031

\bibitem[Miller Bertolami(2016)]{mil16} Miller Bertolami, M.~M.\ 2016,
  \aap, 588, A25. %doi:10.1051/0004-6361/201526577

\bibitem[Mohamed \& Podsiadlowski(2007)]{moh07} Mohamed, S. \&
  Podsiadlowski, P.\ 2007, 15th European Workshop on White Dwarfs,
  372, 397
  
\bibitem[Olofsson et al.(2019)]{olo19} Olofsson, H., Khouri, T.,
  Maercker, M., et al.\ 2019, \aap, 623,
  A153. %doi:10.1051/0004-6361/201834897

\bibitem[Oudmaijer et al.(2022)]{oud22} Oudmaijer, R.~D., Jones,
  E.~R.~M., \& Vioque, M.\ 2022, \mnras, 516,
  L61. %doi:10.1093/mnrasl/slac088

%\bibitem[Osterbrock \& Ferland(2006)]{ost06} Osterbrock, D.~E. \& Ferland, G.~J.\ 2006, Astrophysics of gaseous nebulae and active galactic nuclei, 2nd. ed. by D.E. Osterbrock and G.J. Ferland. Sausalito, CA: University Science Books, 2006

  
  \bibitem[Panagia \& Felli(1975)]{pana75} Panagia, N. \& Felli, M.\ 1975, \aap, 39, 1
  
\bibitem[Purton et al.(1975)]{pur75} Purton, C.~R., Feldman, P.~A., \&
  Marsh, K.~A.\ 1975, \apj, 195, 479 %% M2-9

\bibitem[Ramos-Medina et al.(2018)]{ram18} Ramos-Medina, J.,
  S{\'a}nchez Contreras, C., Garc{\'\i}a-Lario, P., et al.\ 2018,
  \aap, 619, C2. %doi:10.1051/0004-6361/201833177e

\bibitem[Riera et al.(2006)]{rie06} Riera, A., Binette, L., \& Raga,
  A.~C.\ 2006, \aap, 455, 203. %doi:10.1051/0004-6361:20053499

%%\bibitem[Rubin(1968)]{rub68} Rubin, R.~H.\ 1968, \apj, 154, 391. %doi:10.1086/149766

\bibitem[Rybicki \& Lightman(1986)]{Rybicki1986} Rybicki, G.~B. \&
  Lightman, A.~P.\ 1986, Radiative Processes in Astrophysics, by
  George B. Rybicki, Alan P. Lightman, pp. 400. ISBN
  0-471-82759-2. Wiley-VCH , June 1986., 400

  
\bibitem[Sahai et al.(2006)]{sah06} Sahai, R., Young, K., Patel,
  N.~A., et al.\ 2006, \apj, 653, 1241. %doi:10.1086/508507

  \bibitem[S{\'a}nchez Contreras et al.(1998)]{san98} S{\'a}nchez
    Contreras, C., Alcolea, J., Bujarrabal, V., et al.\ 1998, \aap,
    337, 233 % % mm-dust / large grains

\bibitem[S{\'a}nchez Contreras et al.(2017)]{san17} S{\'a}nchez
  Contreras, C., B{\'a}ez-Rubio, A., Alcolea, J., et al.\ 2017, \aap,
  603, A67. %doi:10.1051/0004-6361/201730385

\bibitem[S{\'a}nchez Contreras et al.(2019)]{san19} S{\'a}nchez
  Contreras, C., B{\'a}ez-Rubio, A., Alcolea, J., et al.\ 2019, \aap,
  629, A136. %doi:10.1051/0004-6361/201936057 %%MWC 922.

\bibitem[S{\'a}nchez Contreras et al.(2022)]{san22} S{\'a}nchez
  Contreras, C., Alcolea, J., Rodriguez Cardoso, R., et al.\ 2022,
  \aap, 665, A88. %doi:10.1051/0004-6361/202243623
  
\bibitem[Smith \& Gehrz(2005)]{smg05} Smith, N. \& Gehrz, R.~D.\ 2005,
  \aj, 129, 969. %doi:10.1086/426919
  
\bibitem[Sch{\"o}ier et al.(2004)]{sch04} Sch{\"o}ier, F.~L., Olofsson, H., Wong, T., Lindqvist,
 M., \& Kerschbaum, F.\ 2004, \aap, 422, 651 %%turbulence

\bibitem[Sch{\"o}nberner(1983)]{sch83} Sch{\"o}nberner, D.\ 1983,
  \apj, 272, 708

\bibitem[Schwarz et al.(1997)]{sch97} Schwarz, H.~E., Aspin, C.,
  Corradi, R.~L.~M., et al.\ 1997, \aap, 319, 267

\bibitem[Soker \& Rappaport(2000)]{sok00} Soker, N. \& Rappaport,
  S.\ 2000, \apj, 538, 241. %doi:10.1086/309112

\bibitem[Solf(2000)]{sol00} Solf, J.\ 2000, \aap, 354, 674

\bibitem[Storey \& Hummer(1995)]{sto95} Storey, P.~J., \& Hummer,
  D.~G.\ 1995, \mnras, 272, 41 %% bn departure coefficients
    
\bibitem[Swings \& Andrillat(1979)]{swi79} Swings, J.~P., \&
  Andrillat, Y.\ 1979, \aap, 74, 85 %% B[e]-type

%%\bibitem[Urquhart et al.(2013)]{urq13} Urquhart, J.~S., Thompson, M.~A., Moore, T.~J.~T., et al.\ 2013, \mnras, 435, 400. %doi:10.1093/mnras/stt1310   %% NLyc formula: 

\bibitem[Torres-Peimbert et al.(2010)]{tor10} Torres-Peimbert, S.,
  Arrieta, A., \& Bautista, M.\ 2010, \rmxaa, 46, 221
  
\bibitem[van den Bergh(1974)]{vdB74} van den Bergh, S.\ 1974, \aap,
  32, 351
  
\bibitem[van der Tak et al.(2007)]{radex} van der Tak, F.~F.~S.,
  Black, J.~H., Sch{\"o}ier, F.~L., et al.\ 2007, \aap, 468,
  627. %doi:10.1051/0004-6361:20066820

%%\bibitem[Walmsley(1990)]{wal90} Walmsley, C.~M.\ 1990, \aaps, 82, 201 %% bn departure coefficients 
\bibitem[Vassiliadis \& Wood(1993)]{vas93} Vassiliadis, E., \& Wood,
  P.~R.\ 1993, \apj, 413, 641

\bibitem[Welsh et al.(2013)]{wel13} Welsh, B.~Y., Wheatley, J.,
  Dickinson, N.~J., et al.\ 2013, \pasp, 125, 644. %doi:10.1086/671190
  
\bibitem[Wilson et al.(2009)]{Wilson2009} Wilson, T.~L., Rohlfs, K.,
  \& H{\"u}ttemeister, S.\ 2009, Tools of Radio Astronomy, by Thomas
  L. Wilson; Kristen Rohlfs and Susanne H{\"u}ttemeister. ISBN
  978-3-540-85121-9. Published by Springer-Verlag, Berlin, Germany,
  2009.. %doi:10.1007/978-3-540-85122-6

\bibitem[Vlemmings(2018)]{vlem18} Vlemmings, W.~H.~T.\ 2018, Contributions of the Astronomical Observatory Skalnate Pleso, 48, 187

 
\end{thebibliography}

\appendix

\section{Additional Figures}
\label{ap-figures}

% Fig.---------------------------------------
   \begin{figure*} %[htbp!]
   \centering
   \includegraphics[width=0.475\hsize]{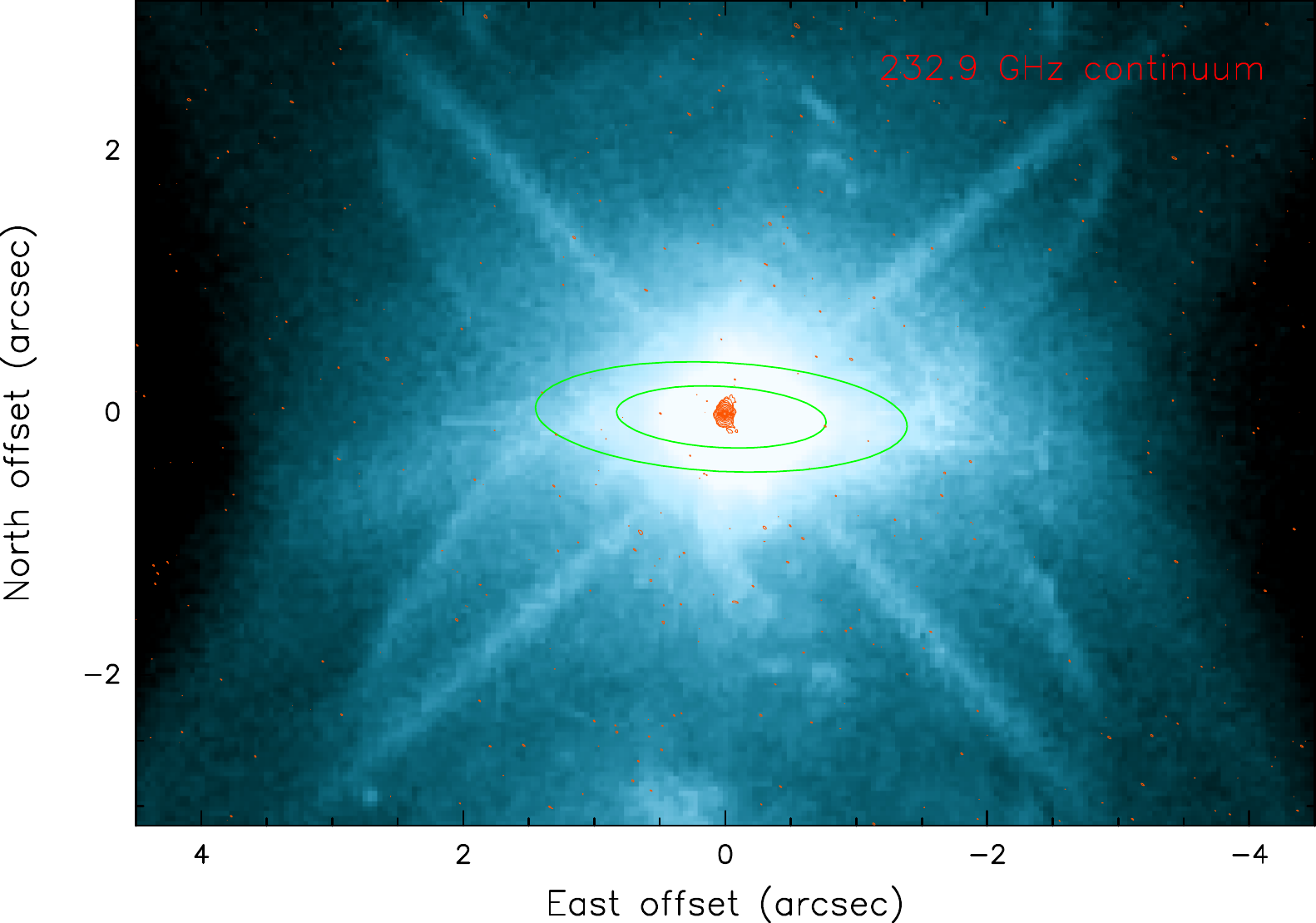}
   \includegraphics[width=0.475\hsize]{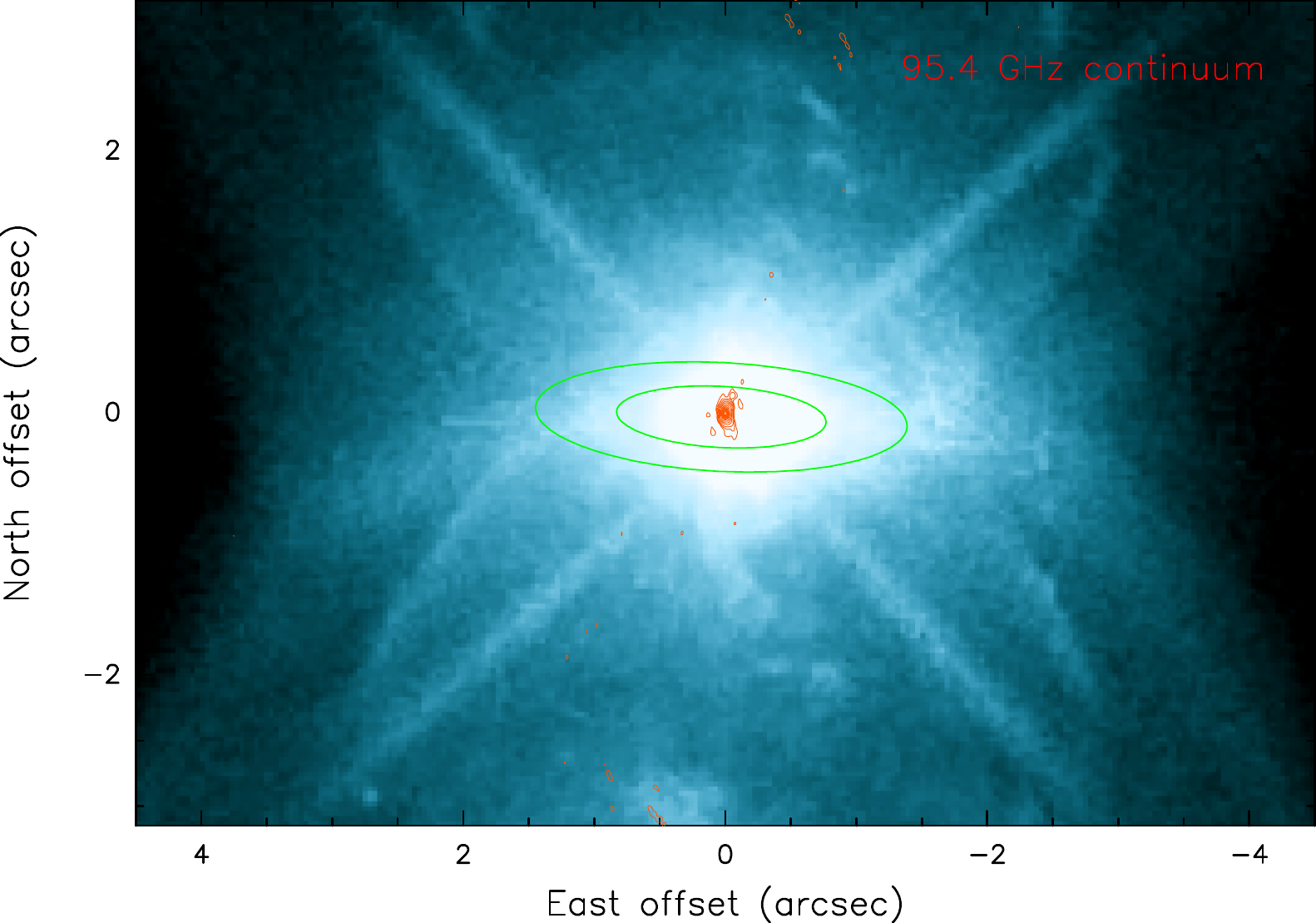}
     \caption{ALMA continuum emission maps of M\,2-9 (as in
       Fig.\,\ref{f-cont}-bottom panels) overlaid on the optical $HST$
       image (Credit: $Hubble$ Legacy Archive 2013, NASA, ESA --
       Processing: Judt Schmidt). The field of view is limited to the
       core and the base of the lobes of the optical nebula.  Contours are at levels of
       0.15\%, 0.25\%, 0.5\%, 1\%, 2.5\%, 5\%, and from 10\% to 100\%
       by 10\%\ steps for the 1\,mm map and 0.25\%, 0.5\%, 1\%, 2.5\%,
       5\%, and from 10\% to 100\% by 10\%\ steps for the 3\,mm map,
       relative to the peak emission (1\,mm: 147.4\,mJy/beam and
       3\,mm: 48\,mJy/beam).  The ellipses delimiting the rims of the
       inner CO equatorial ring reported in \cite{cc17} are also
       plotted as a reference. The relative position between the
       optical image, the CO rings, and the continuum map is as in
       \cite{cc17} -- see their Fig.\,8.
    \label{f-hstcont}
     }
     %% /pcdisk/jbell3/csanchez/m2-9/ACC-hst/@m29_hst_cont_4rrls.map
   \end{figure*}
%% Fig.------------------------------------------------------------------------------------

% Fig.---------------------------------------
   \begin{figure*} %[htbp!]
   \centering
   \includegraphics[width=0.99\hsize]{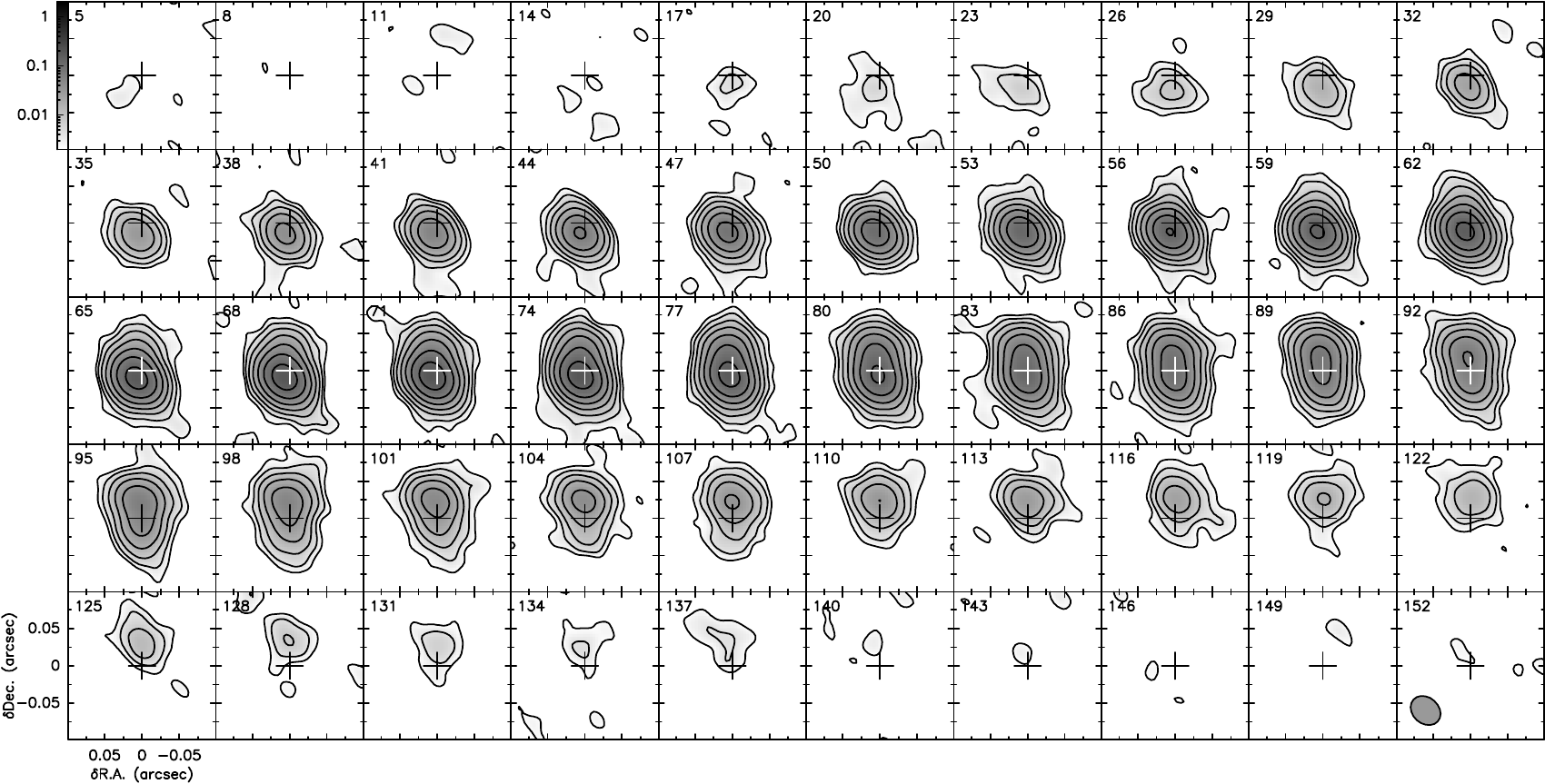}
   \caption{ALMA velocity-channel maps of the \htal\ line in
     M\,2-9. The beam is 0\farc044$\times$0\farc035,
     PA=48\degr\ (bottom-left corner of the last panel). Level
     contours are 2.5$\times$2$^{(i-1)}$\,m\jb, for $i$=1,2,3,... The
     rms noise of these maps is $\sigma$=1\,m\jb\ (dv=3\,\kms).
          \label{f-cubeH30a}
        }
   \end{figure*}
%% Fig.------------------------------------------------------------------------------------
% Fig.---------------------------------------
   \begin{figure*}[htbp!]
   \centering
   \includegraphics[width=0.99\hsize]{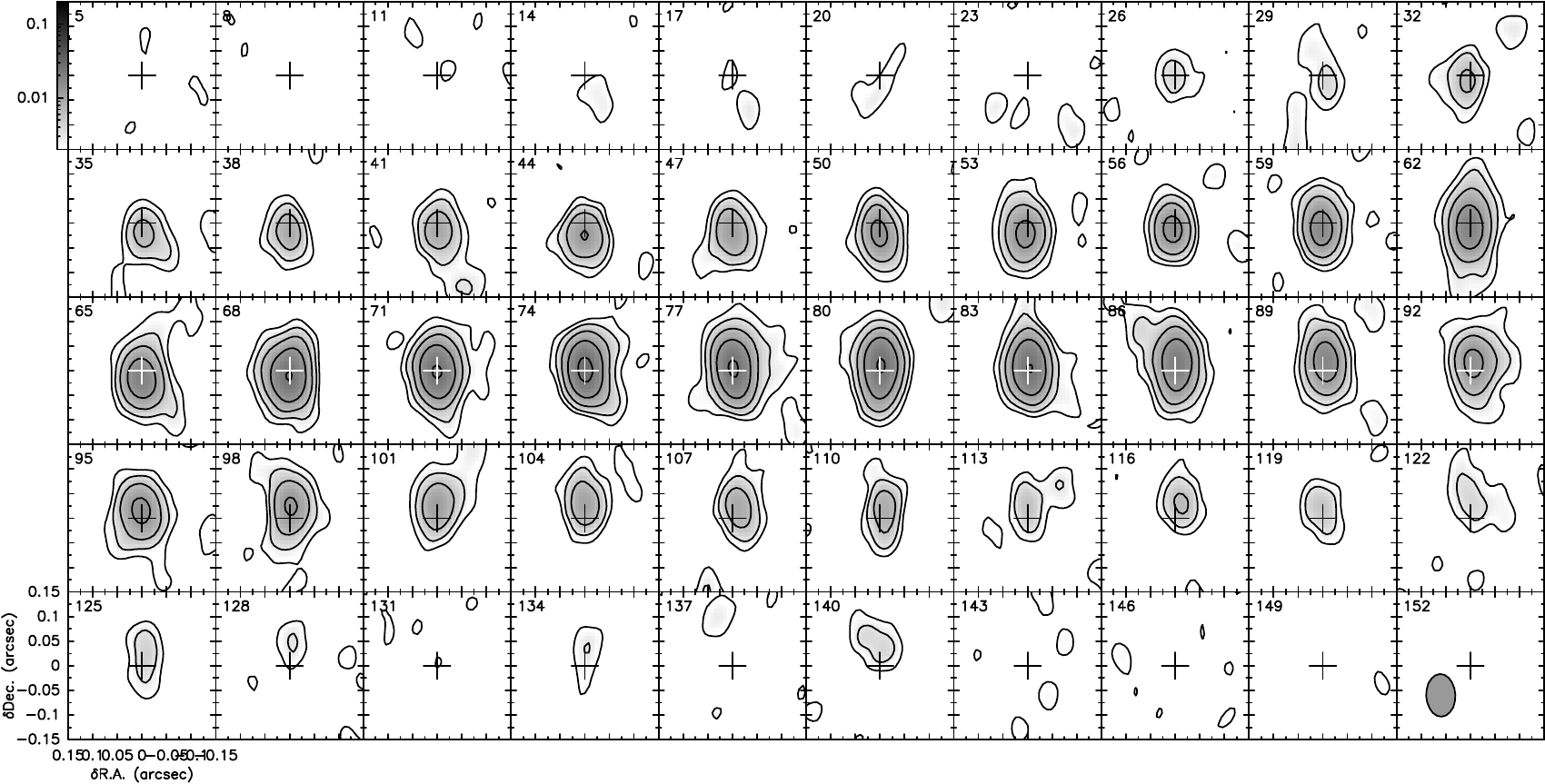}
   \caption{ALMA velocity-channel maps of the \htnal\ line in
     M\,2-9. The beam is 0\farc087$\times$0\farc059,
     PA=2.5\degr\ (bottom-left corner of the last panel). Level
     contours are 1.5$\times$2$^{(i-1)}$\,m\jb, for $i$=1,2,3,... The
     rms noise of these maps is $\sigma$=0.8\,m\jb\ (dv=3\,\kms).
          \label{f-cubeH39a}
        }
   \end{figure*}
   %% Fig.------------------------------------------------------------------------------------
   
%% ========================== Fig. He30a
   \begin{figure*}[htbp!]
     \centering
     \includegraphics*[width=0.30\hsize]{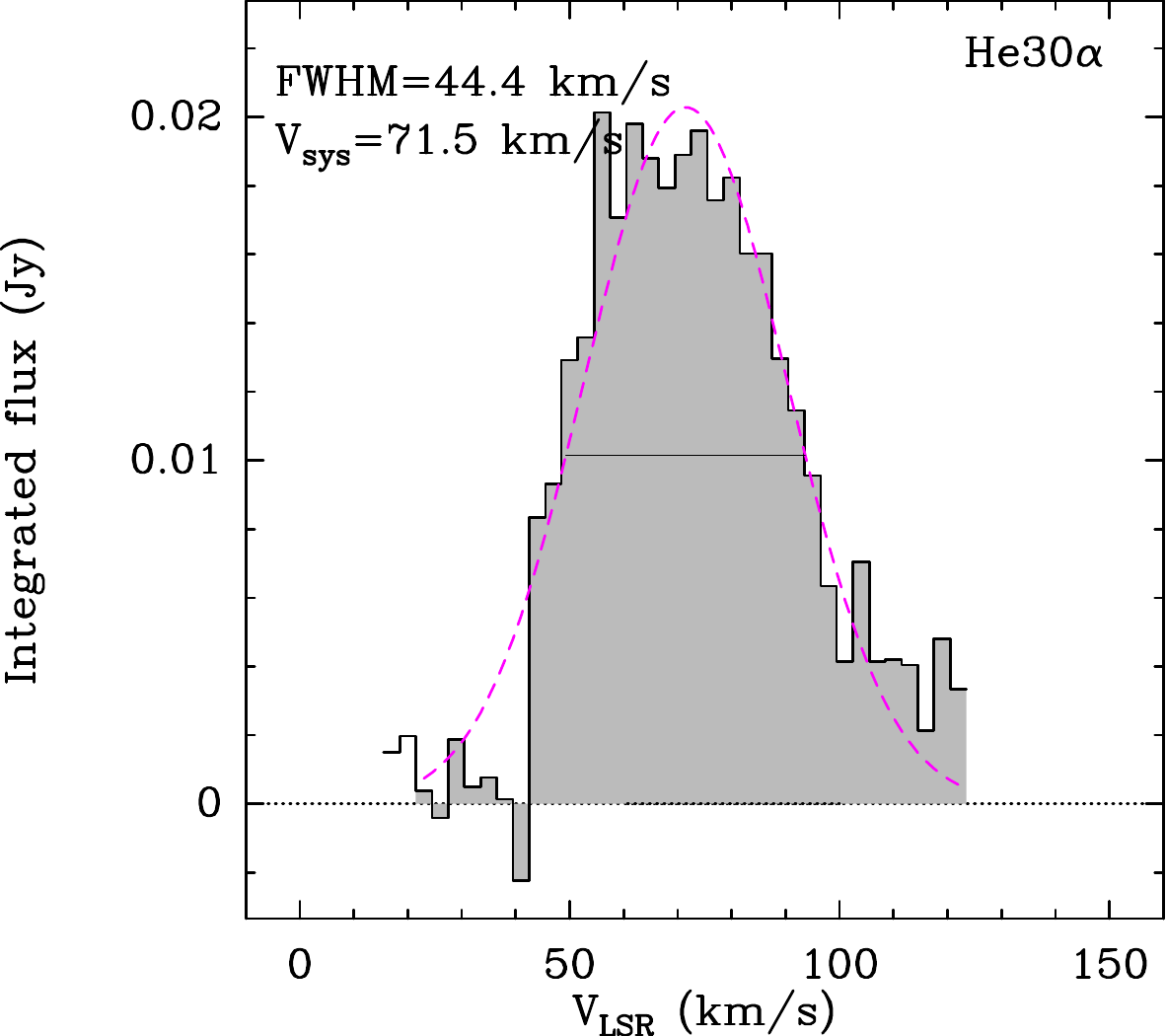}
     %1dspec
     \includegraphics*[width=0.33\hsize]{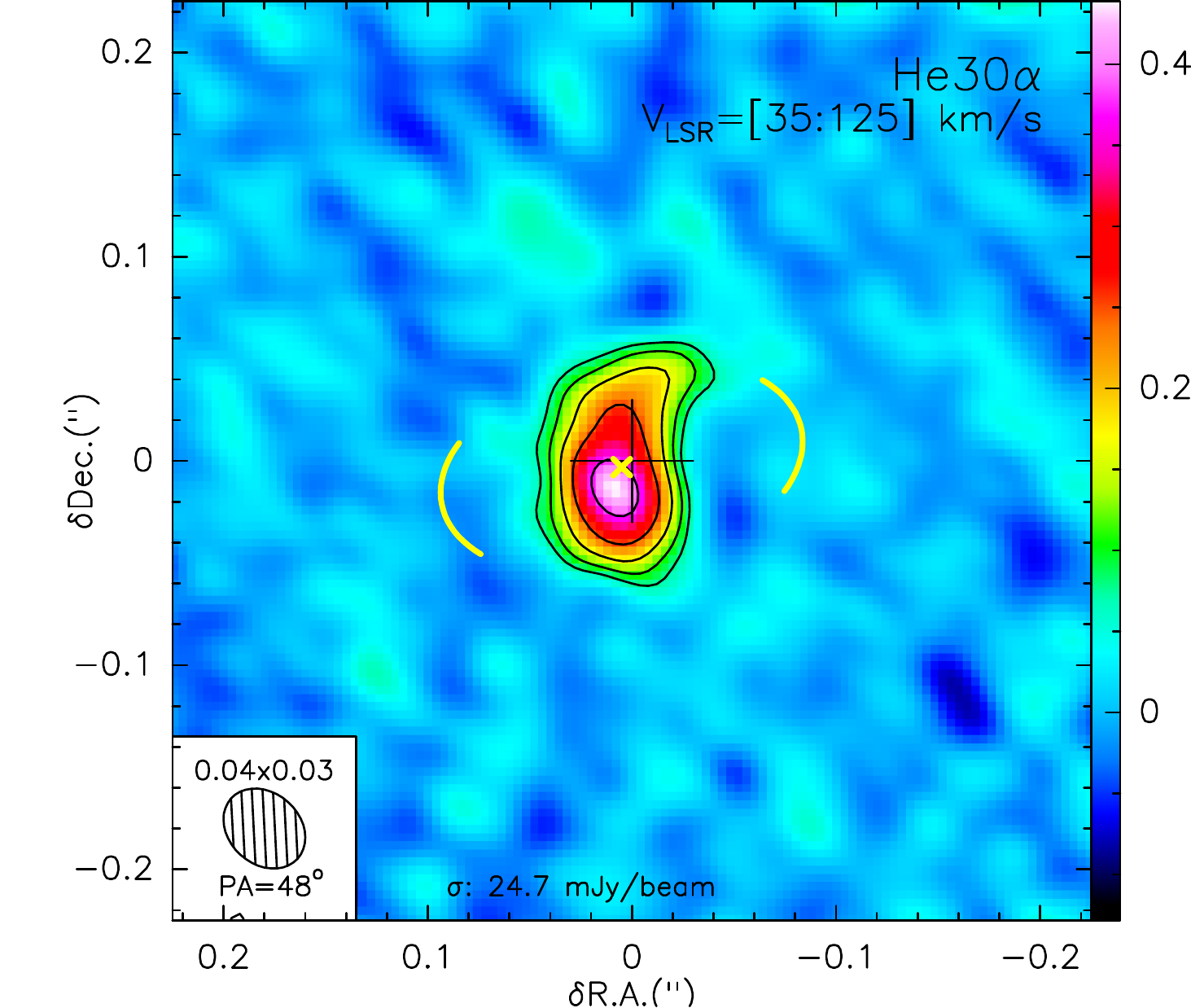}
     %area
     \includegraphics*[width=0.34\hsize]{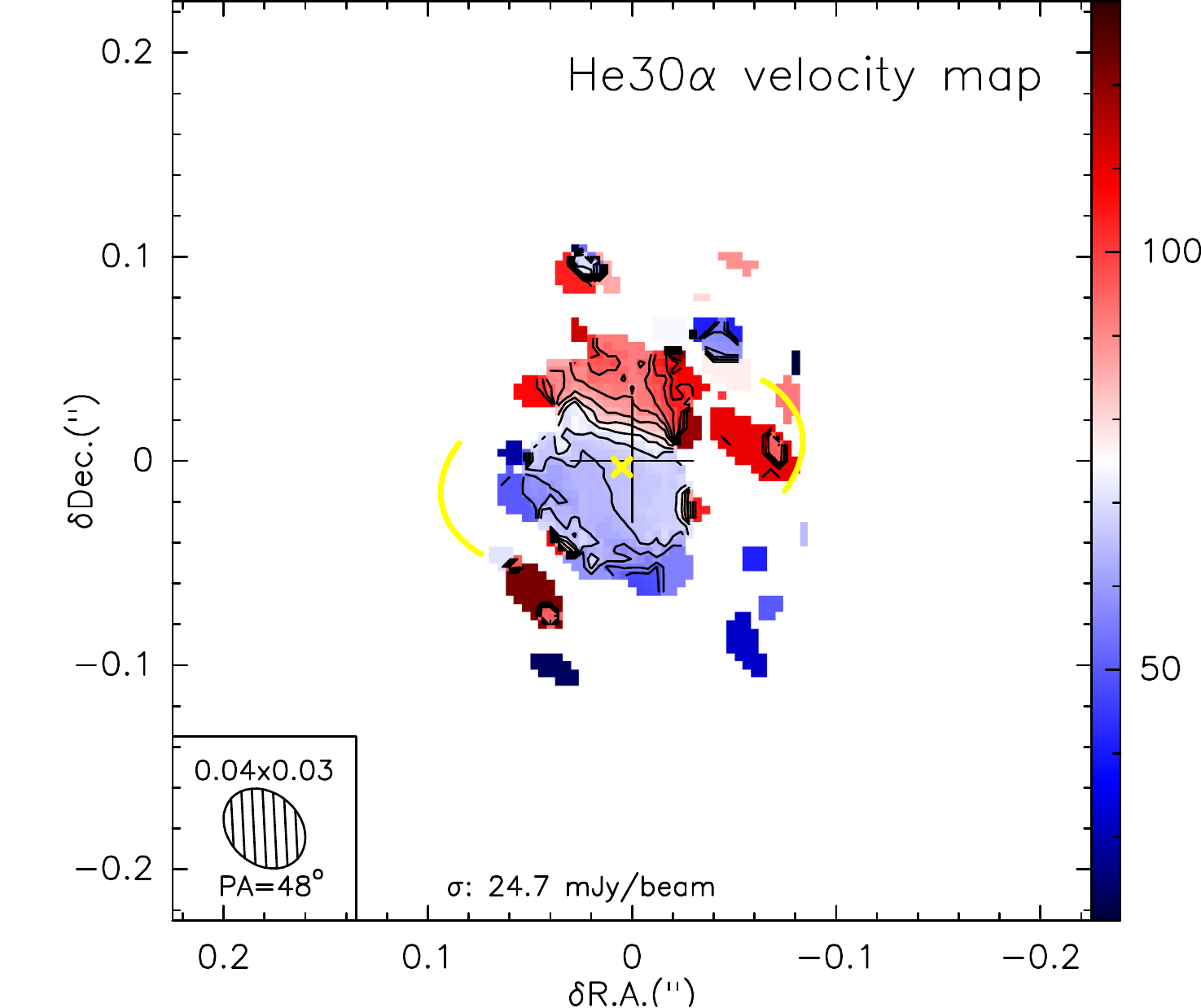}
     %velo
         \caption{Summary of ALMA data for the He30$\alpha$ line as in
           Fig.\,\ref{f-h30yh39}.
           \label{f-he30a}}
   \end{figure*}
   %% /pcdisk/jbell3/csanchez/m2-9/selfcal-csc2/:   @1dspec.greg   @H30area.greg  @He30aspec.greg 

%% ========================== Fig. He30a
%% ========================== Fig. H55g 
   \begin{figure*}[htbp!]
     \centering
     \includegraphics*[width=0.30\hsize]{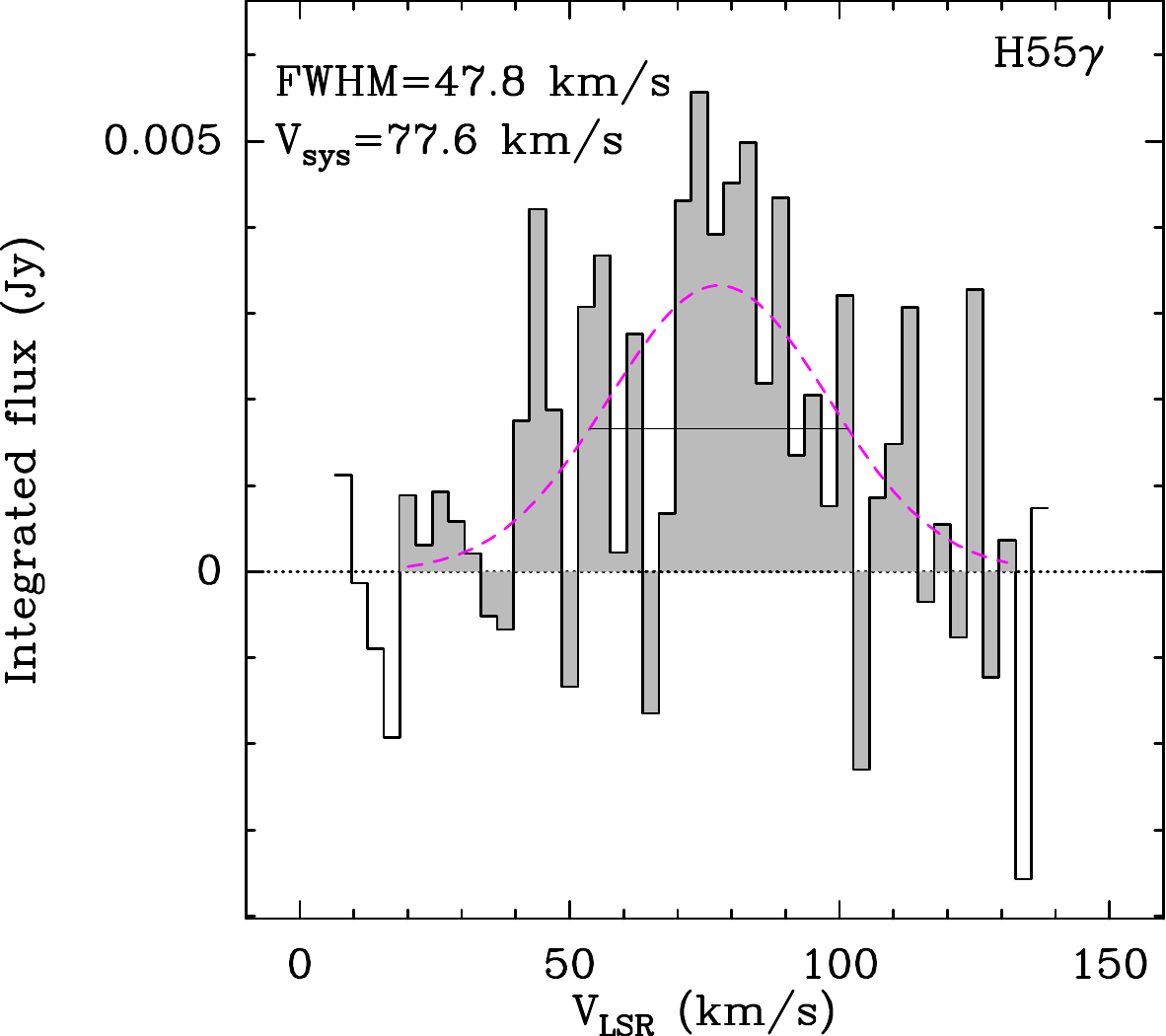}
     %1dspec
     \includegraphics*[width=0.33\hsize]{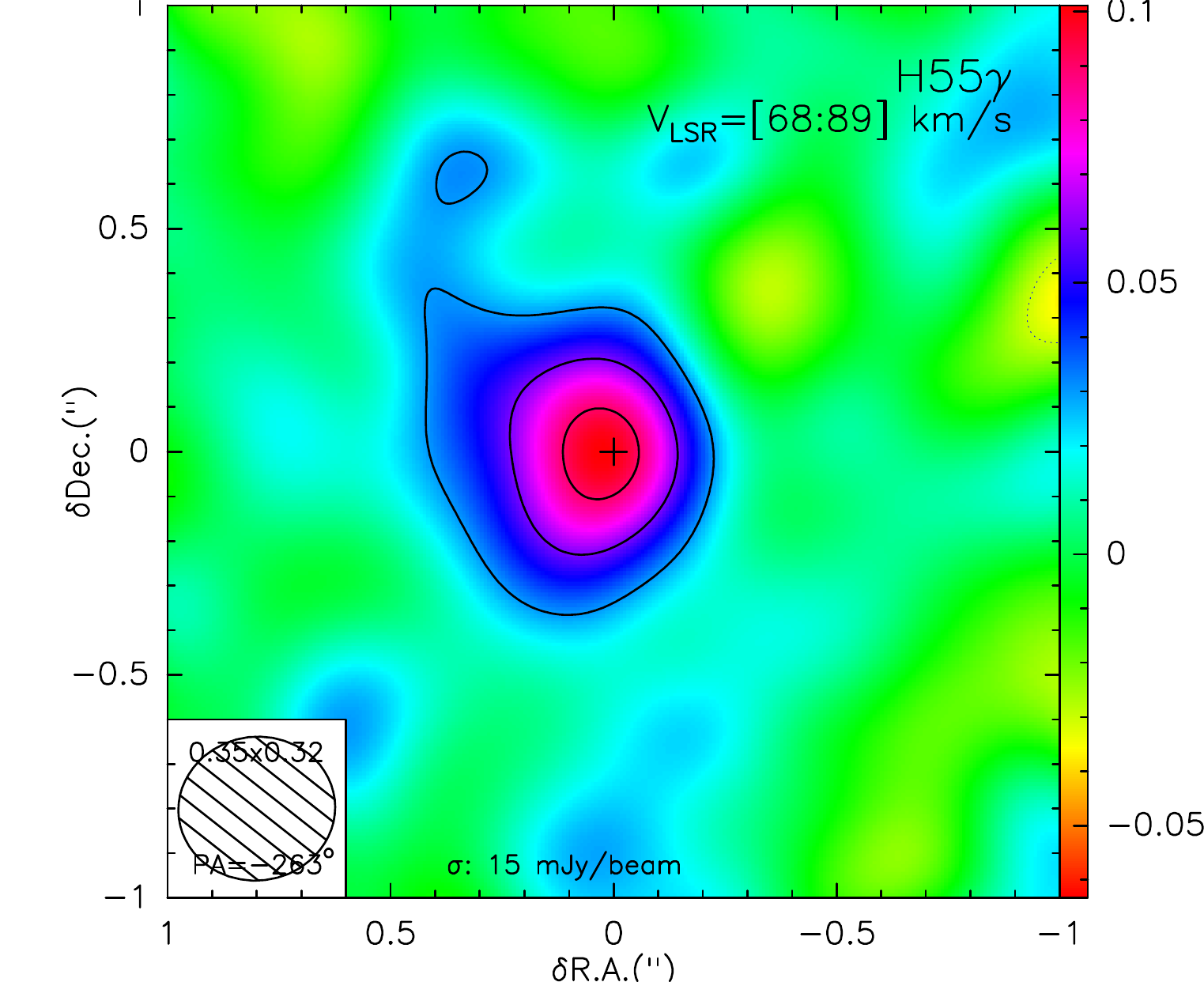}
     %area
         \caption{Summary of ALMA data (integrated 1d spectrum and integrated intensity map) for the \hcg\ line. 
           \label{f-h55g}}
         %% /pcdisk/jbell3/csanchez/m2-9/selfcal-csc/   @H55gspec.greg 
   \end{figure*}
%% ========================== Fig. H55g 
%% ========================== Fig. 13CO
   \begin{figure}[htbp!]
     \centering
     \includegraphics[width=0.85\hsize]{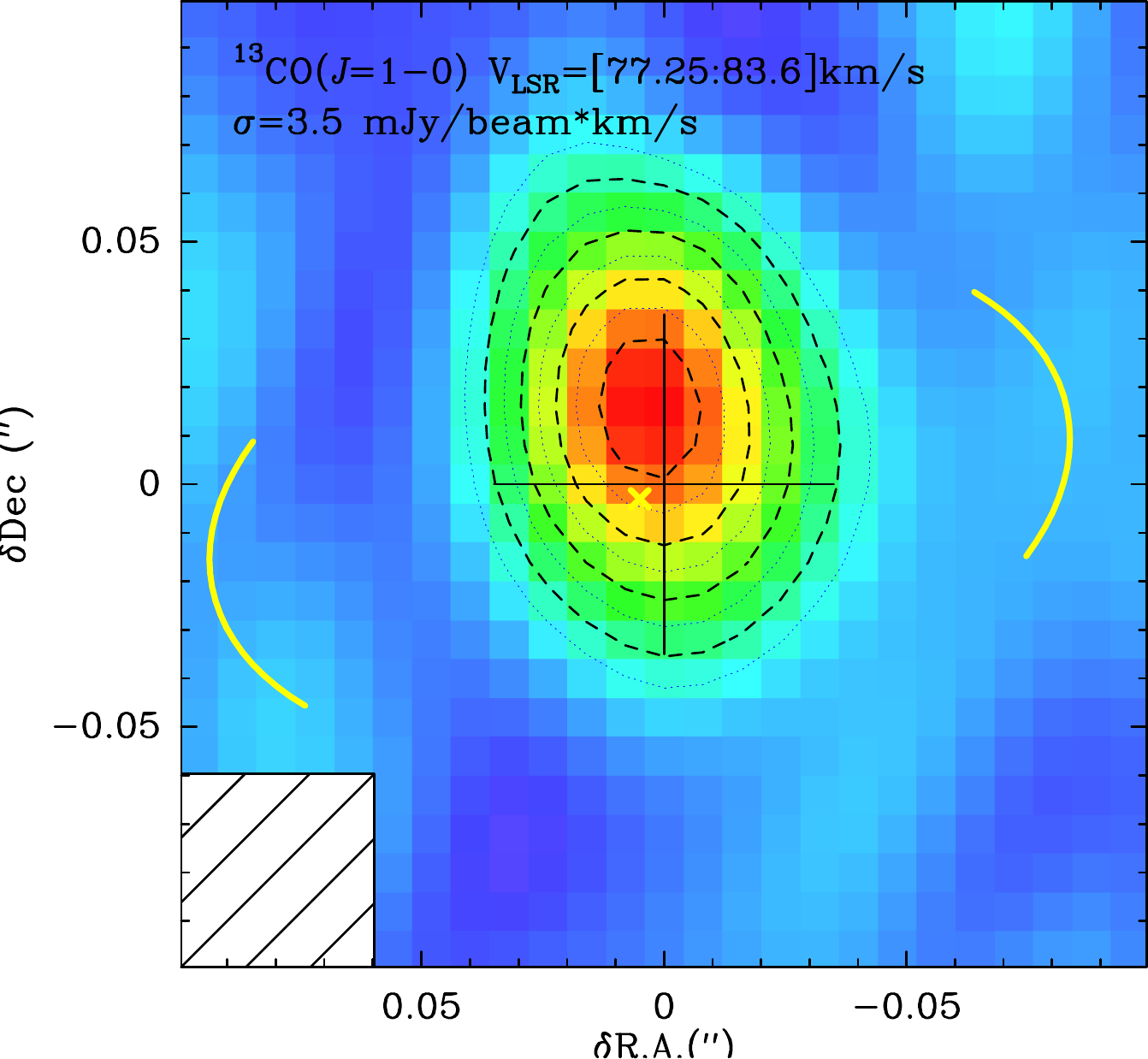}
     \includegraphics[width=0.85\hsize]{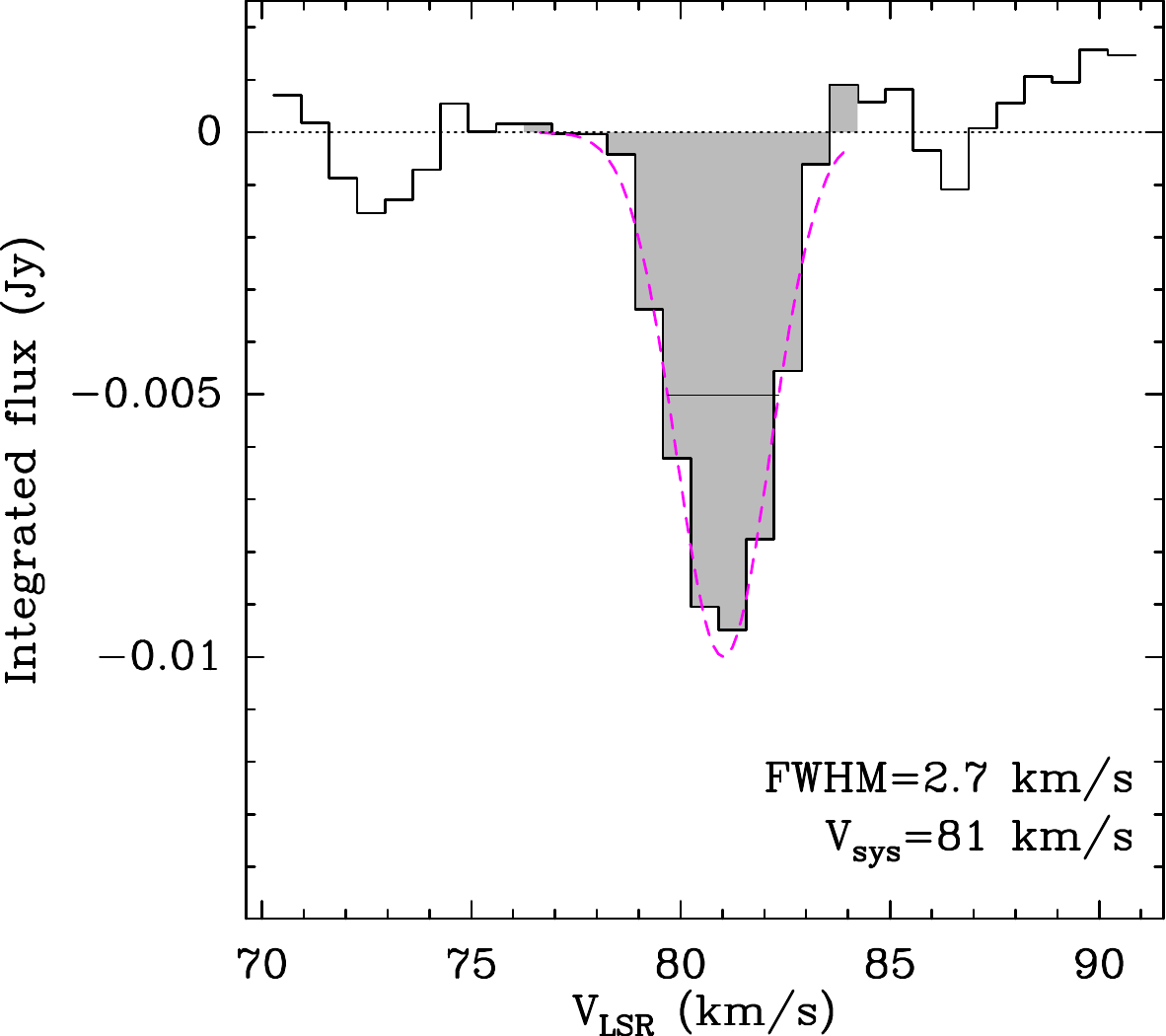}
      \caption{As in Fig.\,\ref{f-co} but for
        \trecem\,1--0. \label{f-13co}}
   \end{figure}
   %% /pcdisk/jbell3/csanchez/m2-9/selfcal-csc/13COspec.greg  -- 13COarea.greg 
%% ========================== Fig. 13CO

   %% ========================== Fig. VELO sin rotacion
   \begin{figure}[htbp!]
     \centering
     \includegraphics*[width=0.85\hsize]{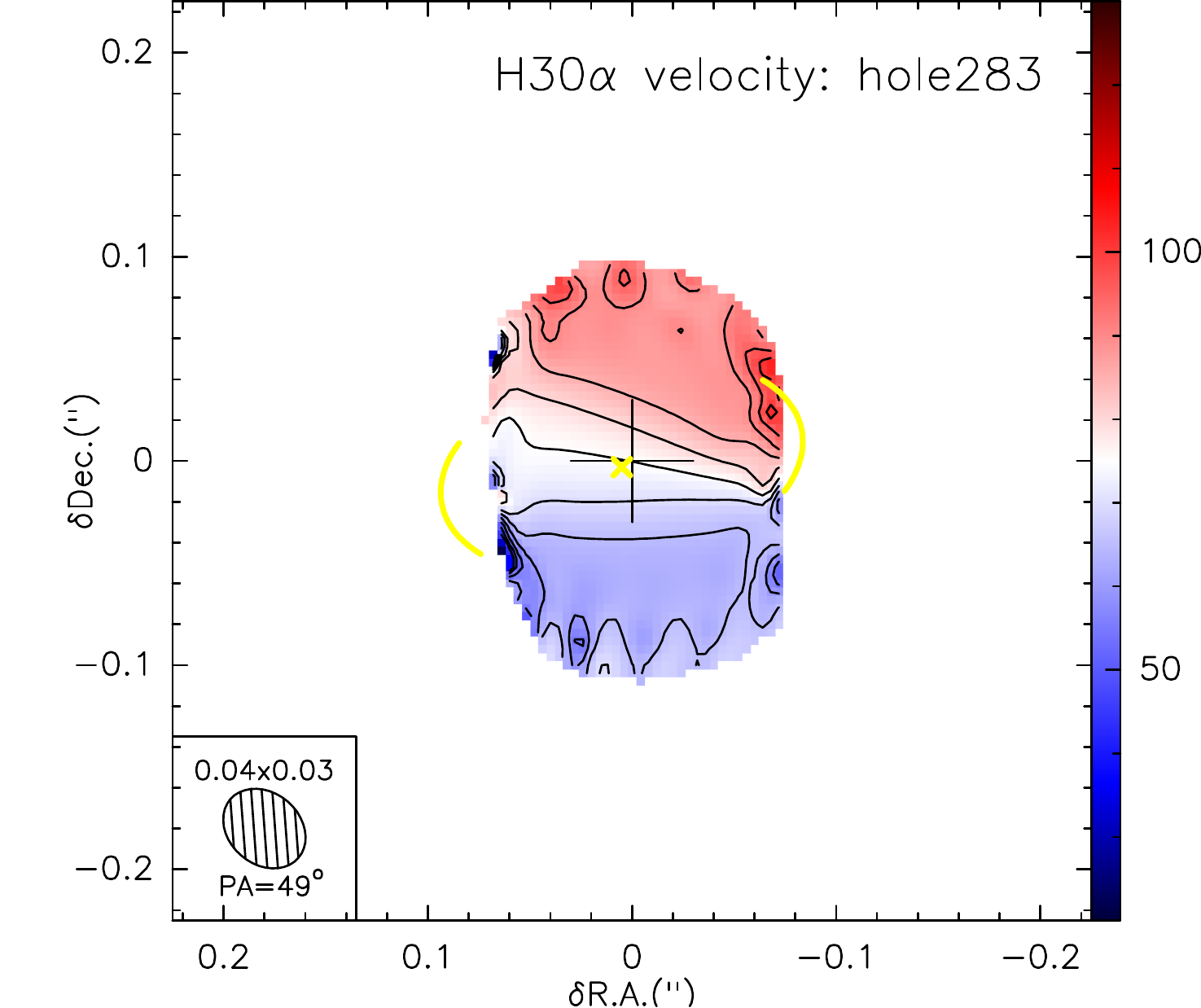}
     \includegraphics*[width=0.85\hsize]{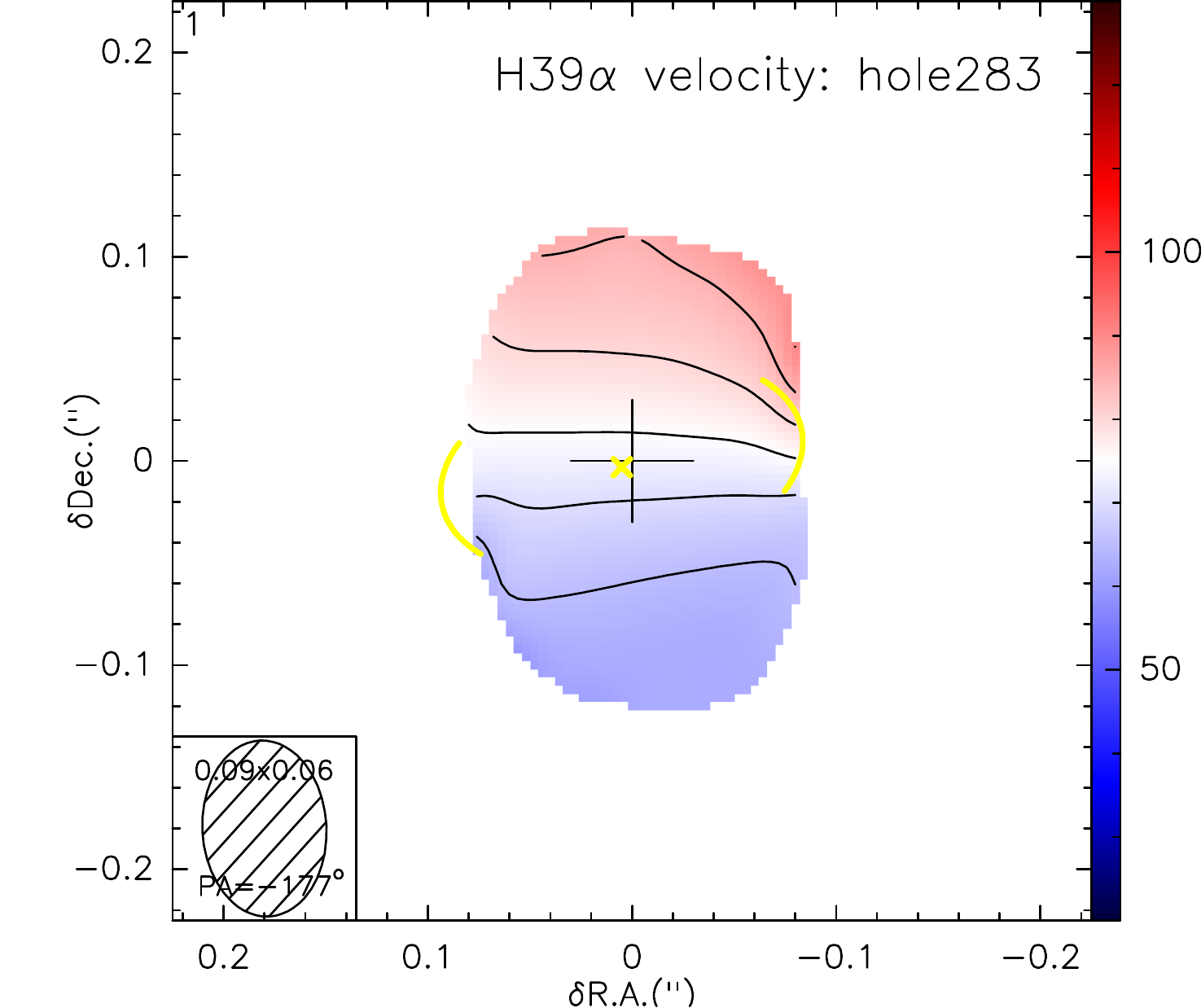}
 \caption{Synthetic first moment maps of \htal\ (top) and
   \htnal\ (bottom) for our best-fit model but without the rotation
   velocity component (as in Fig.\ref{f-h30yh39}). The inclination of
   the iso-velocity contours observed in the data is not reproduced in
   the absence of rotation, even after introducing some C-curvature in
   the wind (data-model mismatch is most notable in the south lobe and
   in the \htnal\ line).}
   \end{figure}
%%   sin ROTATION = modelo hole282 
   %% ========================== Fig. VELO sin rotacion

\FloatBarrier % Ensures all figures from Appendix A are processed before moving on

%%   \newpage 

\section{The non-ETL free-free continuum and mRRL radiative transfer code}
\label{ap-co3ral}

Here, we provide a detailed description of our code used to model the
free-free continuum and radio recombination line ALMA data presented
in this work.

\subsection{Numerical calculations of the solution of the radiative transfer equation}

As is well known, the solution of the radiative transfer equation in a
plane-parallel, homogeous, and isotropic medium can be expressed as:

\begin{equation}\label{eq:4}
I_\nu(\tau_{\nu}) = I_\nu(0){\rm e}^{-\tau_{\nu}} +
\mathcal{S}_{\nu}(1-{\rm e}^{-\tau_{\nu}}), 
\end{equation}

\noindent
involving the frequency (\( \nu \)), the optical depth (\(
\tau_{\nu}\)), the initial specific radiation at that frequency at the
back surface of the source (\( I_\nu(0) \)), the specific intensity
outgoing from the front surface of the source (\( I_\nu(\tau_{\nu}) \)),
and the source function (\( \mathcal{S}_{\nu}\)). 

%involving the frequency (\( \nu \)), the optical depth (\(
%\tau_{\nu}\)), the initial specific radiation at that frequency at the
%back surface of the source \( I_\nu(0) \), where the optical depth is
%0, the specific intensity outgoing from the front surface of the
%source \( I_\nu(\tau_{\nu}) \), where the optical depth is
%$\tau_{\nu}$, and the source function \( \mathcal{S}_{\nu}\). }

\begin{figure}
	\centering
        \includegraphics[width=0.45\textwidth]{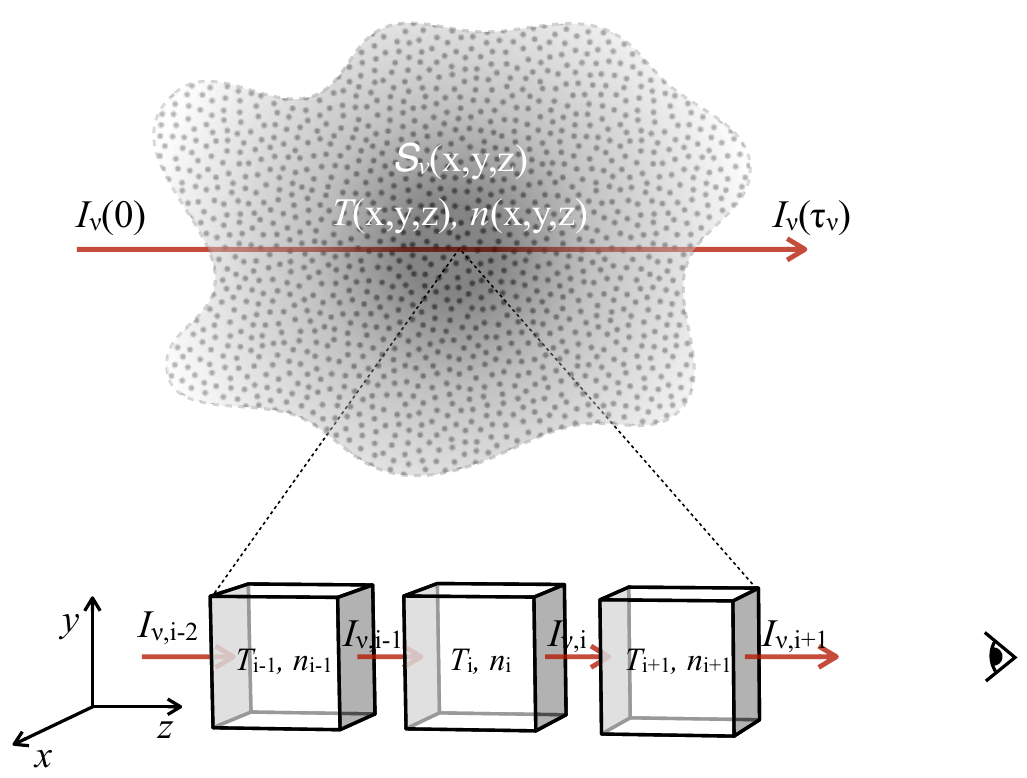}
	\caption{Division of a non homogeneous source into homogeneous
          cells. The physical conditions, density and temperature, can
          be considered constant within each cell. }
	\label{fig:2}
\end{figure}

Astronomical sources are not completely homogeneous and the physical
conditions vary from place to place. To address this, a common
approach for computing the outgoing intensity of a realistically
heterogeneous source is to partition it into small cells, each
characterised by approximately constant physical conditions (refer to
Figure~\ref{fig:2}).
%
%\noindent
Once the source is subdivided into small cells with uniform
temperature and density, Equation~(\ref{eq:4}) can be used to compute
the outgoing intensity of each cell. As depicted in
Figure~\ref{fig:2}, the outgoing intensity \( I_{\nu,i-1} \) of cell
\( i-1 \) serves as the initial intensity \( I_\nu(0) \) for cell \( i
\), and similarly, the outgoing intensity of cell \( i \) is used as
the initial intensity for cell \( i+1 \), and so forth. The intensity
of the pixel with coordinates (\( x, y \)), denoted \( I_\nu(x,y) \),
is determined through successive computations of the radiative
transfer equation solution (Equation~(\ref{eq:4})) for cells along a
single line of sight (i.e. along the \( z \)-axis). As a result, \(
I_\nu(x,y) \) represents the resulting outgoing intensity of the cell
positioned at the forefront of the source, closest to the observer.\\

\subsection{Code for Computing Continuum and Radio-recombination Lines: Co$^{3}$RaL}

%\noindent
The Code for Computing Continuum and Radio-recombination Lines
(Co$^{3}$RaL) is a C-based code designed to calculate the intensity of
the free-free continuum and recombination line emission originating
from an ionised nebula. It operates on user-defined geometrical shapes
that can be specified analytically and relies on provided density,
temperature, and velocity fields given by analytical functions. The
outputs are ASCII tables containing information needed to generate
continuum images, spectral energy distribution (SED) plots, spectral
cubes, and spectral line profiles.

\subsubsection{Definition of the geometry}

%\noindent
The grid of cells considered by \coral, within which the geometry of
the emitting region is defined, is created in a coordinate system
$\mathcal{C}$. The coordinates of each cell are given within the
intervals $[-x_{\rm max}, x_{\rm max}]$, $[-y_{\rm max}, y_{\rm
    max}]$, and $[-z_{\rm max}, z_{\rm max}]$ for the $x$, $y$, and
$z$ axes, respectively. The $x$-$y$ plane corresponds to the
plane-of-the-sky, while the $z$-axis corresponds to the
line-of-sight. Each interval is evenly divided into n$_{x}$, n$_{y}$,
and n$_{z}$ cells along the respective axis. Co$^{3}$RaL sweeps
through the cells of the grid along the $z$-axis and calculates the
outgoing intensity for each cell until it reaches the cell positioned
at the forefront of the source.

\begin{figure*}[htbp!]
	\centering
        \includegraphics*[width=0.95\textwidth]{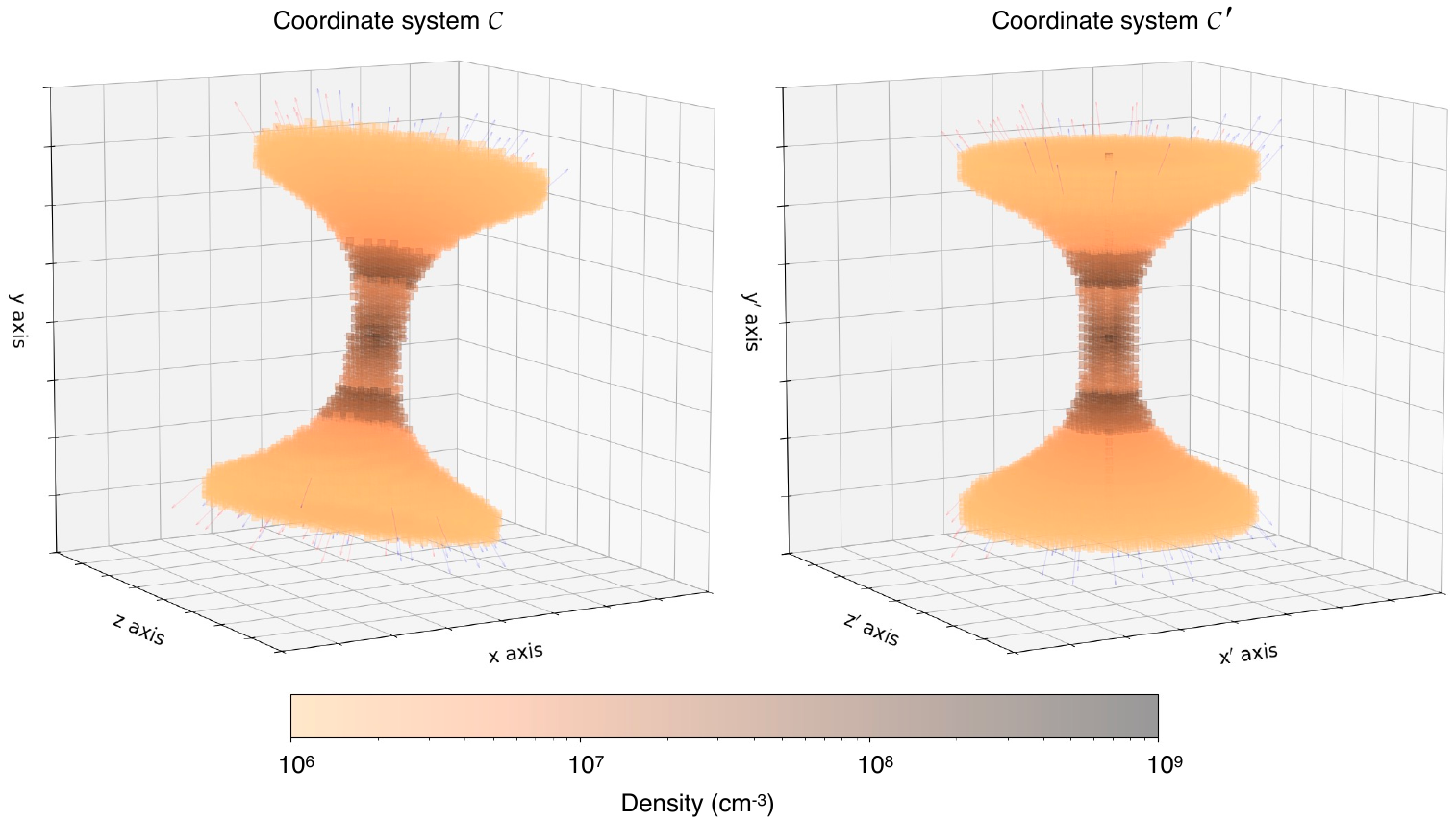}
	\caption{Geometry of the emitting H~II region as seen in the
          coordinate system $\mathcal{C}$ (left) and coordinate system
          $\mathcal{C}^{\prime}$ (right). The source can be rotated in
          the $\mathcal{C}$ system, but by definition it is not
          rotated in the $\mathcal{C}^{\prime}$ system.  }
	\label{fig:3}
\end{figure*}  

%\noindent
The boundaries of the emitting region (i.e., the cells where the
density and temperature are non-zero) are defined by an analytical
function in a coordinate system $\mathcal{C}^{\prime}$. In general,
the coordinate system $\mathcal{C}^{\prime}$ is rotated with respect
to the coordinate system $\mathcal{C}$ (see, for example,
Figure~\ref{fig:3}). Consequently, the emitting region typically
appears rotated relative to the $\mathcal{C}^{\prime}$
system. Co$^{3}$RaL sweeps through the cells of the grid along the
$z$-axis of the coordinate system $\mathcal{C}$. When it reaches a
cell with coordinates ($x$, $y$, $z$), it transforms the coordinates
from $\mathcal{C}$ to $\mathcal{C}^{\prime}$ using the following
rotation matrix:

\begin{equation}\label{eq:5-1}
\begin{bmatrix}
    x' \\ y' \\ z'
\end{bmatrix}
=
\begin{bmatrix}
    \cos(\phi) & \sin(\phi) & 0 \\ -\cos(\theta) \sin(\phi) &
    \cos(\theta) \cos(\phi) & -\sin(\theta) \\ -\sin(\theta)
    \sin(\phi) & \cos(\phi) \sin(\theta) & \cos(\theta)
\end{bmatrix}
\begin{bmatrix}
    x \\ y \\ z
\end{bmatrix},
\end{equation}

\noindent
where $\theta$ and $\phi$ are the inclination angle with respect to
the plane of the sky and the position angle of the source,
respectively. Once the coordinates of the cell are transformed into
the coordinate system $\mathcal{C}^{\prime}$, Co$^{3}$RaL checks
whether the cell is within the emitting region. If the cell is within
the emitting region, the radiative transfer is performed using
Equation~(\ref{eq:4}). If the cell is outside the boundary of the
emitting region, it is skipped, meaning it is assumed that the cell
does not affect the intensity $I_{\nu}$, and the code moves to the
next cell along the line-of-sight. The resulting intensity for each
line-of-sight, \( I_\nu(x,y) \), is output using the coordinates of
the coordinate system $\mathcal{C}$. It is worth noting that, since
the radiative transfer is performed along rays in the line-of-sight,
the result of the calculations is sensitive to the inclination of the
source but not to the position angle.\\

\subsubsection{Definition of the physical parameters}

%\noindent
The definition of the physical parameters is more straightforward in
the rotated $\mathcal{C}^{\prime}$ system, where they are expressed as
relatively simple analytical functions dependent on the primed
coordinates: $T$($x^{\prime}, y^{\prime}, z^{\prime}$),
$n$($x^{\prime}, y^{\prime}, z^{\prime}$), and $\vec{v}$($x^{\prime},
y^{\prime}, z^{\prime}$). However, as mentioned above, Co$^{3}$RaL
sweeps through the cells of the grid along the $z$-axis of the
coordinate system $\mathcal{C}$. Thus, for each cell within the
emitting region, Co$^{3}$RaL transforms the coordinates ($x$, $y$,
$z$) to ($x^{\prime}$, $y^{\prime}$, $z^{\prime}$) (using the rotation
matrix in Equation~(\ref{eq:5-1})), and then calculates the
corresponding values of the physical parameters using the analytical
functions defined in the $\mathcal{C}^{\prime}$ system. Subsequently,
the outgoing intensity of the cell is calculated iteratively using
Equation~(\ref{eq:4}) and the physical parameter values specific to
that cell.

\subsubsection{Computation of the radio continuum}
To compute the radio continuum emission, Co$^{3}$RaL uses the
following free-free emission coefficient \citep{Rybicki1986}:

\begin{equation}\label{eq:5}
j_{\nu}^{\rm cont} = 6.8 \times 10^{-38} \left( \frac{n_e n_i}{4 \pi}
\right) Z^{2} T_{e}^{1/2} {\rm e}^{( -h\nu/k T_{e})} {g}(\nu,T_{e}),
\end{equation}

\noindent
where $n_e$ and $n_i$ are the electron and ion densities,
respectively, $Z$ is the electric charge of the ions, $T_e$ is the
electron temperature, $h$ and $k$ are the Planck and Boltzmann
constants, $\nu$ is the frequency of the radiation, and
${g}(\nu,T_{e})$ is the Gaunt factor, which is the sum of the Gaunt
factors for free-free and bound-free transitions,
${g}(\nu,T_{e})=g^{ff}+g^{bf}$.\\

%\noindent
Following \cite{bae13}, the Gaunt factor for free-free transitions is
calculated in the following way:

\begin{equation}
{g}_{ff} = \sqrt{\left( \frac{\sqrt{3} {\rm e}^{x} K_{0}(x)}{\pi}
  \right)^{2} + \left( a - b \log_{10} \left( \frac{h \nu}{k T_{e}}
  \right) \right)^{2}},
\end{equation}

\noindent
where $K_{0}(x)$ is the modified Bessel function of the second kind
and

\begin{flalign*}
& \gamma^{2} = Z^{2}\frac{h \nu_0}{k T_e}, & \\ & x = \left( \frac{h
    \nu} {2 k T_{e}} \right) \left( 1 + \sqrt{10 \gamma^{2}} \right),
  & \\ & a = 1.2 \exp \left( -\left( \frac{\log_{10}(\gamma^{2}) -
    1}{3.7} \right)^{2} \right), & \\ & b = 0.37 \exp \left( -\left(
  \frac{\log_{10}(\gamma^{2}) + 1}{2} \right)^{2} \right), &
\end{flalign*}

\noindent
where $\nu_0$ is the hydrogen ionisation frequency with a numerical
value of $\nu_0$=3.291$\times$10$^{15}$~Hz.\\

%\noindent
The modified Bessel function of the second kind, $K_{0}(x)$, is
computed using the following approximation \citep{Martin2022}:

\begin{multline*}
K_{0}(x) = \frac{1}{\left(1 + 1.9776 x^2\right) \left(1 + 0.2778
  x^2\right)^{5/4} \cosh(x)}\cdot \\ \left(0.1159 - 0.0119x^2 - 0.5
\cdot \ln\left(\frac{x^2}{1 + x^2}\right) \right. \cdot \\ \left(1 +
3.0749 x^2 + 2.6248 x^4 + 0.4999 x^6\right)\biggr).
\end{multline*}

%\noindent
The Gaunt factor for bound-free transitions is calculated in the
following way \citep{Brussaard1962}:

\begin{flalign*}
& {g}_{bf} =
  2\Theta\sum_{n=m}^{\infty}g_{n}(\nu)\frac{{\rm e}^{\Theta/n^{2}}}{n^{3}},
  & \\ & \Theta=\frac{h\nu_{0}}{kT_{e}}, & \\ & m={\rm
    int}\left(\sqrt{\frac{\nu_{0}}{\nu}}\right)+1, &
\end{flalign*}

\noindent
where int($x$) is the nearest integer value (rounded off value) of
$x$. As pointed out by \cite{bae13}, the value of $g_{n}(\nu)$ is
taken as unity for all the considered frequencies, which yields
results with an accuracy of $\sim$10 -- 20\%. However, it is worth
noting that the Gaunt factors for bound-free transitions become
significant only for frequencies $\nu$$\gtrsim$10$^{4}$~GHz (see
Fig.A.1 of \cite{bae13}). Thus, this factor is negligible for RRLs.\\

%\noindent
Subsequently, the absorption coefficient, $\kappa_{\nu}$, is obtained
using Kirchhoff's law as

\begin{equation}\label{eq:6}
\kappa_{\nu}^{\rm cont} = j_{\nu}^{\rm cont}/B_{\nu}(T),
\end{equation}

\noindent
where $B_{\nu}(T)$ is the Planck function given by

\begin{equation}\label{eq:7}
B_{\nu}(T)=\frac{2 h \nu^3}{c^2} \cdot \frac{1}{{\rm e}^{h \nu/k T} - 1}.
\end{equation}

%\noindent
The optical depth, $\tau_{\nu}^{\rm cont}$ that is used in
Equation~(\ref{eq:4}) to compute the intensity of the continuum
emission is thus computed as

\begin{equation}\label{eq:7}
\tau_{\nu}^{\rm cont} = \kappa_{\nu}^{\rm cont} dl,
\end{equation}

\noindent
where $dl$ is the physical size of the cell, given in CGS units.

\subsubsection{Computation of the radio recombination lines in LTE conditions}

Co$^{3}$RaL computes the emission of RRLs in LTE and non-LTE
conditions. The emission of the lines in LTE conditions is computed
using the following absorption emission coefficient
\citep{Wilson2009}:

\begin{equation}\label{eq:10}
\kappa_{\nu}^{\text{line,*}} = \frac{1}{8 \pi} \left(
\frac{c\,\mathsf{n_{u}}}{\nu_{ul}} \right)^{2} \left( \frac{h^2}{2 \pi
  m_e k T_e} \right)^{3/2} {\rm e}^{h \nu_0/n_l^2 k T_e} A_{ul} \left(1 -
{\rm e}^{-h \nu_{ul}/k T_e} \right) n_e n_i V_{\nu},
\end{equation}

\noindent
where $\nu_{ul}$ is the central rest frequency of the line,
$\mathsf{n_{u}}$ is the upper level electronic quantum number, and
$V_{\nu}$ is the Voigt profile of the line, obtained as a convolution
of Gaussian and Lorentzian profiles. Using the notation of
\cite{Kielkopf1973}, the expression of the convolution is given as
follows:

\begin{equation}
V(\beta_{l},\beta_{g}|\nu)=\int_{-\infty}^{+\infty}
\frac{\beta_{l}/\pi}{\beta_{l}^{2}+(\nu-\nu^{\prime})}\frac{1}{\sqrt{\pi}\beta_{g}}{\rm e}^{-(\nu^{\prime}/\beta_{g})^{2}}d\nu^{\prime},
\end{equation}

\noindent
where $\beta_{l}$ and $\beta_{g}$ are the Lorentzian and Gaussian
widths, defined as the half width at half maximum. Given that the
calculation of this convolution needs to be done for every cell, it
turns out to be computationally expensive. Therefore, Co$^{3}$RaL uses
the approximation to the convolution given by \cite{Kielkopf1973}:

\begin{equation}
V(\beta_{l},\beta_{g}|\beta\,x)=I(\beta_{l},\beta_{g})\,U(x),
\end{equation}

\noindent
where $I(\beta_{l},\beta_{g})$ is given as

\begin{equation}\label{eq:13}
I(\beta_{l},\beta_{g})=\frac{a}{\sqrt{\pi}\beta_{l}}f(a),
\end{equation}

\noindent
and the expression for $U(x)$ is the following:

\begin{equation}\label{eq:14}
U(x)=(1-\eta)G(x)+\eta L(x)+\eta (1-\eta)E(x)[G(x)-L(x)].
\end{equation}

%\noindent
The factor $a$ in Equation~(\ref{eq:13}) is defined as the ratio of
the Lorentzian and Gaussian widths, $a=\beta_{l}/\beta_{g}$, and the
form of the function $f(a)$ depends on the value of $a$ as follows:

\begin{equation}
f(a) =
\begin{cases} 
\frac{1}{\sqrt{\pi}} \left(\frac{1}{a + \frac{1/2}{a + \frac{1}{a +
      \frac{3/2}{a + \frac{2}{a + \dots}}}}}\right) & \text{if } a >
1.5, \\ \frac{1}{\sqrt{\pi}}(b_{1}t+b_{2}t^{2}+b_{3}t^{3}) & \text{if
} a < 1.5,
\end{cases}
\end{equation}

\noindent
where

\begin{equation}
t=\frac{1}{1+b_{0}a},
\end{equation}
\noindent
and $b_{0}=0.47047$, $b_{1}=0.61686$, $b_{2}=-0.16994$, and
$b_{3}=1.32554$.\\

%\noindent
The functions $G(x)$, $L(x)$, and $E(x)$ in Equation~(\ref{eq:14}) are
defined as follows:

\begin{align}
& G(x)={\rm e}^{-\text{ln}(2)x^{2}}, \\ & L(x)=\frac{1}{1+x^{2}}, \\ &
  E(x)=\frac{0.8029-0.4207x^{2}}{1+0.2030x^{2}+0.07335x^{4}}.
\end{align}

%\noindent
The factor $\eta$ in Equation~(\ref{eq:14}) is defined as follows:

\begin{equation}
\eta=\frac{l}{l+g^{2}},
\end{equation}

\noindent
where

\begin{align}
& l=\frac{\beta_{l}}{\beta}, \\ & \beta=\frac{1}{2}\beta_{l} \left\{1
  + \epsilon\,\text{ln}(2) + \left[\left(1 - \epsilon
    \,\text{ln}(2)\right)^{2} +
    \left(4.0\,\text{ln}(2)/a^{2}\right)\right]^{1/2}\right\}, \\ &
  g=\left(\frac{1-l}{\text{ln}(2)}\right)^{1/2},
\end{align}

\noindent
and $\epsilon$=0.0990.\\

%\noindent
The Gaussian width of the line, $\beta_{g}$, is calculated using the
following expression \citep{Mezger1967}:

\begin{equation}
\beta_g = \sqrt{\frac{2 k T_e}{m_H} + \frac{2}{3} v_{\text{tur}}^2},
\end{equation}

\noindent
where $m_H$ is the mass of the hydrogen atom and $v_{\text{tur}}$ is
the turbulent velocity due to turbulent motions of the plasma. The expression of the Lorentzian width, $\beta_{l}$, is taken from Table~B.1 of \cite{bae13}.

%\noindent
Finally, the optical depth of the line, $\tau_{\nu}^{\rm line*}$, in LTE conditions is given as

\begin{equation}
\tau_{\nu}^{\rm line*} = \kappa_{\nu}^{\text{line,*}} dl, 
\end{equation}

\noindent
and the total optical depth, $\tau_{\nu}^{*}$, that is used in Equation~(\ref{eq:4}) to compute the intensity of the line plus continuum is the sum of both optical depths: 

\begin{equation}
\tau_{\nu}^{*} = \tau_{\nu}^{\rm line*} + \tau_{\nu}^{\rm cont}.
\end{equation}

%\noindent
After performing the radiative transfer for the line plus continuum, the emission from the continuum is subtracted to obtain the emission from the line only.

\subsubsection{Computation of the spontaneous emission Einstein coefficients}

%\noindent
The relationship between the Einstein coefficient for spontaneous emission, $A_{ul}$, and the thermalised absorption oscillator strength $f(\mathsf{n_{l}},\mathsf{n_{u}})$, with $\mathsf{n_{u}}>\mathsf{n_{l}}$, is given by \citep{Goldwire1968}:

\begin{equation}\label{Eq:1.3}
A_{ul} = 2\pi\alpha^{3}Z^{4}\, c \, \text{R$_{M}$} \frac{{(\mathsf{n_{u}}+\mathsf{n_{l}})^2 \cdot (\mathsf{n_{u}}-\mathsf{n_{l}})^2}}{{\mathsf{n_{l}}^2 \mathsf{n_{u}}^6}} f(\mathsf{n_{l}},\mathsf{n_{u}}),
\end{equation}

\noindent
where $\alpha$ is the fine structure constant, $Z$ is the atomic number, $c$ is the speed of light, and $\text{R$_{M}$}$ is the Rydberg constant defined as

\begin{equation}\label{Eq:1.4}
\text{R${_M}$} = \frac{\text{R${_\infty}$}}{1 + (m_{e}/M)},
\end{equation}

\noindent
with $m_{\rm e}$ being the mass of the electron and $M$ denoting the total mass of the nucleus. Furthermore, $\text{R$_{\infty}$}$ is given by

\begin{equation}\label{Eq:1.5}
R_\infty = \frac{m_{\rm e} e^4}{8 \varepsilon_{0}^{2}} h^3 c,
\end{equation}

\noindent
where $e$ represents the elementary charge, $\varepsilon_{0}$ is the permittivity of free space, and $h$ stands for the Planck constant. The numerical value of $\text{R$_{\infty}$}$ is 109737.315685 cm$^{-1}$.\\

%\noindent
The most straightforward way to calculate the oscillator strengths is using the following expression given by \cite{Kardashev1959}:  

\begin{equation}\label{Eq:1.1}
\begin{aligned}
    f(\mathsf{n_{l}},\mathsf{n_{u}})=&\frac{2}{3}\mathsf{n_{u}}^{2}\frac{(4\mathsf{n_{u}}\mathsf{n_{l}})^{2\mathsf{n_{l}}+2}(\mathsf{n_{u}}-\mathsf{n_{l}})^{2\mathsf{n_{u}}-2\mathsf{n_{l}}-4}}{(\mathsf{n_{u}}+\mathsf{n_{l}})^{2\mathsf{n_{u}}+2\mathsf{n_{l}}+3}}\\
    &\times\binom{\mathsf{n_{u}}-1}{\mathsf{n_{u}}-\mathsf{n_{l}}-1}^{2}F^{2}(-\mathsf{n_{l}},-\mathsf{n_{l}},\mathsf{n_{u}}-\mathsf{n_{l}};x^{-1})\\
    &-\binom{\mathsf{n_{u}}}{\mathsf{n_{u}}-\mathsf{n_{l}}+1}^{2}x^{-2}F^{2}(-\mathsf{n_{l}}+1,-\mathsf{n_{l}}+1,\mathsf{n_{u}}-\mathsf{n_{l}}+2;x^{-1}),
\end{aligned}
\end{equation}

\noindent
where the parenthesis $\binom{a}{b}$ is the binomial coefficient, $F(a,b,c;z)$ is a hyper-geometric function and $x$ is defined as

\begin{equation}\label{Eq:1.2}
    x = -\frac{4 \mathsf{n_{u}} \mathsf{n_{l}}}{(\mathsf{n_{u}} - \mathsf{n_{l}})^{2}}.\\
\end{equation}

%\noindent
Thus, by substituting Equation~(\ref{Eq:1.1}) into
Equation~(\ref{Eq:1.3}), the Einstein coefficient for spontaneous
emission can be calculated. This approach requires evaluating the
hyper-geometric function in Equation~(\ref{Eq:1.1}), which is not
included in Co$^{3}$RaL. Therefore, the Einstein coefficients for
spontaneous emission are calculated separately, and Co$^{3}$RaL reads
the values from a table.

\subsubsection{Computation of the radio recombination lines in non-LTE conditions}

Non-LTE conditions occur when the populations of the energy levels
deviate from the Boltzmann distribution. The computation of
RRL emission can be done by following the procedure introduced by
\cite{Menzel1937}, in which departure coefficients \( b_n \) relate
the true population of a level, \( N_n \), to the population under LTE
conditions, \( N_n^{\text{*}} \), by the relationship \( N_n = b_n
N_n^{\text{*}} \). Under this consideration, the Boltzmann equation
becomes:

\begin{equation}\label{eq:25}
\frac{N_{u}}{N_{l}}=\frac{b_{u}}{b_{l}}\frac{g_{u}}{g_{l}}{\rm e}^{-h\nu_{ul}/kT},
\end{equation}

\noindent
where $g_{i}$ (with $i=u,l$) are the statistical weights of the electronic levels, calculated as $g_{i}=\mathsf{n_{i}}^{2}$, and $b_{i}$ (with $i=u,l$) are the non-LTE departure coefficients mentioned above. The value of the $b_{i}$ factors is always $<$1, since the $A_{ul}$ coefficient for the lower state is larger and the atom is smaller so collisions are less effective,  $b_{i}$$\rightarrow$1 for LTE conditions \cite{Wilson2009}. Using the expression of Equation~(\ref{eq:25}) in the expression for the absorption emission coefficient (Equation~(\ref{eq:10})), the absorption emission coefficient under non-LTE conditions can be written as

\begin{equation}
\kappa_{\nu}^{\text{line}} = \frac{1}{8 \pi} \left( \frac{c\,\mathsf{n_{u}}}{\nu_{ul}} \right)^{2} \left( \frac{h^2}{2 \pi m_e k T_e} \right)^{3/2} {\rm e}^{h \nu_0/n_l^2 k T_e} b_{l} A_{ul} \left(1 - \frac{b_{u}}{b_{l}}{\rm e}^{-h \nu_{ul}/k T_e} \right) n_e n_i  V_{\nu}.
\end{equation}

%\noindent
Thus, the relation between the absorption coefficients under LTE conditions and non-LTE conditions can be written in the following way: 

\begin{equation}\label{eq:28}
\kappa_{\nu}^{\text{line}} = b_{l} \beta_{ul} \kappa_{\nu}^{\text{line,*}} 
\end{equation}

\noindent
where the coefficient $\beta_{ul}$ is defined as

\begin{equation}
\beta_{ul}=\frac{1 - (b_{u}/b_{l}){\rm e}^{-h \nu_{ul}/k T_e}}{1 - {\rm e}^{-h \nu_{ul}/k T_e}}.
\end{equation}

%\noindent
In general, the relationship between the emission and absorption coefficients for the line can be written as follows \citep{Wilson2009}:

\begin{equation}
\frac{j_{\nu}^{\text{line}}}{\kappa_{\nu}^{\text{line}}}=\frac{2 h \nu^3}{c^2}\left(\frac{1}{\frac{g_{u}}{g_{l}}\frac{N_{l}}{N_{u}} - 1}\right),
\end{equation}

\noindent
and using Equation~(\ref{eq:25}) we arrive to the following expression:

\begin{equation}
\frac{j_{\nu}^{\text{line}}}{\kappa_{\nu}^{\text{line}}}=\frac{2 h \nu^3}{c^2}\cdot\frac{1}{(b_{l}/b_{u}){\rm e}^{h \nu_{ul}/k T_e} - 1}.
\end{equation}

%\noindent
Substituting Equation~(\ref{eq:28}) into this expression and using Kirchhoff’s law, \(j_{\nu}^{\text{line,*}}\) \(/\kappa_{\nu}^{\text{line,*}} = B_{\nu}(T)\), it is possible to write the relation between the emission coefficients under LTE conditions and non-LTE conditions in the following way:

\begin{equation}\label{eq:32}
j_{\nu}^{\text{line}} = b_{u} j_{\nu}^{\text{line,*}}.
\end{equation}

%\noindent
The expression for the source function under non-LTE conditions is given by the following expression:

\begin{equation}
\mathcal{S}_{\nu} = \frac{j_{\nu}}{\kappa_{\nu}} = \frac{j_{\nu}^{\text{line}}+j_{\nu}^{\rm cont}}{\kappa_{\nu}^{\text{line}}+\kappa_{\nu}^{\text{cont}}}= \frac{b_{u} j_{\nu}^{\text{line,*}}+j_{\nu}^{\rm cont}}{b_{l} \beta_{ul} \kappa_{\nu}^{\text{line,*}}+\kappa_{\nu}^{\text{cont}}}, 
\end{equation}

\noindent
which can be written as

\begin{equation}
\mathcal{S}_{\nu} = \eta_{\nu} B_{\nu}(T), 
\end{equation}

\noindent
with 

\begin{equation}
\eta_{\nu} = \frac{b_{u} \kappa_{\nu}^{\text{line,*}}+\kappa_{\nu}^{\text{cont}}}{b_{l} \beta_{ul} \kappa_{\nu}^{\text{line,*}}+\kappa_{\nu}^{\text{cont}}}.
\end{equation}

%\noindent
Thus, the radiative transfer equation (Equation~(\ref{eq:4})) for non-LTE conditions becomes

\begin{equation}\label{eq:36}
I_\nu(\tau_{\nu}) = I_\nu(0){\rm e}^{-\tau_{\nu}} + \eta_{\nu} B_{\nu}(T)(1-{\rm e}^{-\tau_{\nu}}).\\
\end{equation}

\begin{figure}
	\centering
		\includegraphics[width=0.50\textwidth]{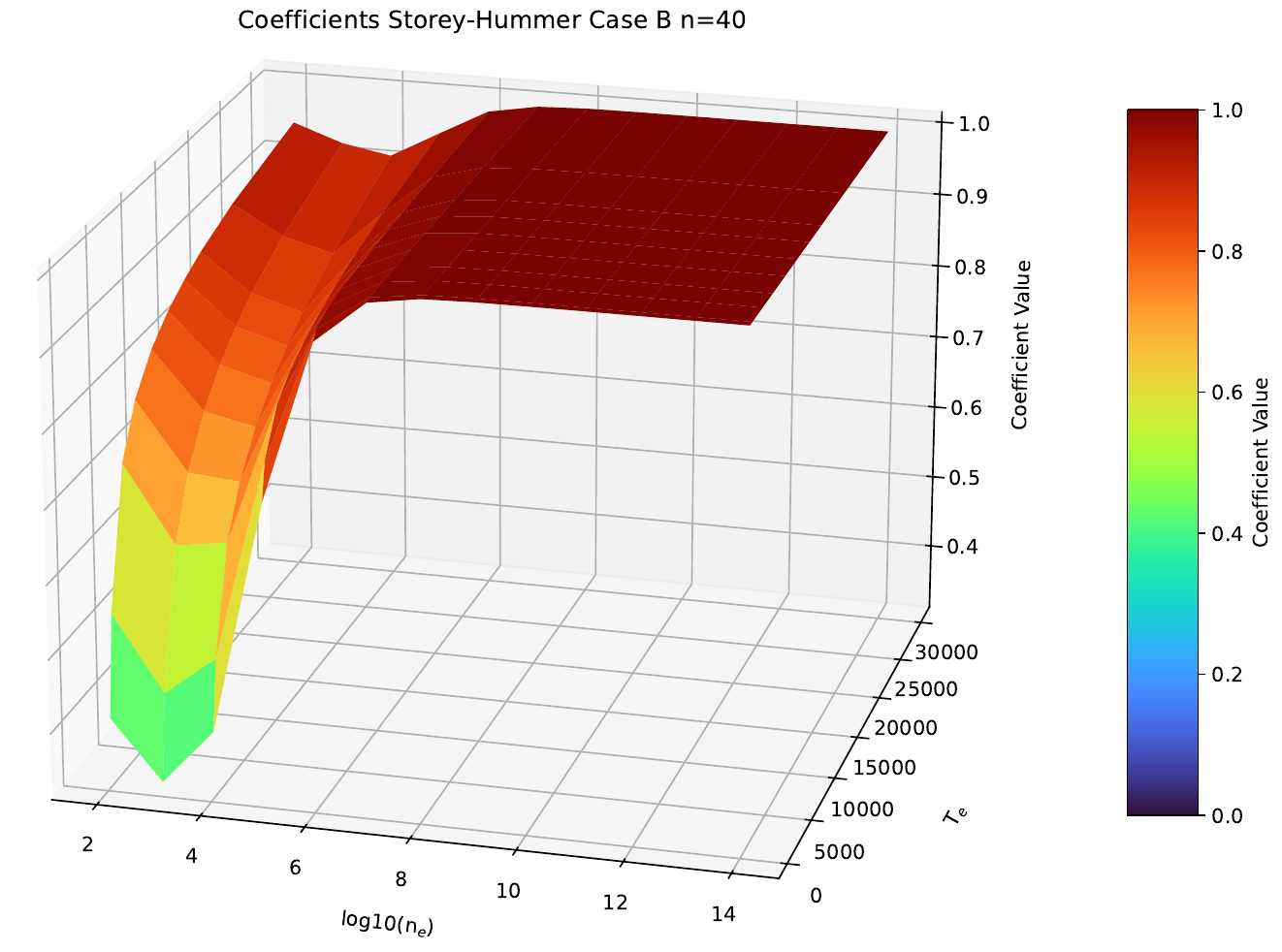}
	\caption{\cite{sto95} non-LTE departure coefficients for the Case B and $\mathsf{n_{u}}$=40.} 
	\label{fig:4}
\end{figure}

%\noindent
The non-LTE departure coefficients, $b_{i}$, are obtained from the tables provided by \cite{sto95} for Case B. These tables include values of non-LTE departure coefficients for a range of temperatures from 500 to 30,000~K and densities from 10$^{2}$ to 10$^{14}$~cm$^{-3}$. Figure~\ref{fig:4} shows the values for $\mathsf{n_{u}}=40$. As can be seen from Figure~\ref{fig:4}, the values are close to 1 for high electron densities and temperatures, and they have lower values for low densities and temperatures.\\

\begin{figure}
	\centering
		\includegraphics[width=0.5\textwidth]{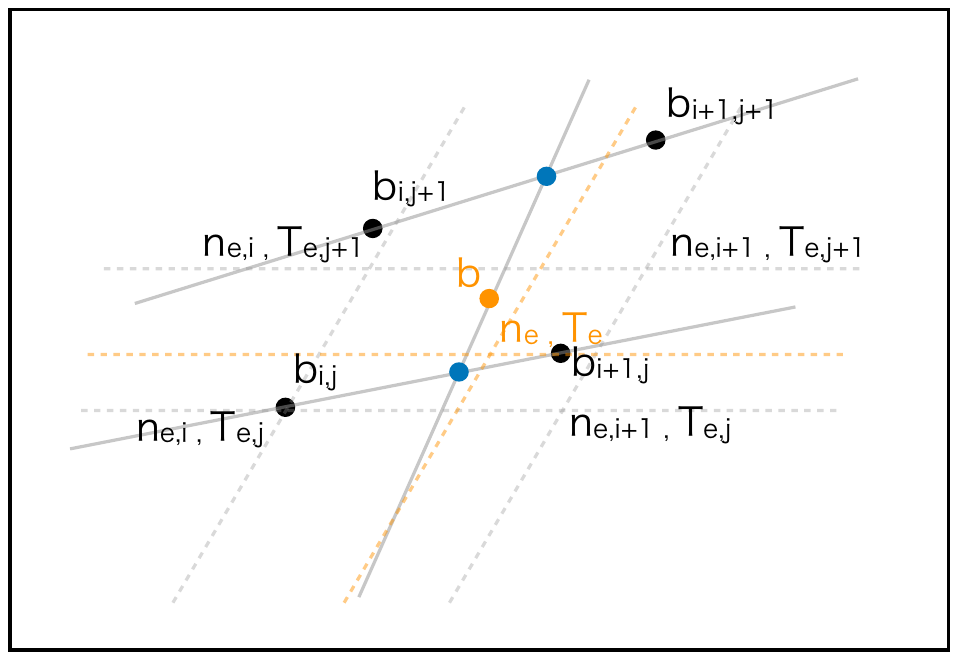}
	\caption{Diagram depicting the procedure to find the value of the non-LTE departure coefficient for any arbitrary electron density and temperature. The gray dashed lines indicate the coordinates of the discrete values of density and temperature from the \cite{sto95} tables. The orange dashed lines indicate the coordinates of an arbitrary value of electron density and temperature. The discrete values of density and temperature from the \cite{sto95} tables are shown as black dots. The solid gray lines indicate interpolation between discrete coefficient values of different densities but the same temperature. The blue dots represent the points connecting the two linear interpolations at an arbitrary electron density. Finally, the orange dot corresponds to the value that Co$^{3}$RaL uses for an arbitrary electron density and temperature.} 
	\label{fig:5}
\end{figure}

%\noindent
To obtain the value of the non-LTE departure coefficients for any value of density and temperature, Co$^{3}$RaL performs a two-dimensional linear interpolation. First, it interpolates the values between densities at the same temperature, and then it interpolates the values between two temperatures corresponding to the electron density in question. Figure~\ref{fig:5} shows a diagram of the procedure to find the value of the non-LTE departure coefficient for any arbitrary electron density and temperature.\\

%\noindent
Various interpolation methods were tested to calculate the non-LTE departure coefficients for different values of density and temperature. However, these methods did not result in significant differences in the outcomes. Therefore, the previously described method was chosen for its simplicity and efficiency.\\

%%% ----------------------------------------------------- %%%%% 
\subsection{\coral\ general description and modeling approach}
\label{ap:approach}

The \coral\ code can consider any given 3D geometry for the ionized
region. The geometry can be analytically described in Cartesian,
spherical or cylindrical coordinates or, alternatively, it can be read
from an input file where the coordinates of the model cells are
provided. In addition to the geometry, the other major inputs for the
model are: the electron density (\dense), the electron temperature
(\te), and the velocity field ($\overrightarrow{v}$).  For simplicity,
\dense\ and \te\ are assumed to follow a power law of the type
$\propto$$r^{\beta}$, where $r$ is the radial distance to the central
source and $\beta$ is a real number positive or negative. Non-uniform
distributions are accommodated by employing different laws, as
necessary, across distinct layers within the ionized wind.
%In other words, the geometry can be divided into various layers, each
%characterized by a different density/temperature law.
The same approach can be applied to describe the velocity field, also
allowing for various combinations of radial, axial, equatorial, and
shear expansion, with equatorial rotation.  The velocity modulus can
be constant or increase/decrease as any given function of $r$.
Latitude- and longitude-dependent variations can be incorporated for
any of the input physical parameters.
%% DECIR QUE NO INCLUYE TURBULENCIA?? es despreciable frente a otros mechanismos de ensanchamiento. 
%Turburlence is not included as an extra
%contributor to line broadening since it is  expected to be negligible compared with  thermal and pressure broadening. 
In addition to the gas macroscopic motions, thermal broadening as well
as electron impact (pressure) broadening are included in the line
profile function.  In the present model of M\,2-9 (Sect.\,\ref{model}), turbulence velocity
dispersion is not considered as an additional form of line broadening
since it is expected to be very small ($\sigma_{\rm
  turb}$$<$2-3\,\kms) in the envelopes of AGB and post-AGB objects
\citep[e.g.][]{sch04,buj05,dec10} and, thus, negligible compared to
other broadening terms.

%%% ----------------------------------------------------- %%%%% 
\section{Comparison of \coral\ with MORELI}
\label{comp}

In the pilot single-dish study of the emerging ionized regions of pPNe by
CSC17, the free-free continuum and mRRLs emission was modelled
using the radiative transfer code MORELI \citep{bae13}. MORELI is a
code previously employed in several studies of ionized regions, including works by
\cite{bae14,san19,mar23}. Here, we present a comparative analysis
between the predictions generated by our code, \coral, and those
produced by MORELI, utilizing identical input models.  This comparison
serves as a mutual benchmarking exercise of both codes.  We separately run
LTE and non-LTE models for M\,2-9 and CRL\,618, respectively, pPNe
that are both modelled by CSC17 using MORELI.

The two input models employed correspond to the configurations
described for M\,2-9 and CRL\,618 in Table\,4 of CSC17,
depicting in both cases a cylindrical outflow radially expanding, with the detailed geometry,
physical structure, and kinematics of the cylindrical outflow being different for, and adapted to, each source.
We refer to these preliminary models based on single-dish data as CSC17's models.
As in CSC17, \coral\ was run  under the LTE approximation for M\,2-9 and non-LTE for CRL\,618. 

\begin{figure*}[htbp!]
     \centering
%m2-9: hole3006 is the final ONE with the best CoRal chupi version debugged, etc...
     \includegraphics*[width=0.395\hsize]{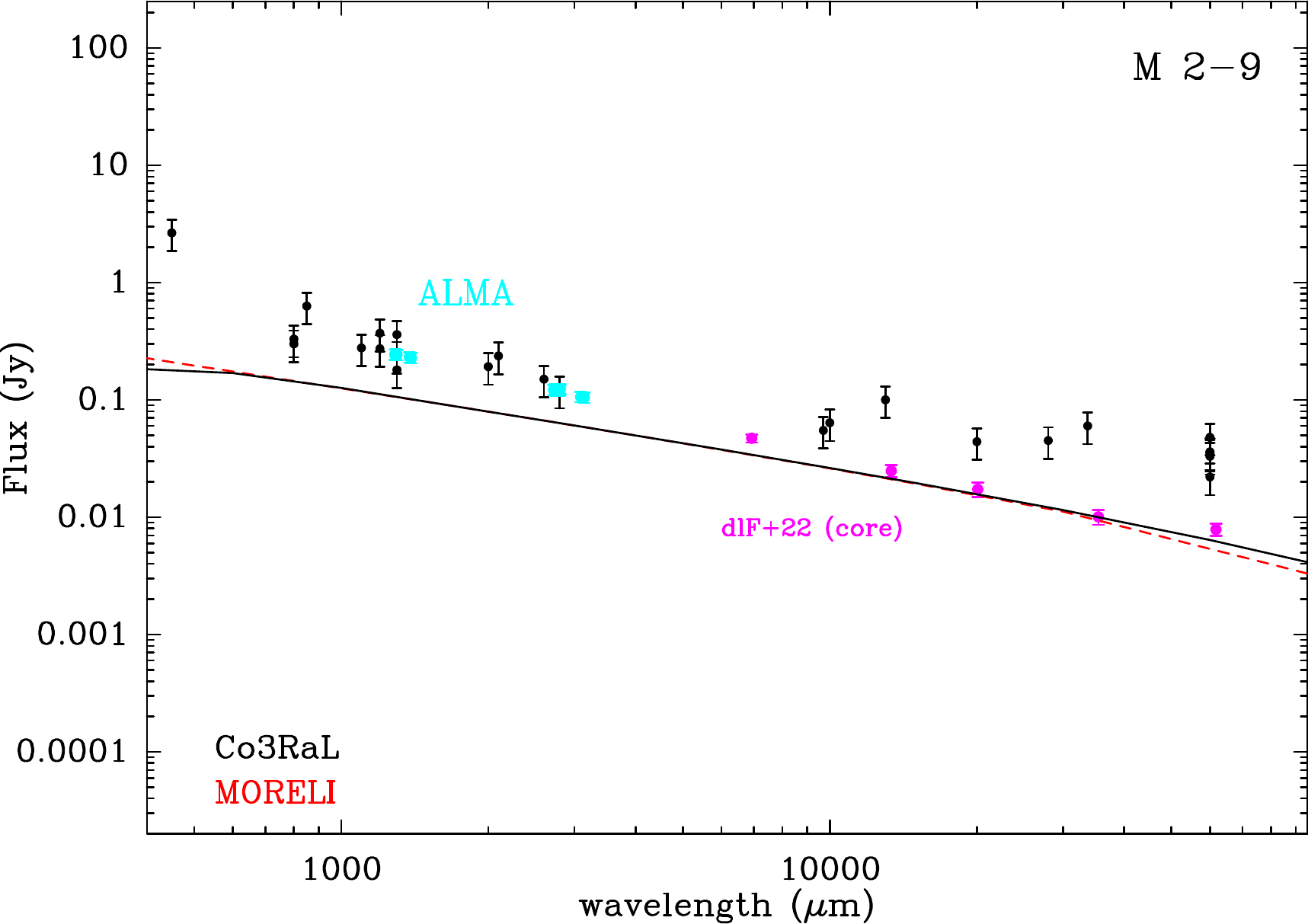} 
     \includegraphics*[width=0.2925\hsize]{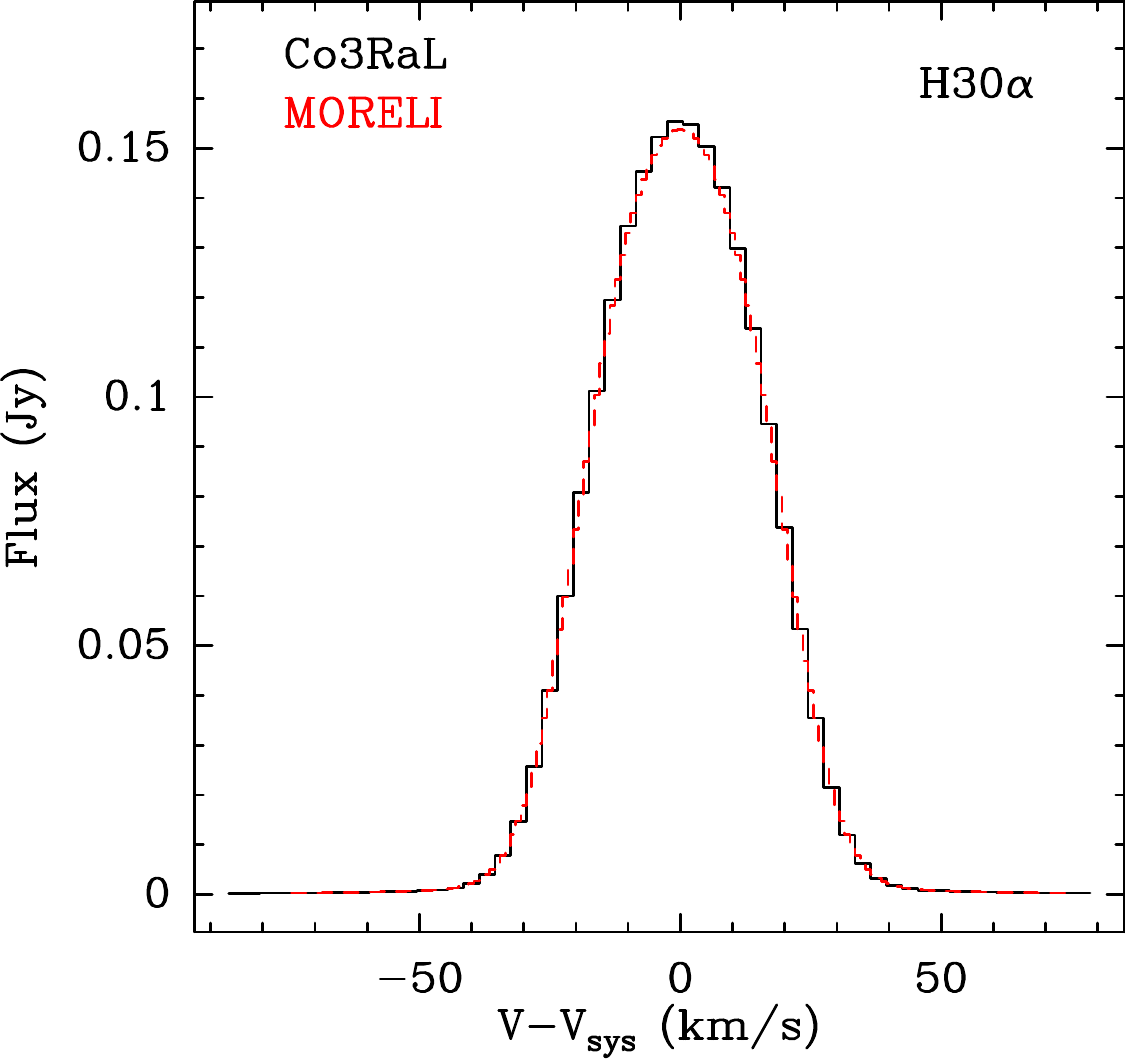}
     \includegraphics*[width=0.2925\hsize]{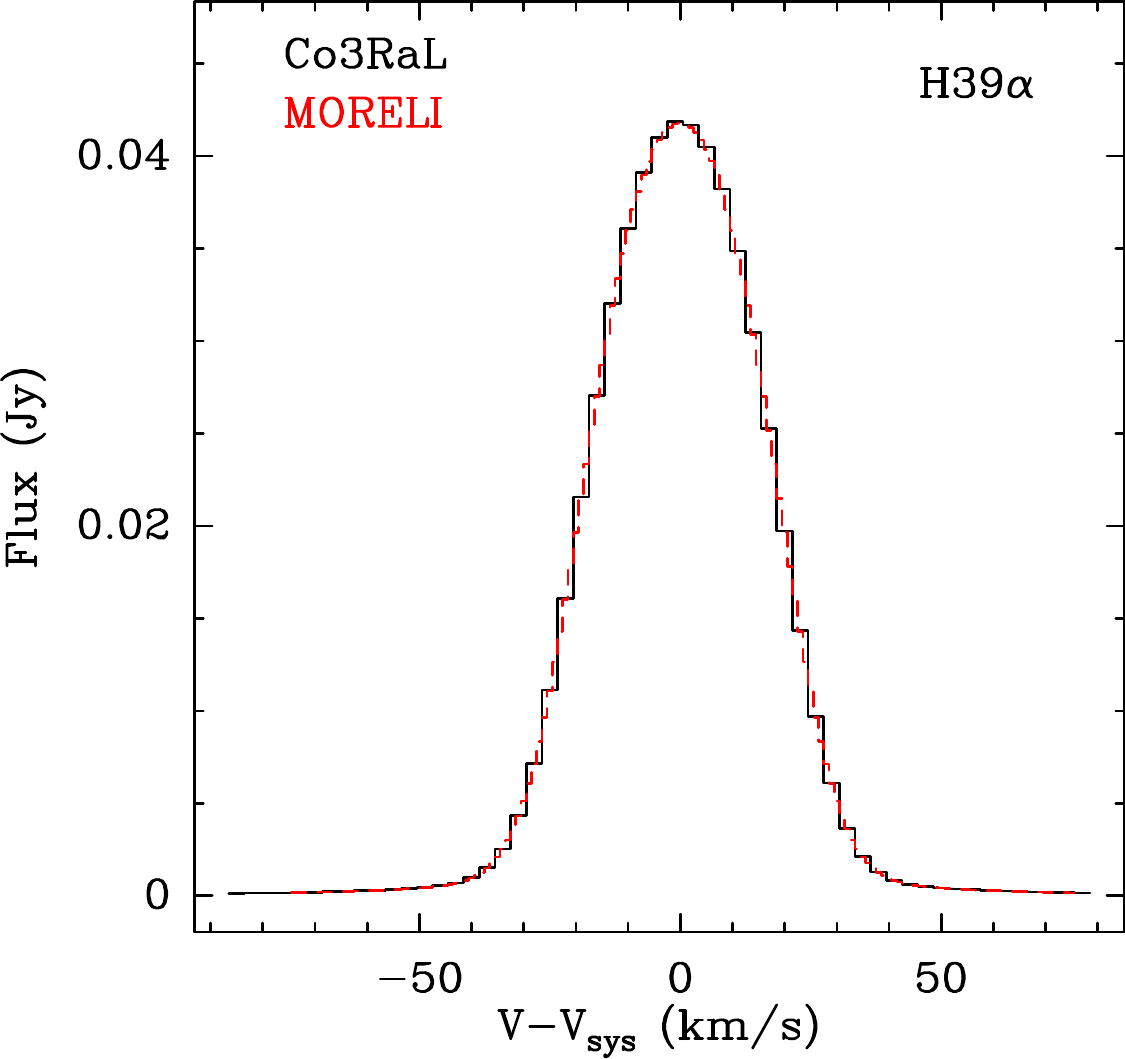} \\ 
%crl618: 
     \includegraphics*[width=0.395\hsize]{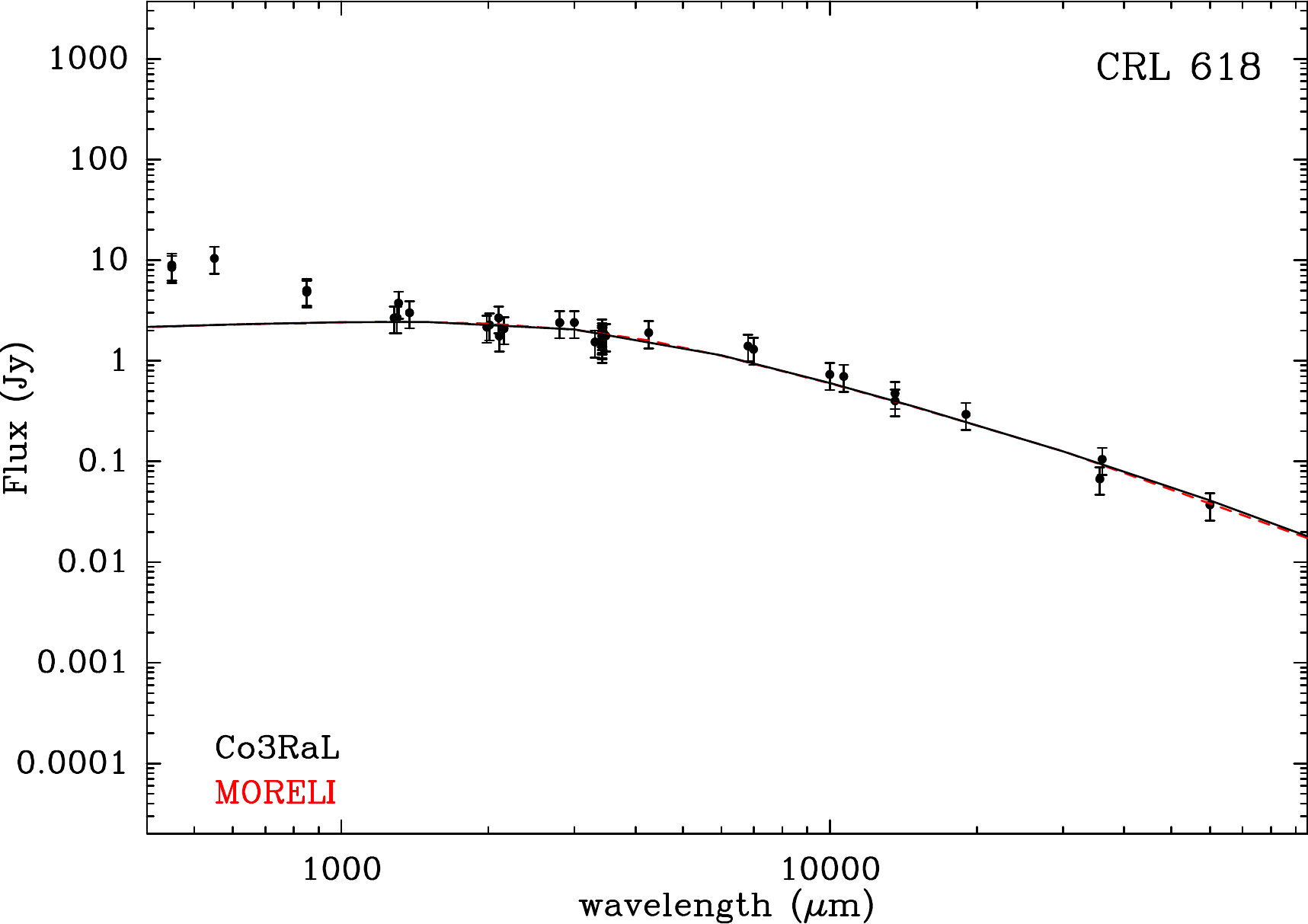}
      \includegraphics*[width=0.2925\hsize]{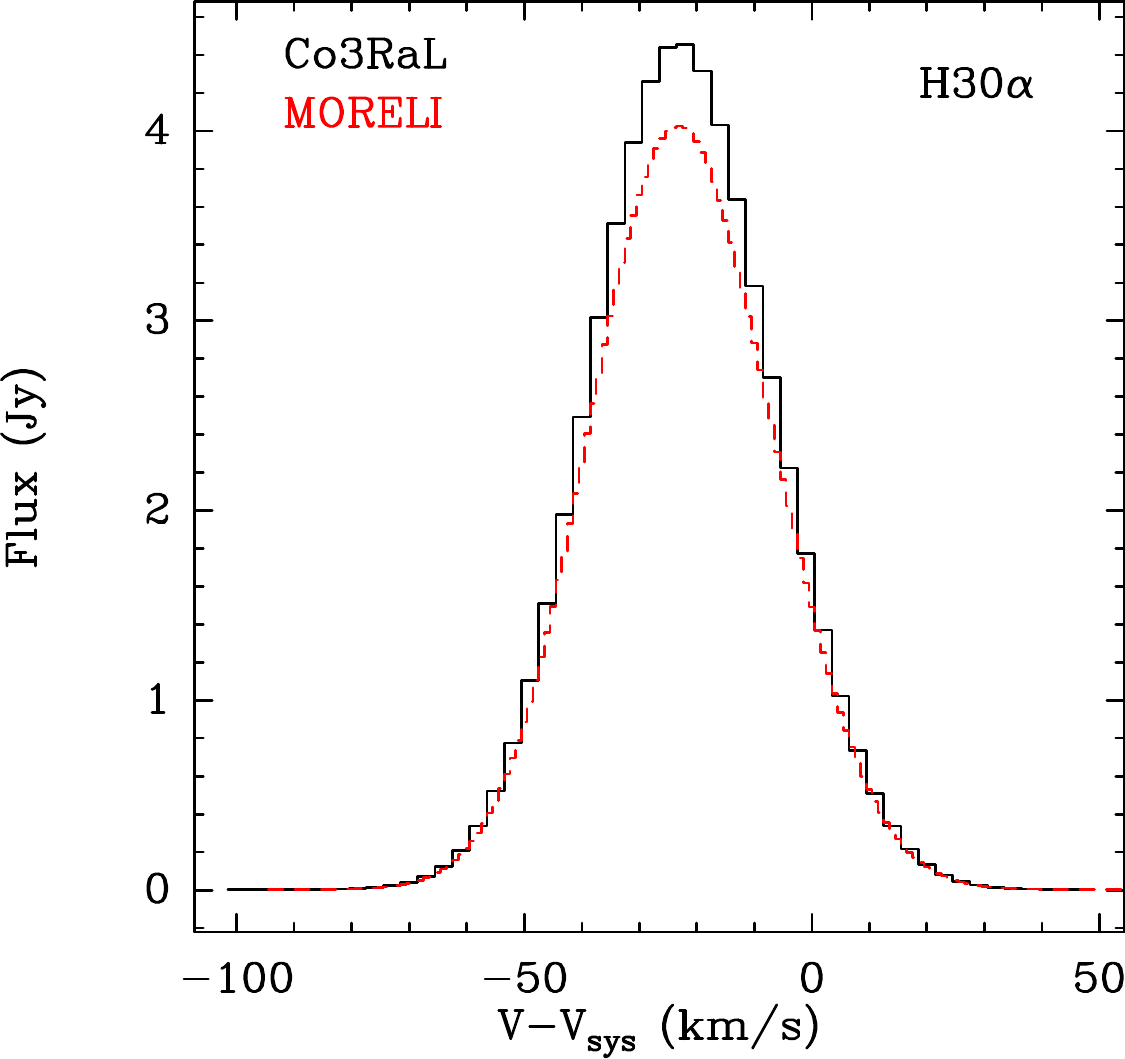}
      \includegraphics*[width=0.2925\hsize]{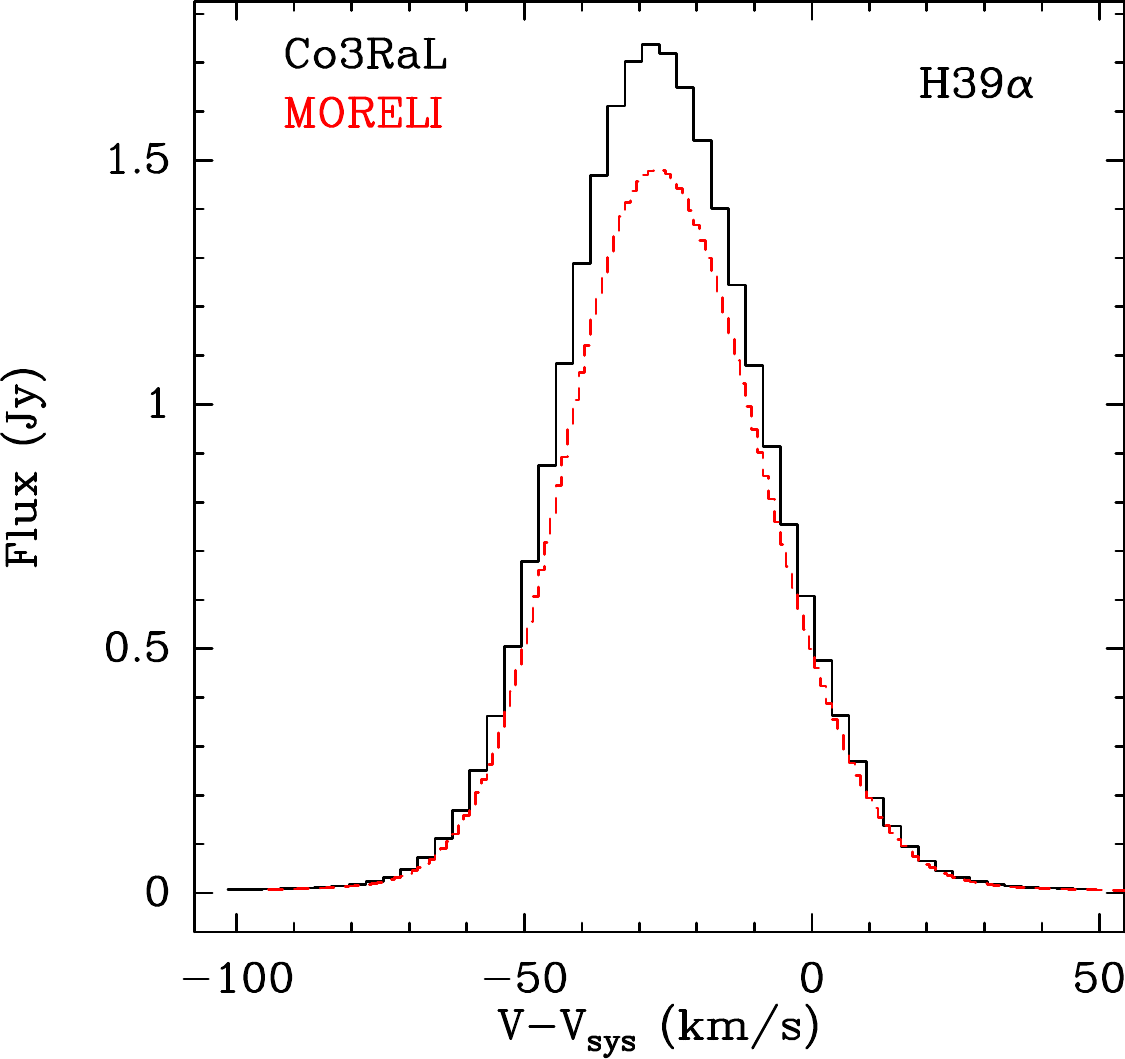}
     \caption{Comparison between the predictions by \coral\ and MORELI
       for the input model of M\,2-9 (ETL, top) and CRL\,618 (non-ETL, bottom) presented in CSC17 that were designed to reproduce
       the \iram\ single-dish data reported by these authors -- see
       Sect.\,\ref{comp}.  Left) Synthetic free-free continuum
       emission spectrum predicted by \coral\ (black solid line)
       and MORELI (red dashed line).  Data points (circles), shown as
       a reference, are as in Fig.\,\ref{f-cont} for M\,2-9 and as in CSC17 for CRL\,618.  Middle and Right)
       Synthetic \htal\ and \htnal\ 1d spectra
       (integrated over the emitting region) predicted by \coral\
       (solid black histogram) and MORELI (dashed red
       histogram).
\label{f:coral_moreli}}
   \end{figure*}

As shown in Fig.\,\ref{f:coral_moreli}, the predictions for the
free-free continuum flux at (sub)mm-to-cm wavelengths, along with the
mRRLs \htal\ and \htnal, for \coral\ and MORELI exhibit a high level of
consistency. The continuum fluxes predicted by MORELI and \coral\ are
indeed almost identical, with differences falling below 1\%-2\%
precision for both objects.
%% ---- 
%: the slightly lower cm-continuum flux
%predicted by MORELI compared with \coral\ is probably attributable to
%the truncation of the ionized region in MORELI at the so-called
%effective radius, which is at a distance a few times larger than the
%radius that encircles $\sim$90\% of the observed mm-continuum
%fluxes.
%% --- 
Under LTE (M\,2-9), the mRRLs line profiles are also
identical. For non-LTE (CRL\,618), \coral\ predicts slightly (10\%--15\%)
higher peak intensities for both lines, resulting in a small
difference in the best-fit electron temperature of \te$^{\rm CoRaL}$--\te$^{\rm MORELI}$$\lsim$700\,K ($<$4\%) when derived
using the two codes on the same data set, while keeping the rest of
CSC17's model parameters the same. Without access to the original MORELI
source code, it is impossible to determine the exact reason for this
small discrepancy. The discrepancy could be related to the various
approximations and interpolations made when estimating the $b_n$ departure
coefficients (available in the literature only for a discrete grid of
electron density and temperature pairs), pressure broadening
coefficients, integration of the Voigt profile, etc (see Appendix \ref{ap-co3ral} for details).

%% ========================== Fig. Co3RaL vs MORELI
   %%MACROS:
   %% --- M2-9: 
   %% /pcdisk/jbell3/csanchez/m2-9/MODELO/hole3006/:   @sedm29_co3ral_moreli.greg
   %% /pcdisk/jbell3/csanchez/m2-9/MODELO/hole300/:  @lines_co3ral_moreli.greg
   %% /pcdisk/jbell3/csanchez/m2-9/continuum/sedm29_co3ral_moreli.greg
   % ---- CRL618: 
   %% /pcdisk/jbell3/csanchez/m2-9/MODELO/c618-modelo-csc17/ @sedcrl618_co3ral_moreli.greg yes yes
   %% ========================== Fig. Co3RaL vs MORELI 

   %% tambien ponemos el cont. y pv de hal para este modelo, para demostrar que NO es consistente con los mapas ALMA.

\section{Comparison of CSC17's model for M\,2-9 with new ALMA maps from this work}
\label{ap:csc17}

\begin{figure*}[htbp!]
     \centering
     \includegraphics*[width=0.30\hsize]{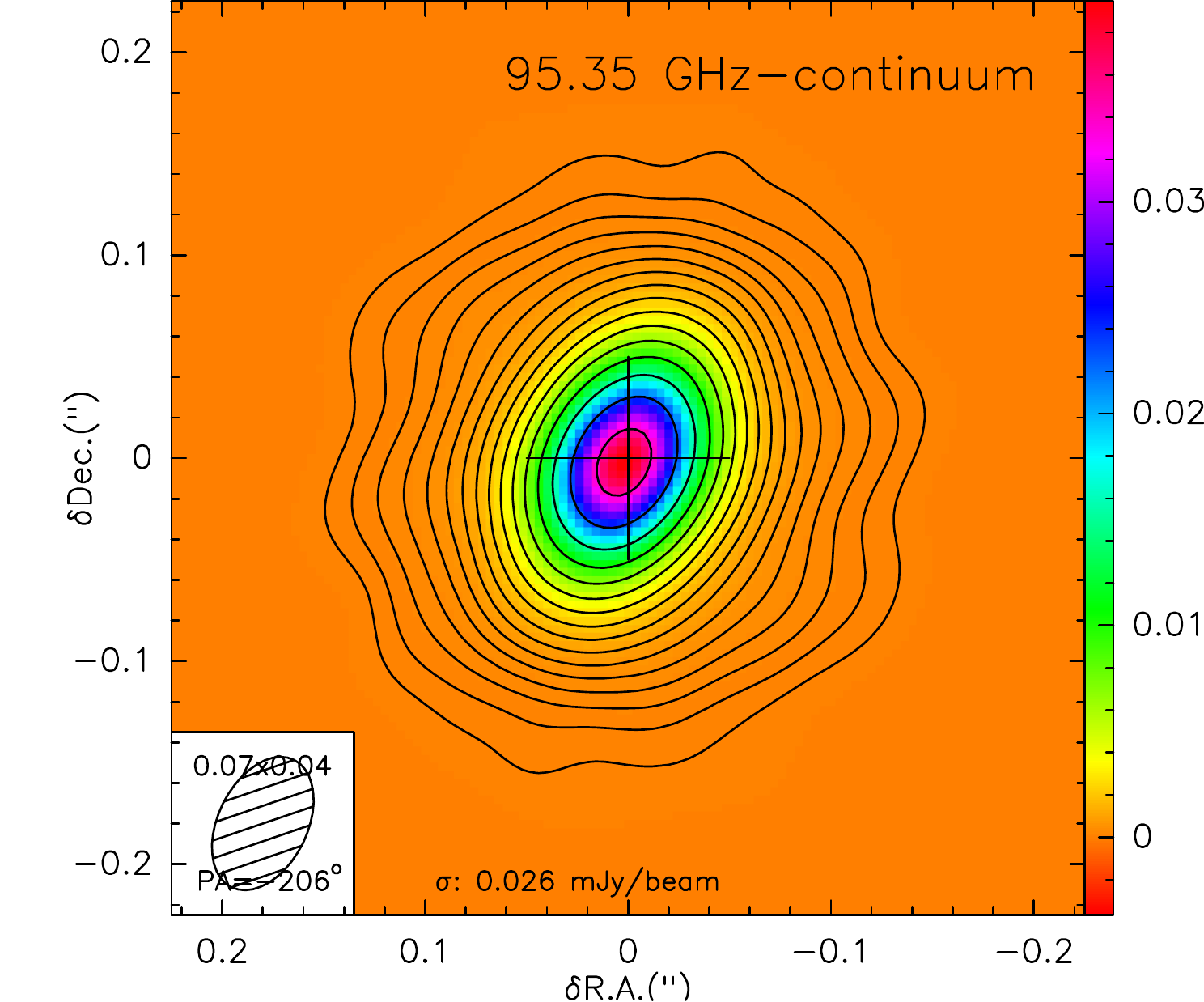}
     \includegraphics*[width=0.345\hsize]{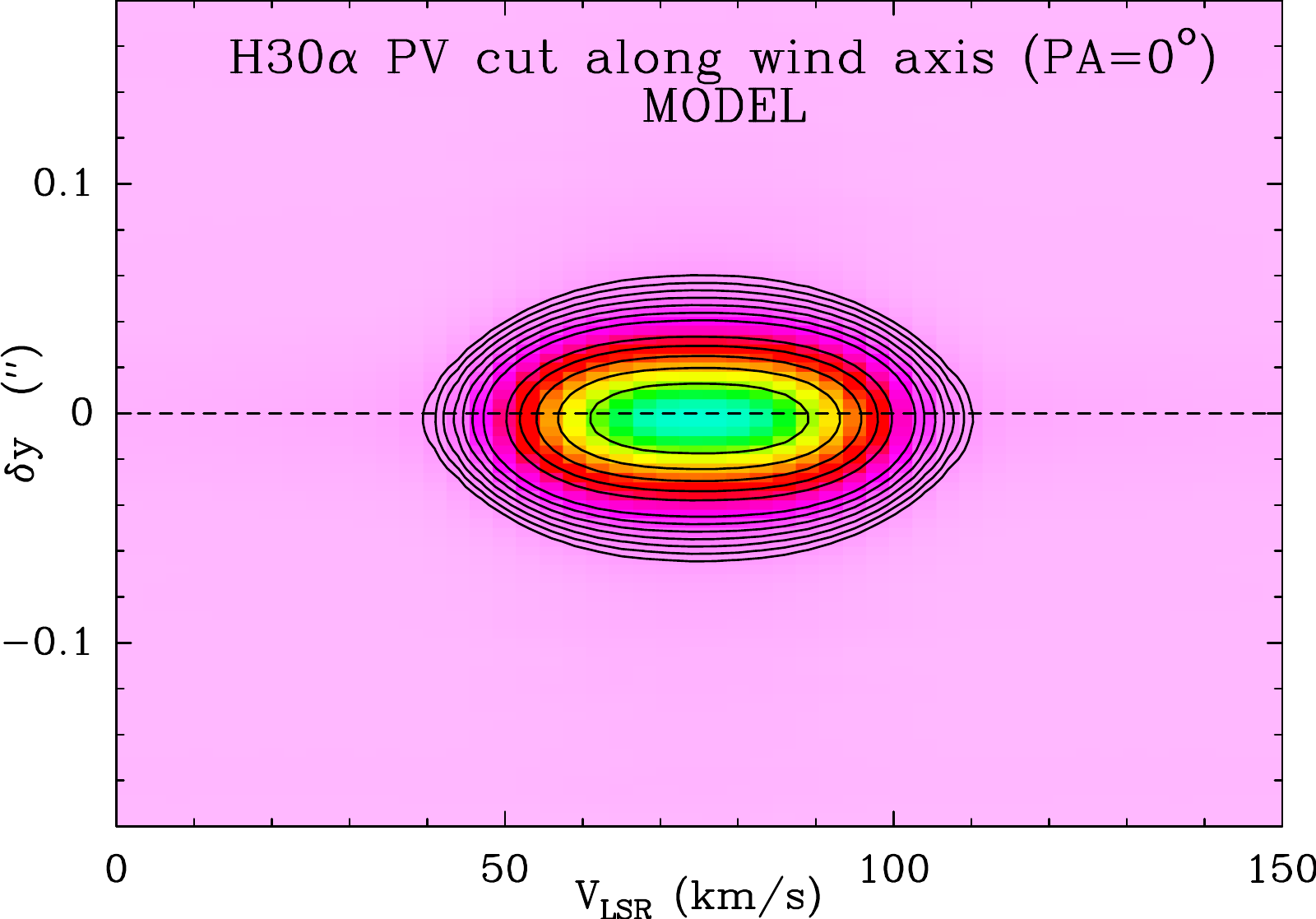}
     \includegraphics*[width=0.345\hsize]{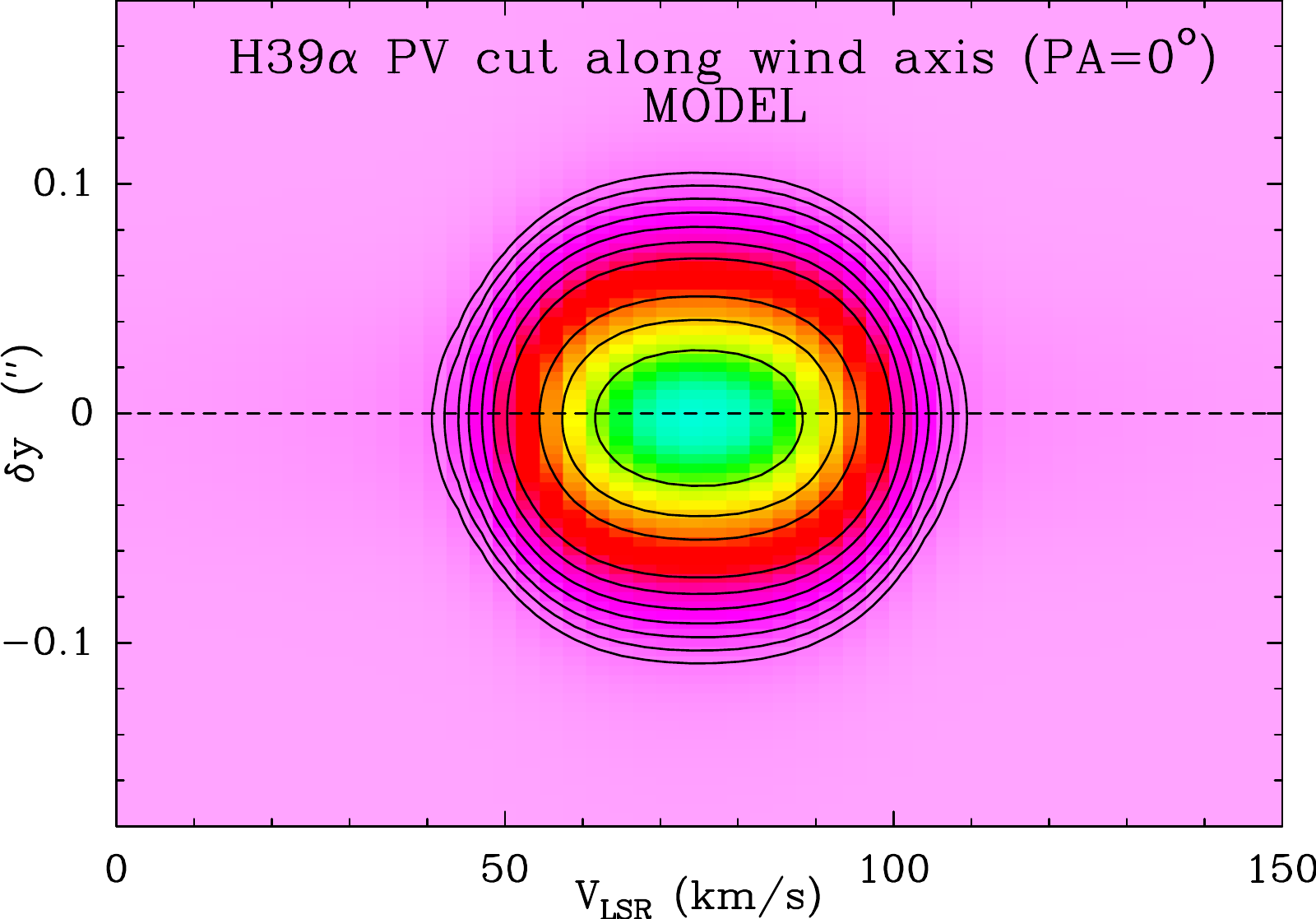}
     \caption{Synthetic ALMA-like cubes of the free-free continuum emission at
       93\,GHz (left) and the \htal\ and \htnal\ lines (middle and
       right, respectively) obtained from the M\,2-9 model in
       CSC17 --- see their Table\,4 --- to be compared with
       analogous datasets obtained with ALMA in Fig.\,\ref{f-cont}
       (top-right) and Fig.\,\ref{f-pv} (top panels). 
       \label{f:csc17}}
   \end{figure*}

   In Fig. \ref{f:csc17}, we show synthetic ALMA-like cubes
   representing emission from 93\,GHz free-free continuum and
   \htal\ and \htnal\ lines, based on the model of M\,2-9 derived from 
   single-dish observations by CSC17. While this
   model accurately reproduces the free-free (mm-to-cm)
   continuum flux and the single-dish line profiles, the corresponding 
   synthetic ALMA images reveal notable differences from the actual ALMA data (Figs.\,\ref{f-cont} and
   \ref{f-pv}). These differences suggest the presence of a collimated wind or jet
   that is narrower than initially assumed and that exhibits a certain C-shaped
   curvature as well as a significant velocity gradient along its
   axis. In Section\,\ref{model}, we introduce an updated model that
   replicates these and other features observed in our ALMA datasets, which has a resolution of a few tens of
   milliarcsecond.

%% ========================== Fig. Co3RaL vs MORELI
   %% MACROS:
   %% /pcdisk/jbell3/csanchez/m2-9/MODELO/hole300/:   @sedm29_co3ral_moreli.greg @lines_co3ral_moreli.greg
   %% /pcdisk/jbell3/csanchez/m2-9/continuum/sedm29_co3ral_moreli.greg

%%   
%% ========================== Fig. Co3RaL vs MORELI 

   \section{Model caveats and final modeling remarks}
\label{ap:caveats}

   It is important to emphasize that the relatively simple model
   presented in Sect.\,\ref{model} is probably not unique and that more
   intricate geometries, kinematics, and density and temperature
   profiles (e.g.\ latitude-dependent) cannot be dismissed as
   possibilities. Although a comprehensive parameter space study has
   not been performed and is beyond the scope of this paper, we are
   confident that the bulk features deduced from the model-- such as the
   presence of dense, high-velocity shell-like regions within a
   narrow-waisted bipolar (horn-like) wind-- are robust and necessary to
   reproduce the main features inferred directly from the data. The
   model's primary role is to reinforce these features and provide a
   more precise characterization of the physical properties in the
   wind. Below, we describe the uncertainties associated with some
   parameters, but we stress that these do not affect our overall
   conclusions, as we have been careful not to overinterpret minor
   details. Instead, we focused on bulk parameters that control the
   main features, whose effects are clearly distinguishable in the
   synthetic maps. While some minor parameters may exhibit hidden
   interdependencies, their impact on the overall scientific
   conclusions derived in our study is very small.

   Uncertainties in the absolute densities and temperatures deduced
   are also moderate, within a factor $\sim$2--3 and 15--20\%,
   respectively. The position and width of the HVSs are accurate down
   to $\sim$2--3\,au. The radius of the wind at its base (waist),
   \req$\sim$7--8\,au, is relatively well constrained by the extent of
   the continuum maps at the center. However, the size of the nucleus,
   its true geometry, and accurate density distribution remain
   unknown. For simplicity, we assume a spherical region with an
   intermediate value for the radius of 4\,au and an average density
   of a few\ex{8}\,\cm3. However, larger\footnote{Up to a maximum
   value of \req$\sim$7--8\,au} (or smaller) dimensions together with
   smaller (or larger) average density are also possible.  The
   systemic velocity of the source is probably in the range
   75--78\,\kms.

%% A cavity inside the nucleus, as in model CSC17, may exist but it is not neccessary-

One shortcoming of our model is its failure to predict the asymmetry
in the \htal\ line profile, particularly noticeable in the axial
position-velocity diagram, where the blue-shifted emission wing from
the south lobe is moderately more intense than the red-shifted one from the north
lobe. This asymmetry, also observed in the 1mm-continuum map, may be
partially attributed to an intrinsic asymmetry in the ionized wind
and/or to partial absorption of the free-free emission from the base
of the northern lobe that is behind the front side of the circumbinary
dust disk (see Sect.\,\ref{disk_co}). These effects are not considered in
our model.
%% La asimetria se ve mas a 1mm que a 3mm, pero la linea H39 tambien es algo asimetrica.
%% si se pone algo de asimetria en el modelo que haga la linea H30a y H39a algo mas intensa en el sur,
%% esto haria que 
Furthermore, the model predicts an \htnal\ line that is too wide and
has a slightly blueshifted centroid compared to the observations. This
issue results from a combination of pressure broadening, opacity, and
non-LTE effects, which could only be alleviated by reducing the
density below 1\ex{7}\,\cm3 in all regions,
including the central ones (nucleus and inner wind). However, we have
not found any satisfactory models with such low densities, as these
would require a much larger emitting volume, incompatible with the
extent of the ionized wind observed in the maps.

It is also crucial to recognize that the accuracy of the ultimate
(best-fit) model is influenced by the chosen values of critical
parameters that remain challenging to precisely quantify, such as
departure coefficients or approximated expressions for pressure
broadening, Gaunt factors, etc.
%% Gaunt factors
For example, departure coefficients $b_n$ indeed vary depending on the radiation field
for a specific assumed geometry and physical structure, which may
differ from those presumed in the models of \cite{sto95} used in this work.
%% Tambien estan los de  Prozesky & Smits (2018)
Furthermore, pressure broadening expressions used in this work,
commonly employed in analytical approaches \citep[Table B.1
  in][]{bae13}, posed modeling challenges at high densities
($\gsim$10$^8$\,cm$^{-3}$), where the line wings, notably those of
\htnal\footnote{Pressure broadening, i.e.\,broadening of the energy
levels due to collisions with electrons, increases as a power-law of
the level principal quantum number $n$.}, significantly exceed
observations in central regions (regions I and II) of
M\,2-9. Throughout the modeling process, efforts have been made to
address this discrepancy by reducing electron density (to alleviate
the effect) and enlarging the dimensions of the emitting region (to
maintain a consistent free-free continuum flux). This was done to
simultaneously reproduce narrower central wings in the \htnal\ line
and the observed dimensions of the emitting region. However, there is
a possibility of inaccuracies in the pressure broadening expressions,
which might have led to uncertainties in the density in these central
regions.

\end{document}